\documentclass[pdflatex,sn-mathphys-num]{sn-jnl}

\usepackage{graphicx}%
\usepackage{multirow}%
\usepackage{amsmath,amssymb,amsfonts}%
\usepackage{amsthm}%
\usepackage{mathrsfs}%
\usepackage[title]{appendix}%
\usepackage{xcolor}%
\usepackage{textcomp}%
\usepackage{manyfoot}%
\usepackage{booktabs}%
\usepackage{algorithm}%
\usepackage{algorithmicx}%
\usepackage{algpseudocode}%
\usepackage{listings}%

\usepackage[T1]{fontenc}
\usepackage{xurl}
\usepackage{hyperref}
\usepackage{breakurl}

\usepackage{csquotes}

\usepackage{anyfontsize}

\catcode`\|=12\relax
\usepackage{orcidlink}

\usepackage{tabularx}
\usepackage{diagbox}
\newcolumntype{C}{>{\centering\arraybackslash}X} 
\usepackage{hhline}

\usepackage{xltabular}
\usepackage{tabularray}

\usepackage{adjustbox}
\usepackage{array}
\newcolumntype{R}[2]{%
    >{\adjustbox{angle=#1,lap=\width-(#2)}\bgroup}%
    l%
    <{\egroup}%
}
\newcommand*\rot{\multicolumn{1}{R{45}{1em}}}

\usepackage{fontawesome5}

\usepackage{threeparttable}

\makeatletter
\newcommand*{\citenums}[2][]{%
  \def\NAT@cmprs{\z@}
  \begingroup
  \renewcommand\NAT@open{}
  \renewcommand\NAT@close{}
  \cite[#1]{#2}%
  \endgroup
}
\makeatother

\usepackage{tikz}
\usepackage{tikz-qtree}
\usetikzlibrary{mindmap,trees,shapes.geometric,positioning,arrows.meta,bending}

\usepackage{mathtools}

\usepackage{enumitem}
\newlist{dimensions}{enumerate}{1}
\setlist[dimensions]{%
  label=\textbf{\arabic*.},
  ref={\arabic*},
  wide,
  listparindent=-\labelwidth,
  topsep=1.0em, 
  itemsep=1.0em,
  labelindent=0pt
}
\NewDocumentCommand{\enumdimension}{ m }{\item \textbf{#1}}

\usepackage[printonlyused, nolist]{acronym}

\chardef\&=`&

\definecolor{JP-red}{RGB}{227,110,93}
\definecolor{JP-red_50p}{RGB}{241,182,174}
\definecolor{JP-orange}{RGB}{242,159,92}
\definecolor{JP-orange_50p}{RGB}{248,207,173}
\definecolor{JP-yellow}{RGB}{244,224,123}
\definecolor{JP-yellow_50p}{RGB}{249,239,189}
\definecolor{JP-green}{RGB}{143,200,142}
\definecolor{JP-turquoise}{RGB}{88,195,184}
\definecolor{JP-light_blue}{RGB}{100,177,208}
\definecolor{JP-blue}{RGB}{58,124,185}
\definecolor{JP-purple}{RGB}{175,128,190}
\definecolor{JP-purple_50p}{RGB}{215,191,222}
\definecolor{JP-lighter_grey}{RGB}{153,153,153}
\definecolor{JP-light_grey}{RGB}{119,119,119}
\definecolor{JP-grey}{RGB}{85,85,85}
\definecolor{JP-dark_gray}{RGB}{51,51,51}
\definecolor{JP-darker_gray}{RGB}{17,17,17}

\newcommand{\tikzcircle}[2][black]{%
    \tikz[baseline=-0.6ex] \node[circle, draw, #1, fill=#1, inner sep=0pt, minimum size=0.8em]{};
}
\newcommand{\tikzoctagon}[2][black]{%
    \tikz[baseline=-0.6ex] \node[regular polygon, regular polygon sides=8, draw, #1, fill=#1, inner sep=0pt, minimum size=0.85em]{};
}

\DeclareMathOperator*{\argmin}{argmin}
\DeclareMathOperator*{\argmax}{argmax}

\begin{document}


\title[Emergent Language: A Survey and Taxonomy]{Emergent Language: A Survey and Taxonomy}


\author*[1]{\fnm{Jannik} \sur{Peters}\,\orcidlink{0000-0003-0822-7202}}\email{jpeters@uni-wuppertal.de}

\author[1]{\fnm{Constantin} \sur{Waubert~de~Puiseau}\,\orcidlink{0000-0001-7764-1322}}\email{waubert@uni-wuppertal.de}

\author[1]{\fnm{Hasan} \sur{Tercan}\,\orcidlink{0000-0003-0080-6285}}\email{tercan@uni-wuppertal.de}

\author[2]{\fnm{Arya} \sur{Gopikrishnan}\,\orcidlink{0009-0001-5359-3546}}\email{ag3974@drexel.edu}
\equalcont{Work done during and after a DAAD RISE internship at Institute of Technologies and Management of Digital Transformation.}

\author[3]{\fnm{Gustavo Adolpho} \sur{Lucas De Carvalho}\,\orcidlink{0009-0000-4728-1231}}\email{lucasdec@usc.edu}
\equalcont{Work done during and after a DAAD RISE internship at Institute of Technologies and Management of Digital Transformation.}

\author[1]{\fnm{Christian} \sur{Bitter}\,\orcidlink{0000-0001-8221-2792}}\email{bitter@uni-wuppertal.de}

\author[1]{\fnm{Tobias} \sur{Meisen}\,\orcidlink{0000-0002-1969-559X}}\email{meisen@uni-wuppertal.de}

\affil[1]{\orgdiv{University of Wuppertal}, \orgname{Institute of Technologies and Management of the Digital Transformation}, \orgaddress{\street{Rainer-Gruenter-Str.~21}, \city{Wuppertal}, \postcode{42119}, \state{NRW}, \country{Germany}}}

\affil[2]{\orgdiv{Drexel University}, \orgname{College of Engineering}, \orgaddress{\street{3141 Chestnut St}, \city{Philadelphia}, \postcode{19104}, \state{PA}, \country{USA}}}

\affil[3]{\orgdiv{University of Southern California}, \orgname{Department of Computer Science}, \orgaddress{\street{941 Bloom Walk}, \city{Los Angeles}, \postcode{90089}, \state{CA}, \country{USA}}}

\newcommand{\litSearchDate}{17\textsuperscript{th} of June 2024}
\newcommand{\litCTotal}{613} 
\newcommand{\litCUnique}{516} 
\newcommand{\litCAdditional}{23} 
\newcommand{\litCCorpus}{539} 
\newcommand{\litSsciencedirect}{5}
\newcommand{\litSieeeexplore}{9}
\newcommand{\litSacmEC}{60}
\newcommand{\litSacmEL}{19}
\newcommand{\litSWOS}{16}
\newcommand{\litSarxivEC}{207}
\newcommand{\litSarxivEL}{66}
\newcommand{\litSarxivEMAC}{208}
\newcommand{\litSsemanticscholar}{23}
\newcommand{\litRsortedout}{327}
\newcommand{\litRremaining}{212} 
\newcommand{\litRpartiallyrelevant}{31}
\newcommand{\litRrelevant}{181}

\abstract{The field of emergent language represents a novel area of research within the domain of artificial intelligence, particularly within the context of multi-agent reinforcement learning. Although the concept of studying language emergence is not new, early approaches were primarily concerned with explaining human language formation, with little consideration given to its potential utility for artificial agents. In contrast, studies based on reinforcement learning aim to develop communicative capabilities in agents that are comparable to or even superior to human language. Thus, they extend beyond the learned statistical representations that are common in natural language processing research. This gives rise to a number of fundamental questions, from the prerequisites for language emergence to the criteria for measuring its success. This paper addresses these questions by providing a comprehensive review of \litRrelevant{} scientific publications on emergent language in artificial intelligence. Its objective is to serve as a reference for researchers interested in or proficient in the field. Consequently, the main contributions are the definition and overview of the prevailing terminology, the analysis of existing evaluation methods and metrics, and the description of the identified research gaps.}

\keywords{emergent language, emergent communication, artificial intelligence, reinforcement learning, multi-agent}

\maketitle

\begin{acronym}[MARL] 
\acro{el}[EL]{emergent language}
\acro{ec}[EC]{emergent communication}
\acro{nl}[NL]{natural language}
\acro{nlp}[NLP]{natural language processing}
\acro{rl}[RL]{reinforcement learning}
\acro{marl}[MARL]{multi-agent reinforcement learning}
\acro{llm}[LLM]{large language model}
\acro{hci}[HCI]{human-computer interaction}
\end{acronym}

\section{Introduction}\label{sec_introduction}
Communication between individual entities is based on conventions and rules that emerge from the necessity or advantage of coordination. Accordingly, Lewis~\cite{lewis.1969} formalized settings that facilitate the emergence of language as \enquote{coordination problems}~\cite{lewis.1969} and introduced a simple signaling game. This game, in which a speaker describes an object and a listener confronted with multiple options has to identify the indicated one, extensively shaped the field of \ac{el} research in computer science. Early works examined narrowly defined questions regarding the characteristics of \ac{ec} via hand-crafted simulations~\cite{wagner.2003, steels.1997, nowak.1999, kirby.2002, cangelosi.2002, christiansen.2003, batali.1998, oliphant.1997, steels.1995, skyrms.2002, smith.2003}. These approaches mostly utilized supervised learning methods and non-situated settings, limiting them in their ability to examine the origins and development of complex linguistic features~\cite{wagner.2003}. However, \ac{el} research experienced an upsurge in the period between 2016 and 2018~\cite{foerster.2016, lazaridou.2016, havrylov.2017, bouchacourt.2018, cao.2018, mordatch.2017, das.2017, sukhbaatar.2016} with a focus on \ac{marl} approaches~\cite{agarwal.2019, blumenkamp.2020, brandizzi.2021, brandizzi.2022, chaabouni.2022, gupta.2021, karten.2022, lo.2022, lowe.2019, vanneste.2022b, verma.2021, yu.2022} to enable the examination of more complex features.

One fundamental goal of \ac{el} research from the \ac{marl} perspective is to have agents autonomously develop a communication form that allows not only agent-to-agent but also agent-to-human communication in \ac{nl} style fashion~\cite{brandizzi.2022, lowe.2019, bogin.2018, bouchacourt.2018, wagner.2003, noukhovitch.2021}. Therefore, \ac{rl} methods are attractive from two points of view. First, successful communication settings might lead to agents that are \enquote{more flexible and useful in everyday life}~\cite{lazaridou.2020}. Furthermore, they may provide insights into the evolution of \ac{nl} itself~\cite{galke.2022}. 
However, encouraging communication alone will not automatically produce a language with natural language characteristics~\cite{mu.2021}. Providing the right incentives for language development is therefore crucial.

\ac{el} is the methodological attempt to enable agents to not only statistically understand and use \ac{nl}, like \ac{nlp} models that learn on text alone~\cite{steinertthrelkeld.2022, bender.2020}, but rather to design, acquire, develop, and learn their own language~\cite{lemon.2022, browning.2022}. The autonomy and independent active experience of \ac{rl} learning settings is a crucial difference to the data-driven approaches in the field of \ac{nlp}~\cite{manning.2005, qiu.2020, wolf.2020} and its \acp{llm}. According to Browning and LeCun, \enquote{we should not confuse the shallow understanding \acp{llm} possess for the deep understanding humans acquire}~\cite{browning.2022} through their experiences in life. In \ac{el} settings, the agents experience the benefits of communication through goal-oriented tasks~\cite{vaneecke.2020} just like it happens naturally~\cite{lewis.1969} and therefore have the opportunity to develop a deeper understanding of the world~\cite{keresztury.2020, bogin.2018}. Hence, advances in \ac{el} research enable novel applications of multi-agent systems and a considerably advanced form of human-centric AI~\cite{lazaridou.2020}.

In the current state of \ac{el} research, numerous different methods and metrics are already established but they are complex to structure and important issues remain regarding the analysis and comparison of achieved results~\cite{lazaridou.2020, hernandezleal.2019, lowe.2019}. Therefore, we see a need for a taxonomy to prevent misunderstandings and incorrect use of established metrics. In this paper, we address these issues by providing a comprehensive overview of publications in \ac{el} research and by introducing a taxonomy for discrete \ac{el} that encompasses key concepts and terminologies of this field. Additionally, we present established and recent metrics for discrete \ac{el} categorized according to the taxonomy and discuss their utility. Our goal is to provide a clear and concise description that researchers can use as a shared resource for guidance. Finally, we create a summary of \ac{el} research that highlights its achievements and provides an outlook on future research directions.
We base our work on a comprehensive and systematic literature search with reproducible search terms on well-known databases. We follow the PRISMA~\cite{page.2021} specifications and show a corresponding flow diagram in Figure~\ref{fig_PRISMA_flowchart} in Appendix~\ref{sec:appendix_b}. The literature search and review process as well as its results are described in detail in Section~\ref{sec_studymethodology}. All identified work has been reviewed and categorized according to an extensive list of specific characteristics, e.g. regarding communication setting, game composition, environment configuration, language design, language metrics, and more.

Previous surveys of \ac{el} in computer science focused only on a subgroup of characteristics or specific parts of this research area.
Some of these earlier surveys focus on specific learning settings~\cite{vaneecke.2020, lipowska.2022, denamganai.2020}, on methodological summaries and criticism~\cite{korbak.2020, lacroix.2019, lemon.2022, lowe.2019, mihai.2021, galke.2024, vanneste.2022}, or provide a more general overview~\cite{hernandezleal.2019, brandizzi.2022, moulinfrier.2020, galke.2022, fernando.2020, lazaridou.2020, brandizzi.2023b}. The most similar ones to our work are~\cite{lazaridou.2020} and~\cite{brandizzi.2023b}. \cite{lazaridou.2020}~gives an introduction and overview of the \ac{el} field before 2021, however, it is mostly a summary of previous work and does not provide a taxonomy or review of existing metrics in the field as we do. \cite{brandizzi.2023b}~focuses on common characteristics in \ac{ec} research and the development of emergent human-machine communication strategies. They discuss distinctions and connections of \ac{ec} research to linguistics, cognitive science, computer science, and sociology, while we focus on emergent language and its analysis. We describe and discuss all relevant surveys in more detail in Section~\ref{sec_relatedsurveys}.

Based on this preliminary work, the current state of research on \ac{el} misses an overarching review and a comprehensive compilation and alignment of proposed quantification and comparability methods. Accordingly, the key contributions of the present survey are:

\begin{itemize}
\item A taxonomy of the \ac{el} field, in particular regarding the properties of discrete \ac{el}, see Section~\ref{sec_taxonomyofemergentlanguage}.
\item A list of categorized quantification approaches and metrics in a consistent notation, see Section~\ref{sec_metrics}.
\item A summary of open questions and an outlook on potential future work, see Section~\ref{sec_map}.
\end{itemize}

In addition, we introduce the fundamental concepts of \ac{nl} and \ac{ec} that underlie our survey in Section~\ref{sec_background}. As mentioned, we provide a detailed summary of related surveys in Section~\ref{sec_relatedsurveys}. Section~\ref{sec_studymethodology} describes our study methodology, including the keywords and terms of our systematic literature search. Finally, Section~\ref{sec_discussion} offers a concluding discussion and final remarks.

\section{Background}\label{sec_background}
To contextualize the presented taxonomy and analysis, this section summarizes the key concepts of communication and linguistics and provides an overview of \ac{el} research.

\subsection{Communication}\label{sec_background_communication}
\begin{figure}[tbp]%
\centering
\includegraphics[width=0.9\textwidth]{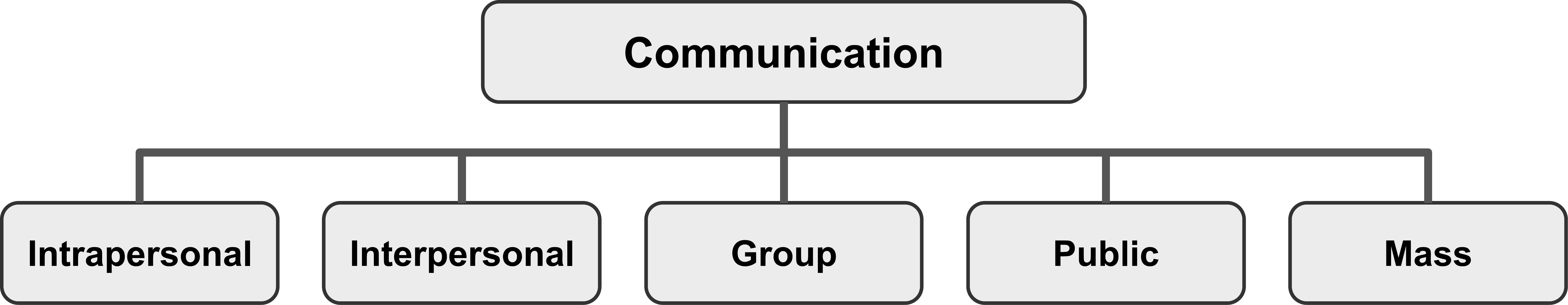}
\caption{The different forms of communication. They are divided by type of recipients and purpose. \emph{Intrapersonal communication} encompasses self-centered communication like internal vocalization. The remaining forms of communication are directed externally and are utilized to transmit information to individuals, in the \emph{interpersonal setting}, or groups of addressees. In \emph{group communication} the participants usually have a common goal, whereas \emph{public communication} focuses on the general transfer of information to a group of interested but not necessarily goal-aligned entities. Finally, \emph{mass communication} is used to describe any form of communication that is directed towards a general audience and focuses availability, for example, through the use of various media, including the internet. Adapted from~\protect\cite{jones.2018}.}
\label{fig_forms_of_communication}
\end{figure}

Communication at its very basis is the transfer or exchange of signals, which can be interpreted to form some information. These signals include both intended, such as deliberate utterances, and unintended, such as uncontrolled bodily reactions, and include both explicit and implicit parts~\cite{carston.2009}. According to Watzlawik's \enquote{Interactional View}~\cite{watzlawick.1967}, \enquote{one cannot not communicate}. In this regard, communication is ubiquitous and necessary, occurring through various channels and modes~\cite{andersen.1991, antos.2008, witt.2016, browning.2022, hartley.1993}. Depending on the specific channel and purpose, communication can be roughly divided into the five forms depicted in Figure~\ref{fig_forms_of_communication}. 

\begin{figure}[btp]%
\centering
\includegraphics[width=0.9\textwidth]{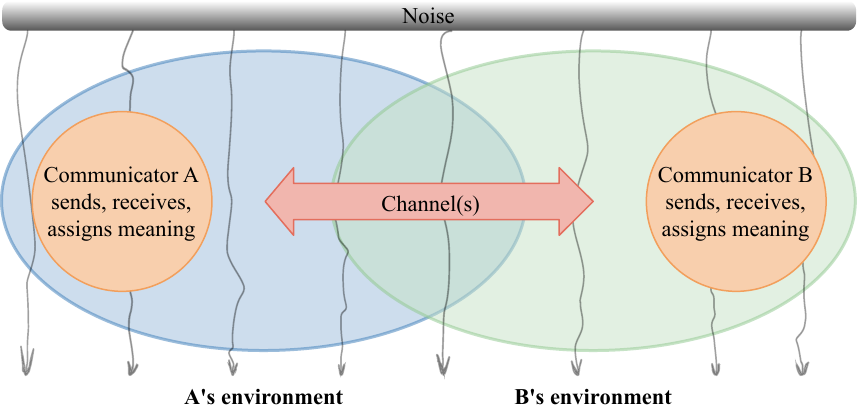}
\caption{Interpersonal communication. Actors are communicator A and B, each depicted by orange circles. They are each situated in their individual environment, depicted by the blue and green ellipses. At the overlap point the read arrow indicates the available communication channel. The potential environmental noise, influencing the communication, is represented by grey arrows going through the entire image. Adapted from~\protect\cite{adler.2012}.}
\label{fig_interpersonal_communication}
\end{figure}

In the context of \ac{el}, two of these forms are actively studied, namely interpersonal communication and group communication. Interpersonal communication is communication between entities that mutually influence each other, and its general setting is depicted in Figure~\ref{fig_interpersonal_communication}. This form of communication is based on individual entities, each within its perceivable environment. Although these environments are agent-specific, they overlap and allow communication through a common channel. In addition, there may be noise in this process that affects the perception of the environment or the communication itself.
Group communication, on the other hand, differs only in the number of entities involved and the communication goal. Usually, group communication is more formal and focuses on a common goal or group task while interpersonal communication has a social character and might only relate to a goal or task of one of the participants. Accordingly, the group communication setting can be found in most population-based \ac{el} research.
Intrapersonal communication (e.g., internal vocalization), public communication (e.g., lectures), and mass communication (e.g., blog entries) are not currently examined in the \ac{el} literature.

Communication has been studied in many different disciplines from many different perspectives, including animals~\cite{bossert.1963, sales.1974, rauschecker.2009}, pre-linguistic infants~\cite{tronick.1989, grosse.2010}, and sign language~\cite{stokoe.1980}. However, in order to keep the present work concise, we will refer mainly to research in the fields of linguistics and computer science.
Generally, communication can be seen as a utility to coordinate with others~\cite{choi.2018, austin.1975, clark.1996, wittgenstein.1989}. Conversely, the necessity for collaboration within a collective may be a fundamental precursor to the evolution and sustained functionality of explicit communication~\cite{nowak.1999, mihai.2021}. This theory leads to an essential differentiation regarding context-dependent communication. Meaningful communication might emerge in a cooperative but not in a fully competitive or manipulative setting. However, a partially competitive setting might be vital for the emergence of resilient and comprehensive communication, e.g. to enable the detection and use of lies~\cite{noukhovitch.2021}. Accordingly, the level of cooperation is a defining element of the communication setting in \ac{el} research.

\ac{nl} is a tool that allows us to encode very complex information within a discrete and humanly manageable amount of utterances. A lot of artificial intelligence research aims to develop \ac{nl} models, with applications ranging from translation to coherent full-text generation based on single-word input~\cite{wolf.2020, brown.2020, lauriola.2022, khurana.2022}. 
%
However, current research is mostly based on \acp{llm} which \enquote{achieve a sophisticated level of inductive learning and inference}~\cite{lappin.2023} but are also \enquote{far from human abilities in natural language inference, analogical reasoning, and interpretation}~\cite{lappin.2023}. A lot of research from the \ac{el} community is based on the theory, that models which learn language statistically based on static datasets are limited in their communicative and cognitive abilities~\cite{lazaridou.2021, browning.2022, merrill.2021, choi.2018, mordatch.2017}. Recent publications have shown that \acp{llm} have \enquote{weak reasoning and decision-making abilities}~\cite{liu.2024b}, their \enquote{reasoning is fragile}~\cite{mirzadeh.2024}, and that current \acp{llm} face \enquote{reliability issues}~\cite{zhou.2024c}. 
Correspondingly, the field of \ac{el} research in AI aims to enable agents to utilize intended communication in the same way humans use it to increase cooperation, performance, and generalization and, in the long run, enable direct meaningful communication between humans and artificial systems~\cite{mordatch.2017, noukhovitch.2021}. In line with this, multiple explicit forms of \ac{ec} in artificial intelligence research have been investigated as shown in Section~\ref{sec_taxonomy_phonology}. In contrast, work focusing on implicit communication, like the information content of spatial positioning of agents in a multi-agent setting~\cite{grupen.2021}, is not part of the present survey.

\subsection{Natural Language}\label{sec_background_naturallanguage}
\ac{nl} is a prime example of a versatile and comprehensive form of communication designed to convey meaning~\cite{dor.2014}. The flexibility of \ac{nl} allows humans to be exact but also deliberately ambiguous in their communication~\cite{pinker.1990}. It is a vital feature that distinguishes us from other species and gives us a great advantage in terms of knowledge storage, sharing, and acquisition~\cite{pinker.1990}. However, the origin and evolution of language is still a mystery~\cite{hauser.2014, hock.2019}. In the field of linguistics, many conflicting theories have been introduced so far~\cite{chomsky.1986, hock.2019, ney.1989, tomasello.2010, lakoff.2003}, ranging from behavioral to biological explanations. Additionally, accompanying research in the field of computer science has a long history~\cite{verma.2021} with a comparable range of theories. Even though there is still a debate around this topic, it is commonly agreed upon that a very intricate evolutionary process was involved~\cite{hock.2019, pinker.1990}. This evolution most likely took place in two different areas simultaneously, biologically and linguistically. On the biological side, the human brain most likely developed specific areas and functionalities specifically for more complex language-based communication, that are studied in the scientific field of neurolinguistics~\cite{locke.1997}. On the linguistics side, this evolution can be seen in language development itself, which is a constantly ongoing process~\cite{pinker.1990} that might be strongly connected to the development of cognitive skills~\cite{tomasello.2009} and the social environment~\cite{lakoff.1990}. Similarly, \ac{el} is concerned with the research of suitable model structures for the processing of language, while concurrently developing and evaluating language.

\begin{figure}[btp]%
\centering
\includegraphics[width=0.9\textwidth]{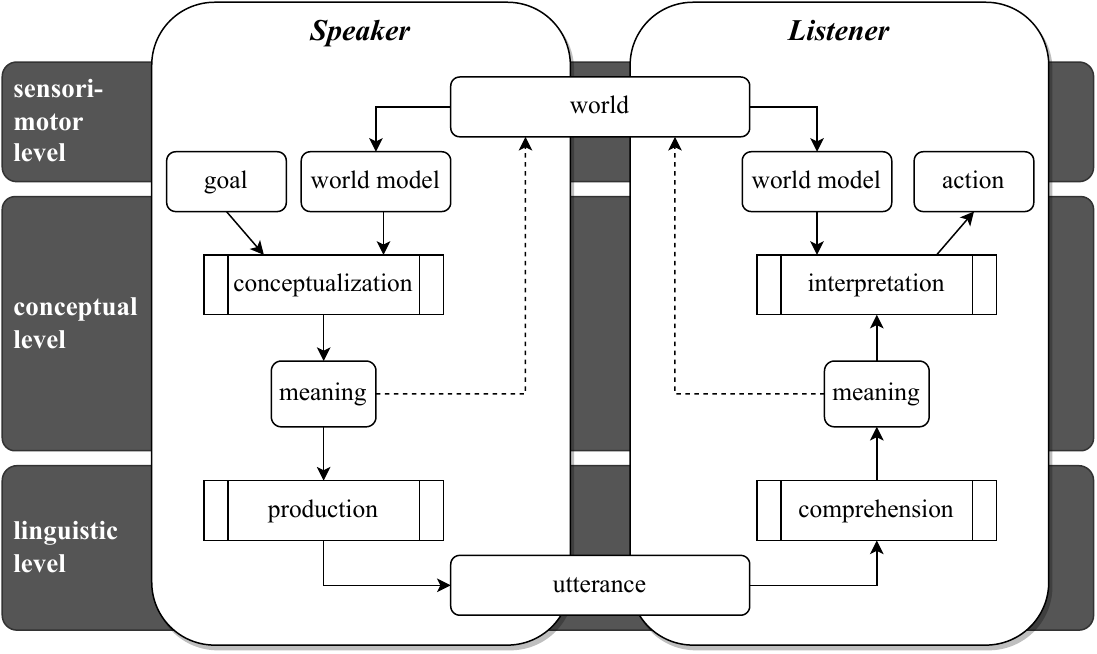}
\caption{The semiotic cycle. This framework of a language-based exchange between two separate entities, called speaker and listener, categorizes the process into three levels. The \emph{sensori-motor level} encompasses sensor and world-oriented components, the \emph{conceptual level} includes internal and intangible parts like the individual world model and conceptualization capabilities, and the \emph{linguistic level} consists of the production and comprehension of the linguistic exchange, which is the externalized connection between speaker and listener. Adapted from~\protect\cite{bleys.2015, vaneecke.2020}.}
\label{fig_semiotic_cycle}
\end{figure}

While the exact origin of language is highly debatable, the actual communication process via \ac{nl} is generally easier to conceptualize. For example, it can be modeled by the semiotic cycle depicted in Figure~\ref{fig_semiotic_cycle}~\cite{bleys.2015, vaneecke.2020}. This depiction applies to multiple expressive channels, e.g. speech and writing. 
It assumes at least two involved parties, a speaker and a listener. The speaker produces an utterance based on the meaning to be conveyed. This meaning results from the combined conceptualization of the speaker's goal and model of the world. On the other hand, the listener receives the utterance and comprehends it to derive a meaning, which is not a direct copy of the initial one by the speaker but it still refers to the shared world. The interpretation of the meaning, which the listener's world model informs, leads to some action by the listener.
At the center of this process are the shared world and the respective world models of speaker and listener that function as grounding for the information exchange via language. Further, both linguistic level components of production and comprehension allow the respective agent to participate in the language process.

\begin{figure}[btp]%
\centering
\includegraphics[width=0.6\textwidth]{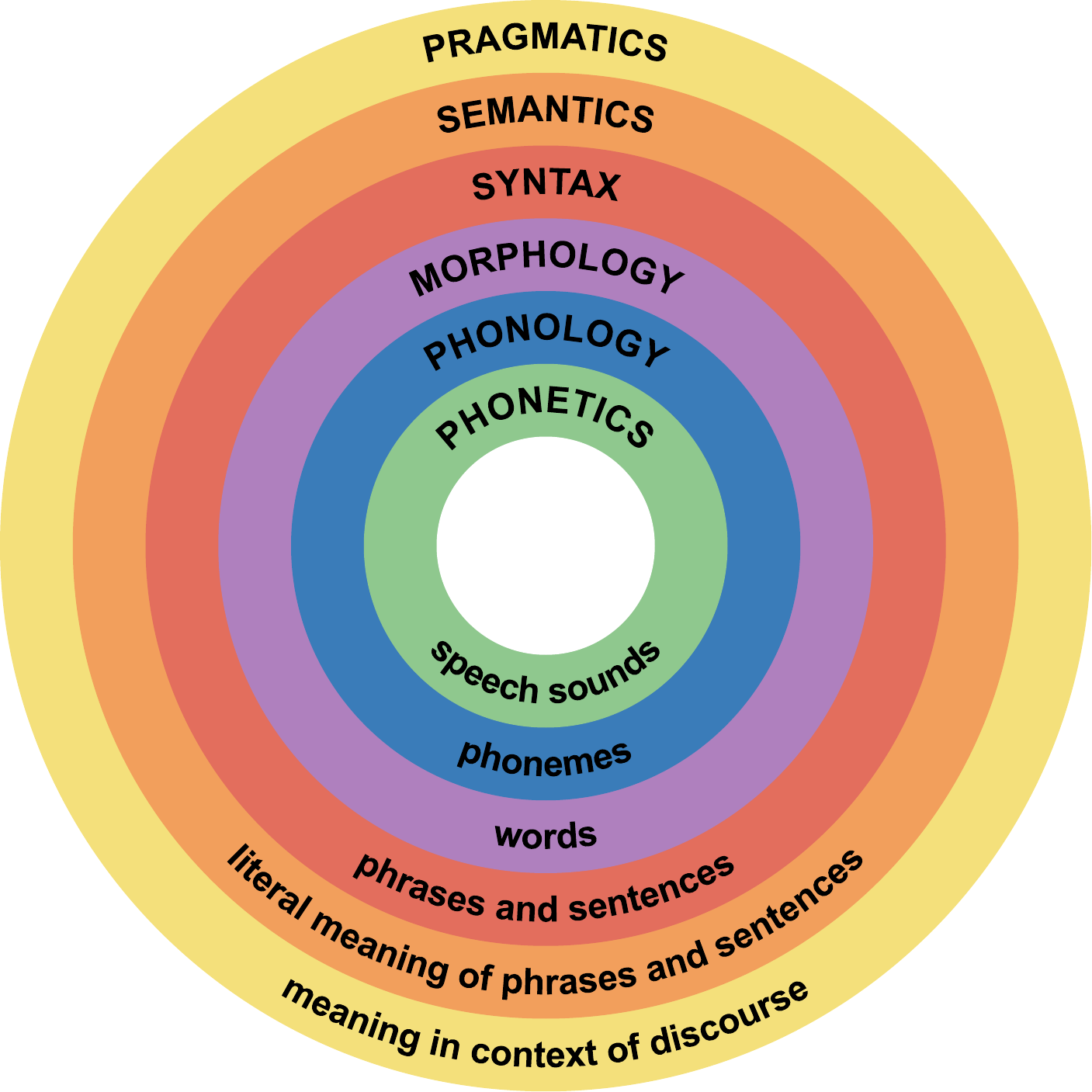}
\caption{Major levels of linguistic structure. This conceptual structure depicts the elements of a language as concentric rings that build upon each other. From the inner circle, \emph{phonetics}, through \emph{phonology}, \emph{morphology}, \emph{syntax}, and \emph{semantics}, to \emph{pragmatics}. Adapted from~\protect\cite{ieeecomputersociety.2005}.}
\label{fig_major_levels_of_linguistic_structure}
\end{figure}

The semiotic cycle puts the utterance as an externalized information carrier into focus. While the other components are internalized and thus difficult to define and measure, the utterance itself is external and available for analysis. Fundamentally, this specific utterance is based on the underlying communication process and specifically, the language used. Accordingly, most research papers investigate characteristics of the utilized language to analyze the communication possibilities and capabilities of users.
To this end, linguistics subdivides the language structure into six major levels~\cite{ieeecomputersociety.2005, chandler.2007, brinton.2010}, as illustrated in Figure~\ref{fig_major_levels_of_linguistic_structure}. This structure was originally developed for spoken language, as indicated by the terms \enquote*{phonetics} and \enquote*{phonology} derived from the Greek word \enquote*{phon} meaning \enquote*{sound}. However, the levels are also applicable to written language in the context of \ac{el}. Therefore, the following description will address both spoken and written language within this framework.

The \emph{phonetics} level includes the entire bandwidth of the chosen, often continuous, language channel. For example, it comprises the full range of possible speech sounds available to humans. Consequently, it is fundamental for the general transfer range and describes it without any limitation.
At the \emph{phonology} level are the atomic building blocks of the spoken or written language, defined as phonemes or graphemes. A phoneme or grapheme enables the creation of meaning as well as the necessary distinction at the lowest level of language. However, in a \ac{nl} with an alphabetic writing system, phonemes and graphemes, which in this case correspond to letters, are often not a direct match and are only roughly related. Nevertheless, these individual units comprise the set of used elements from the continuous channel range for a specific language.
These are used and combined at the \emph{morphology} level to create and assign meaning by making words, in linguistics called lexemes. In this context, word-forming rules and underlying structures are of interest.
Utilizing these meaningful building blocks, sentences can be realized at the \emph{syntax} level. This level only concerns the structure of sentences and in particular, their assembly rules and the word categories that are used. The meaning of these sentences is relevant at the next level, \emph{semantics}. At this level, the literal meaning of language constructions is of interest while the final level, \emph{pragmatics}, focuses on how context contributes to the meaning. Accordingly, it analyzes how language is used in interactions and the relationship between the involved parties. Overall, the presented levels are not only important to describe language functionally and structurally but also to distinguish language characteristics and metrics. Thus, we use them to organize parts of the taxonomy in Section~\ref{sec_taxonomyofemergentlanguage} and the metrics in Section~\ref{sec_metrics}.

\subsection{Emergent Language}\label{sec_background_emergentlanguage}
\ac{el} refers to a form of communication that develops among artificial agents through interaction, without being explicitly pre-programmed. Thus, it is a bottom-up approach, arising from the agents' need to cooperate and solve tasks within a given environment~\cite{brandizzi.2023b}. This process involves the agents creating, adapting, and refining linguistic structures and meanings to enhance their ability to exchange information effectively and efficiently~\cite{lipowska.2022}. \ac{el} research aims to understand the principles and mechanisms underlying this spontaneous development of communication. It explores how linguistic elements such as syntax~\cite{ueda.2022}, semantics~\cite{colas.2020, qiu.2021}, and pragmatics~\cite{lowe.2019} can arise from the interaction of artificial agents and how these elements contribute to the agents' performance and cooperation.

A \ac{nl}-like communication form would make artificial agents and computer systems, in general, more accessible, simpler to comprehend, and altogether more powerful~\cite{lazaridou.2020, brandizzi.2022, noukhovitch.2021}. \ac{el} research originally focused on the question of language origin~\cite{steels.1997}. Recently, this focus shifted to the more functional aspect of \ac{el}, focusing on how to enable agent systems to benefit from a mechanism that helped humanity thrive and how to achieve communication capabilities as close as possible to \ac{nl}~\cite{lazaridou.2020}. Today, \ac{el} within computer science is about self-learned~\cite{nowak.1999}, reusable~\cite{keresztury.2020}, teachable~\cite{cogswell.2019, li.2019}, interpretable~\cite{lazaridou.2016}, and powerful~\cite{mordatch.2017} communication protocols. In the long run, \ac{el} aims to enable machines to communicate with each other and with humans in a more seamless and extendable manner~\cite{lazaridou.2020, yao.2022}.

Accordingly, various research questions and areas were derived. 
For example, recent papers have addressed issues around the nature of the setting, which can be semi-cooperative~\cite{liang.2020, noukhovitch.2021}, include adversaries~\cite{blumenkamp.2020, yu.2022}, have message-influencing noise~\cite{cope.2020}, or incorporate social structures~\cite{dubova.2020, fitzgerald.2020}.  
Moreover, some are concerned with the challenge of grounding \ac{el}, e.g. using representation learning as basis~\cite{lin.2021}, combining supervised learning and self-play~\cite{lowe.2020}, or utilizing \ac{el} agents as the basis for \ac{nl} finetuning approaches~\cite{yao.2022}.
Others tackle the direct emergence of language with \ac{nl} characteristics, e.g. looking at internal and external pressures~\cite{luna.2020, kalinowska.2022, dagan.2020, ren.2020, rita.2022}, evaluating factors to enforce semantic conveyance~\cite{mihai.2021}, looking at compositionality~\cite{resnick.2020}, generalization~\cite{chaabouni.2022}, or expressivity~\cite{guo.2021b}, or questioning the importance of characteristics like compositionality~\cite{kharitonov.2020} and the connection between compositionality and generalization~\cite{chaabouni.2020}.

Based on these examples and the introduced goals and approaches, the difference in comparison to \ac{nlp} research becomes apparent. Current approaches in \ac{nlp}, namely \acp{llm}, learn language imitation via statistics, but they might not capture the functional aspects and the purpose of communication itself~\cite{mordatch.2017, browning.2022}. In contrast, \ac{el} uses language not as the sole objective but as a means to achieve something with meaning~\cite{brandizzi.2021}. Accordingly, agents have to learn their own \ac{el} to enable functionality beyond simple statistical reproduction. Specifically, agents should learn communication by necessity or benefits~\cite{luna.2020} and they need a setting that rewards or encourages communication, e.g., an at least partially cooperative setting~\cite{noukhovitch.2021}.

While the \ac{el} concept sounds simple, it comes with many challenges. Encouraging communication alone can lead to simple gibberish that helps with task completion but does not represent the intended natural language characteristics~\cite{mu.2021}. Providing the right incentives for language development is therefore crucial. In addition, it is important to examine how agents use communication and the opportunity to send and receive information, raising the question of how to measure successful communication~\cite{lowe.2019}. The measurability of language properties such as syntax, semantics, and pragmatics is also important for assessing the emergence of desirable language properties~\cite{chaabouni.2020}. The following sections explore these challenges and related constructs and approaches in detail.

\section{Related Surveys}\label{sec_relatedsurveys}
%
\newcolumntype{L}{>{\raggedright\arraybackslash}X}
\begin{table*}[!bt]
\caption{Previously published surveys on \ac{el}, organized according to their primary focus.}
\label{tab:previous_surveys}
\begin{tabularx}{\textwidth}{lL}
\toprule
Settings & 
\mbox{van Eecke and Beuls~\cite{vaneecke.2020}},
\mbox{Lipowska and Lipowski~\cite{lipowska.2022}}, 
\mbox{Denamgana\"{i} and Walker~\cite{denamganai.2020}}
\\
Methods &
\mbox{Korbak et~al.~\cite{korbak.2020}},
\mbox{LaCroix~\cite{lacroix.2019}},
\mbox{Lemon~\cite{lemon.2022}},
\mbox{Lowe et~al.~\cite{lowe.2019}},
\mbox{Mihai and Hare~\cite{mihai.2021}},
\mbox{Galke and Raviv~\cite{galke.2024}},
\mbox{Vanneste et~al.~\cite{vanneste.2022}}
\\
General &
\mbox{Hernandez-Leal et~al.~\cite{hernandezleal.2019}},
\mbox{Brandizzi and Iocchi~\cite{brandizzi.2022}},
\mbox{Moulin-Frier and Oudeyer~\cite{moulinfrier.2020}},
\mbox{Galke et~al.~\cite{galke.2022}},
\mbox{Fernando et~al.~\cite{fernando.2020}},
\mbox{Suglia et~al.~\cite{suglia.2024}},
\mbox{Zhu et~al.~\cite{zhu.2024}},
\mbox{Lazaridou and Baroni~\cite{lazaridou.2020}},
\mbox{Brandizzi~\cite{brandizzi.2023b}}\\
\bottomrule
\end{tabularx}
\end{table*}

As briefly mentioned in Section~\ref{sec_introduction}, our literature review identified 19 publications that we classified as surveys. We adopted a broad definition of what constitutes a survey, categorizing any publication as a survey if it either explicitly described itself as such or provided a particularly comprehensive and structured review of previous research. These publications conduct similar investigations on \ac{el} research but with different scopes. We focus on discrete language emergence, associated taxonomy, characteristics, metrics, and research gaps. In contrast, in our review of the existing survey work, three distinct interpretive directions emerge, which we categorize as summarized in Table~\ref{tab:previous_surveys}: Surveys that focus on the learning \emph{settings} ~\cite{vaneecke.2020,lipowska.2022,denamganai.2020}, surveys that summarize and review utilized \emph{methods}~\cite{korbak.2020,lacroix.2019,lemon.2022,lowe.2019,mihai.2021,galke.2024,vanneste.2022}, and surveys that provide a \emph{general} discussion or overview of the \ac{el} field~\cite{hernandezleal.2019,brandizzi.2022,moulinfrier.2020,galke.2022,fernando.2020,lazaridou.2020, brandizzi.2023b}. The following section briefly summarizes these surveys within these categories.

\paragraph*{Settings}
Surveys within the \emph{settings} category primarily focus on the design of language learning environments and the general structure of learning problems. For instance, van Eecke and Beuls~\cite{vaneecke.2020} explored the language game paradigm, categorizing experiments and identifying properties critical for \ac{marl} research, such as symmetric agent roles and autonomous behavior. While their work offers a foundational perspective, our survey extends beyond the language game paradigm to analyze a broader range of approaches in greater depth (see Section~\ref{sec_map}). Similarly, Lipowska and Lipowski~\cite{lipowska.2022} emphasized sociocultural aspects, such as migration and teachability, within simple naming games. While these are part of our analysis, our review situates them within a unified framework, providing a more comprehensive perspective. Denamgana\"{i} and Walker~\cite{denamganai.2020} introduced ReferentialGym as a tool for studying referential games and their associated metrics, like positive signaling and positive listening~\cite{lowe.2019}. In contrast, our survey goes beyond referential games, offering a broader exploration of \ac{el} metrics and their applications across diverse frameworks.

\paragraph*{Methods}
The \emph{methods} category encompasses surveys that primarily examine learning and evaluation methodologies in \ac{el}, each offering unique perspectives on key challenges. A recurring theme in this category is the need for more comprehensive evaluation metrics that capture the complexity of emergent communication~\cite{lacroix.2019, korbak.2020, mihai.2021}. For example, Korbak et~al.~\cite{korbak.2020} highlighted the limitations of existing compositionality metrics, introducing the \emph{tree reconstruction error} to address semantic compositionality, a challenge we contextualize further in Section~\ref{sec_taxonomy_compositionality}. In contrast, LaCroix~\cite{lacroix.2019} critiqued the overemphasis on compositionality, advocating for a shift towards reflexivity, though metrics for this remain unexplored. This gap underscores the fragmented nature of current evaluation practices.

Grounding and utility also feature prominently in this category. Lemon~\cite{lemon.2022} emphasized the interplay of symbolic and conversational grounding, highlighting data-related challenges, while Lowe et~al.~\cite{lowe.2019} proposed pragmatic metrics such as positive signaling and positive listening, which inspired our taxonomy of language utility in Section~\ref{sec_taxonomy_pragmatics}. These works collectively underscore the necessity of balancing semantic depth with practical utility, a balance our survey seeks to achieve by integrating diverse perspectives into a unified framework.

Further, recent studies like those by Galke and Raviv~\cite{galke.2024} explored the role of linguistic pressures and biases in bridging the gap between \ac{el} and human \ac{nl}, providing insights into the origins of \ac{nl} phenomena in \ac{el}. Vanneste et~al.~\cite{vanneste.2022} tackled discretization methods critical for \ac{el} learning, offering a comparative analysis that complements our work.

\paragraph*{General}
The \emph{general} category includes surveys that provide broad overviews or address themes not confined to specific \emph{settings} or \emph{methods}. Key contributions in this category highlight the interdisciplinary perspectives, interaction paradigms, and structural dimensions of emergent communication research.

Hernandez-Leal et~al.~\cite{hernandezleal.2019} offered a foundational overview of multi-agent deep reinforcement learning (MARL), including emergent behavior and communication. While their survey provides valuable historical context and practical challenges, our work builds upon this by focusing specifically on emergent language (\ac{el}) within MARL, analyzing it with finer granularity and from a metrics-driven perspective. Similarly, Brandizzi and Iocchi~\cite{brandizzi.2022} emphasized the underrepresentation of human-in-the-loop paradigms, proposing novel interaction settings but lacking the systematic categorization and metrics-oriented discussion presented in our survey.

Moulin-Frier and Oudeyer~\cite{moulinfrier.2020}, Fernando et~al.~\cite{fernando.2020}, and Galke et~al.~\cite{galke.2022} explored interdisciplinary connections and cognitive constraints in \ac{el}, with the latter focusing on the perceived gaps between \ac{el} and \ac{nl}. While these works underscore key challenges, our survey contextualizes such limitations across a broader set of metrics and emergent properties, providing actionable insights for bridging these gaps.

Other surveys, such as those by Suglia et~al.~\cite{suglia.2024} and Zhu et~al.~\cite{zhu.2024}, structured \ac{el} research into multimodal and dimensional frameworks, respectively. These works serve as useful complements to our survey, which introduces an extensive taxonomy (Section~\ref{sec_taxonomyofemergentlanguage}) that synthesizes and organizes findings from diverse sources. Similarly, Lazaridou and Baroni~\cite{lazaridou.2020} and Brandizzi~\cite{brandizzi.2023b} provided comprehensive overviews of the field but lacked the detailed quantification and taxonomy of metrics that form the core of our work.

By synthesizing these contributions, our survey is distinguished by its focus on emergent language metrics and quantification, complemented by a systematic taxonomy to address fragmentation in the field. This integrated approach provides a structured roadmap for advancing \ac{el} research, with an emphasis on both practical measurability and interdisciplinary relevance.

\section{Study Methodology}\label{sec_studymethodology}
The literature search that resulted in the body of work surveyed in this paper was conducted on the \litSearchDate{}. The used libraries and databases are:
\href{https://www.sciencedirect.com/}{ScienceDirect}, \href{https://ieeexplore.ieee.org/}{IEEE Xplore}, \href{https://dl.acm.org/}{ACM Digital Library}, \href{https://www.webofscience.com/}{WebOfScience}, \href{https://arxiv.org/}{arXiv}, and \href{https://www.semanticscholar.org/}{SemanticScholar}. \href{https://www.semanticscholar.org/}{SemanticScholar} is a special case, due to the nature of the provided search machine that does not allow complex queries and filtering like the others. Consequently, we hand-picked suitable papers from the first 50 entries of the search result list. A PRISMA~\cite{page.2021} flow diagram of the publication selection process is provided in Figure~\ref{fig_PRISMA_flowchart} in Appendix~\ref{sec:appendix_b}. Additionally, the individual queries and results of all services are summarized in Table~\ref{tab:literature_search}.
The queries delivered \litCTotal{} hits in total which resulted in \litCUnique{} unique papers. A first quick read of these papers led to \litCAdditional{} additional papers, referenced by some of the originally found work. Accordingly, the literature review started with a corpus consisting of \litCCorpus{} individual papers.

\begin{table}[btp]
\begin{center}
\begin{minipage}{\textwidth}
\caption{Literature databases and search queries used for the present survey and the number of results obtained for each.}\label{tab:literature_search}
\begin{tabularx}{\textwidth}{lXl}
\textbf{Source}    & \textbf{Query}    & \textbf{Results} \\
\hline

\href{https://www.sciencedirect.com/}{ScienceDirect}
                    & \enquote{emergent language} \newline Subject areas: Computer Science & \litSsciencedirect{}   \\

\href{https://ieeexplore.ieee.org/}{IEEE Xplore}
                    & \enquote{emergent language}                   & \litSieeeexplore{}    \\ 

\href{https://dl.acm.org/}{ACM Digital Library}
                    & All: \enquote{emergent communication}         & \litSacmEC{}   \\
                    & All: \enquote{emergent language}               & \litSacmEL{}   \\ 

\href{https://www.webofscience.com/}{WebOfScience}
                    & TS=(\enquote{emerg* communication} or \enquote{emerg* language}) NOT TS=emergency NOT TS=5G NOT TS=wireless \newline Refined by: WEB OF SCIENCE CATEGORIES: ( COMPUTER SCIENCE ARTIFICIAL INTELLIGENCE ) Timespan: All years. 
                                                             & \litSWOS{}  \\

\href{https://arxiv.org/}{arXiv}
                    & all=\enquote{emergent communication}           & \litSarxivEC{}              \\
                    & all=\enquote{emergent language}                & \litSarxivEL{}               \\
                    & all=Emergent Multi-Agent Communication & \litSarxivEMAC{}             \\

\href{https://www.semanticscholar.org/}{SemanticScholar}
                    & multi agent emergent language \newline Filtered by topic Computer Science
                                                             & \litSsemanticscholar{}        \\

References          & -
                                                             & \litCAdditional{} 
\end{tabularx}
\end{minipage}
\end{center}
\end{table}

Of the \litCCorpus{} papers, \litRsortedout{} were sorted out due to the substantial divergence from the searched topic, often focusing on domains like 5G, networking, and radio. Of the remaining \litRremaining{} papers, \litRrelevant{} directly address the field of interest, while \litRpartiallyrelevant{} are only partially relevant. Papers were deemed partially relevant if they mentioned the surveyed topic but primarily focused on different areas such as datasets, language theory, simulation, or unrelated case studies. In conclusion, this survey mainly reviews  \litRrelevant{} papers that directly discuss or contribute to the topic of \ac{el} in computer science. 

\begin{figure}[btp]%
\centering
\includegraphics[width=\textwidth]{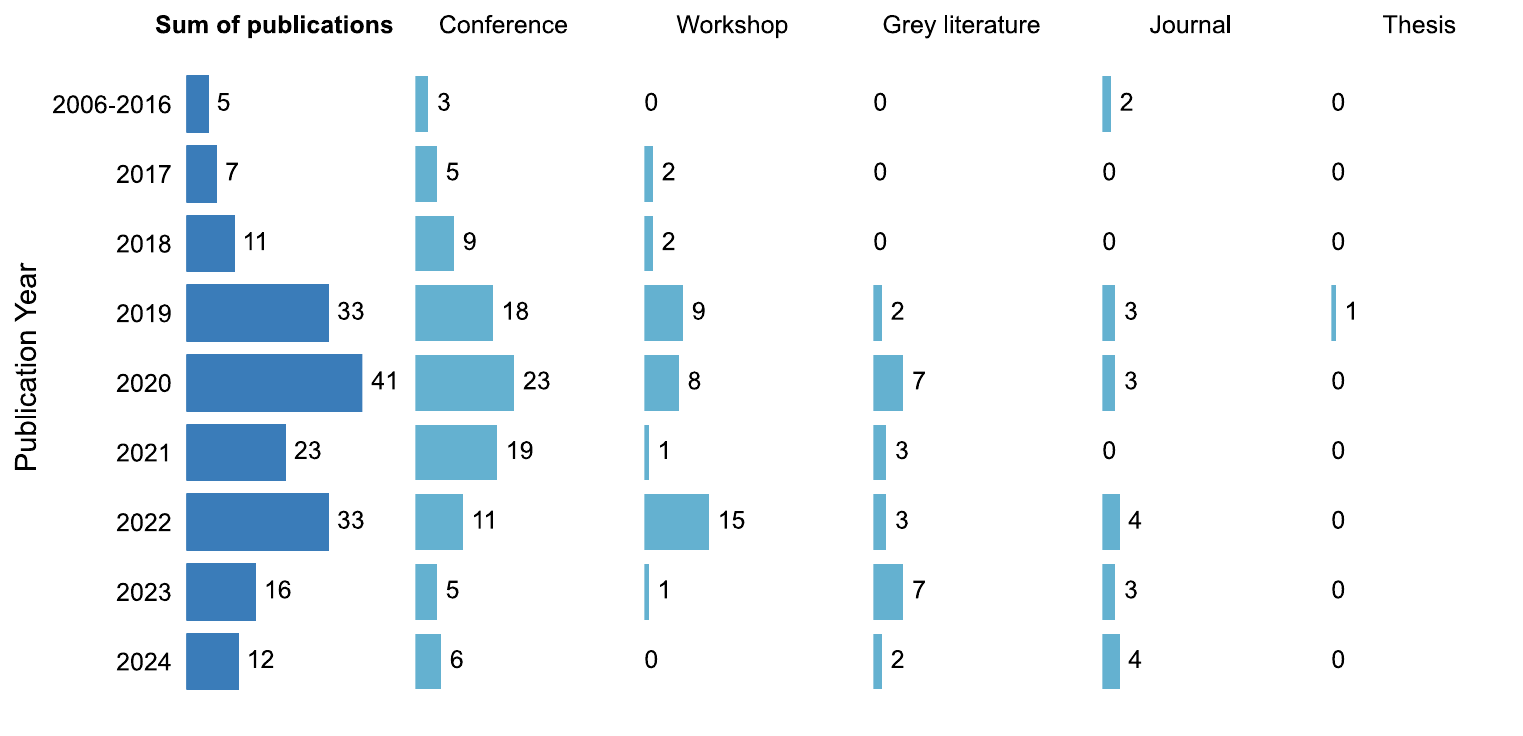}
\caption{Number of publications per year and by type. The number of publications per year is provided in the leftmost column, and the distribution of different publication types is shown in the remaining columns.}
\label{fig_publication_category_year_panel_bar_chart}
\end{figure}

Figure~\ref{fig_publication_category_year_panel_bar_chart} presents the distribution of the \litRrelevant{} relevant publications over the years, categorized by publication type. The topic of \ac{el} has maintained a steady presence in conference publications, peaking in 2020. The subsequent decline in total publications may be attributed to the absence of recent topic-specific workshops. Additionally, the surge in interest in \ac{llm} technologies might have diverted attention from \ac{el} research. It is also worth noting that some recent studies may not have been openly published at the time of our literature search. We therefore expect the publication count to increase by 2024.

\section{Taxonomy of Emergent Language}\label{sec_taxonomyofemergentlanguage}
In the course of our comprehensive literature review, we identified recurrent instances of taxonomic inconsistencies due to missing standardization~\cite{boldt.2022b} and \enquote{ill-adapted metrics}~\cite{chaabouni.2022}. Particular concern arises from the discrepancy between the concepts intended for measurement and their corresponding metrics, or the absence of such metrics~\cite{chen.2023, keresztury.2020, korbak.2020, kottur.2017b, perkins.2021b}. This section is dedicated to the formulation of a systematic taxonomy aimed at enhancing comparability and mitigating confusion within the field. This taxonomy forms the basis for the following sections and is designed to ensure consistent representation throughout the survey. It is created with the hope that it will serve as a cornerstone for future research, promoting the use of standardized terminology, particularly in the domain of language characteristics.

The taxonomy first describes the main factors influencing the \ac{el}, before categorizing the language characteristics. These influencing factors have a significant impact on the investigative possibilities of \ac{el} research and are therefore of particular importance when analyzing \ac{el}. 
Thus, the taxonomy introduces a classification system for the communication setting (Section~\ref{sec_taxnonomy_communicationsetting}) and communication games (Section~\ref{sec_taxnonomy_languagegames}) that agents encounter during language emergence. The communication setting encompasses factors such as the number of agents and the type of communication available to them. The communication game involves the environmental configuration and crucial factors influencing challenges and the complexity of multi-task learning. Furthermore, a short discussion on the concept of language priors is provided in Section~\ref{sec_taxnonomy_languageprior}, considering that the presence of a prior significantly influences the characteristics of the emerging language~\cite{lazaridou.2016, lowe.2020}. We conclude this section with a comprehensive overview of the concepts and characteristics examined within \ac{el} research (Section~\ref{sec_taxonomy_languagecharacteristics}). The taxonomy adheres to the six major linguistic structural levels introduced in Section~\ref{sec_background_naturallanguage} and illustrated in Figure~\ref{fig_major_levels_of_linguistic_structure}.

\subsection{Communication Setting}\label{sec_taxnonomy_communicationsetting}
In the literature, several communication settings are represented. One distinguishing factor is the number of agents involved. We derived three classes - the single agent, dual agent, and population setting. While the single agent setting is rare, the other two are well represented in the examined literature, as shown in Table~\ref{tab:communication_settings_overview}. A single agent is typically used to train human-machine interfaces~\cite{buck.2018} or fine-tune existing models~\cite{steinertthrelkeld.2022}. In contrast, dual-agent settings are more common and often involve a pair of speaker-listener agents, with one agent designated as the speaker and the other as the listener exclusively~\cite{lazaridou.2016}. The population setting involves larger groups of agents in the language emergence process. This requires more computational resources but also enables more possibilities for regularization~\cite{agarwal.2019} and language evolution~\cite{ren.2020}. Accordingly, the population setting offers more opportunities to actively shape the process~\cite{agarwal.2019, hardinggraesser.2019, vaneecke.2020}.

\renewcommand*\rot{\multicolumn{1}{R{45}{1em}}}
\begin{table*}[ptb]
\begin{threeparttable}
\caption{Classification of the communication settings in the literature reviewed.}
\label{tab:communication_settings_overview}
\begin{tabularx}{\textwidth}{llllX}
\rot{Agents} & \rot{Cooperation} & \rot{Symmetry} & \rot{Recipients} & Paper \\
\midrule
\multirow[t]{2}{*}{\faUser} & \multirow[t]{2}{*}{\faSmileBeam[regular]} & \faTimes & \faBullseye & \citenums{chevalierboisvert.2018, colas.2020, steinertthrelkeld.2022} \\
\cmidrule{3-5}
 &  & \faCheck & \faBullseye & \citenums{buck.2018, chaabouni.2019b, woodward.2019b} \\
\cmidrule{1-5}
\multirow[t]{5}{*}{\faUserFriends} & \multirow[t]{2}{*}{\faSmileBeam[regular]} & \faTimes & \faBullseye & \citenums{andreas.2019, auersperger.2022, bogin.2018, boldt.2022, boldt.2022d, bosc.2022, bouchacourt.2018, bullard.2021, carmeli.2022, carmeli.2024, chaabouni.2019, chaabouni.2020, chaabouni.2022, chen.2023, choi.2018, chowdhury.2020, chowdhury.2020b, chowdhury.2020c, cope.2020, cowenrivers.2020, denamganai.2020b, denamganai.2023, denamganai.2023b, dessi.2019, dessi.2021, downey.2023, eloff.2021, feng.2024, guo.2019, guo.2019b, guo.2020, guo.2021b, hagiwara.2021, havrylov.2017, hazra.2020, kalinowska.2022, kang.2020, karten.2023, keresztury.2020, kharitonov.2019, kharitonov.2019b, kharitonov.2020, korbak.2019, kubricht.2023, kucinski.2020, kucinski.2021, lazaridou.2016, lazaridou.2018, lei.2023b, li.2019, li.2020b, lipinski.2024, lobostsunekawa.2022, luna.2020, mihai.2019, mihai.2021b, mu.2021, mu.2023, mul.2019, noukhovitch.2021, ohmer.2022, ohmer.2022b, ossenkopf.2022, perkins.2021, portelance.2021, qiu.2021, ren.2020, resnick.2020, ri.2023, rita.2022b, santamariapang.2019, santamariapang.2020, sowik.2020, sowik.2020b, steinertthrelkeld.2019, tucker.2021, tucker.2022, ueda.2022, ueda.2023, unger.2020, vanderwal.2020b, vani.2021, vanneste.2022b, verma.2019, verma.2021, villanger.2024, xu.2022, yao.2022, yu.2023, yuan.2019} \\
\cmidrule{3-5}
 &  & \faCheck & \faBullseye & \citenums{ampatzis.2008, bachwerk.2011, bouchacourt.2019, chaabouni.2019b, das.2017, evtimova.2017, hagiwara.2019, hardinggraesser.2019, kolb.2019, kottur.2017b, lee.2017b, lin.2021, lobostsunekawa.2022, lowe.2020, ohmer.2022b, patel.2021, ropke.2021, saha.2019, villanger.2024, woodward.2019b, yuan.2019b} \\
\cmidrule{2-5}
 & \multirow[t]{2}{*}{\faMeh[regular]} & \faTimes & \faBullseye & \citenums{noukhovitch.2021} \\
\cmidrule{3-5}
 &  & \faCheck & \faBullseye & \citenums{ropke.2021} \\
\cmidrule{2-5}
 & \faAngry[regular] & \faTimes & \faBullseye & \citenums{noukhovitch.2021} \\
\cmidrule{1-5}
\multirow[t]{6}{*}{\faUsers} & \multirow[t]{4}{*}{\faSmileBeam[regular]} & \multirow[t]{2}{*}{\faTimes} & \faBullseye & \citenums{bullard.2020, chaabouni.2022, dagan.2020, fitzgerald.2019, gupta.2021, kajic.2020, korbak.2019, li.2019, nevens.2020, ohmer.2022b, ossenkopf.2022, ren.2020, rita.2022, sirota.2019, tieleman.2019, vani.2021, vanneste.2022b, verma.2021} \\
\cmidrule{4-5}
 &  &  & \faIcon{volume-up} & \citenums{gupta.2020b, li.2019, ren.2020, thomas.2021, vanneste.2022b, verma.2021} \\
\cmidrule{3-5}
 &  & \multirow[t]{2}{*}{\faCheck} & \faBullseye & \citenums{agarwal.2019, bachwerk.2011, baronchelli.2006, blumenkamp.2020, botokoekila.2024, botokoekila.2024b, cogswell.2019, das.2018, dubova.2020, fitzgerald.2020, hardinggraesser.2019, hildreth.2019, lee.2017b, lin.2021, loreto.2016, lorkiewicz.2011, ohmer.2022b, simoes.2020, simoes.2020b, taylor.2021, wang.2019b} \\
\cmidrule{4-5}
 &  &  & \faIcon{volume-up} & \citenums{bachwerk.2011, brandizzi.2021, das.2018, eccles.2019, gupta.2020, jimenezromero.2023, karten.2022, karten.2023b, lee.2017b, li.2021b, lin.2021, lipinski.2022, lo.2022, mordatch.2017, pesce.2020b, resnick.2018, simoes.2020, simoes.2020b, sukhbaatar.2016, wang.2019b, wu.2021, yuan.2024} \\
\cmidrule{2-5}
 & \multirow[t]{2}{*}{\faMeh[regular]} & \faTimes & \faIcon{volume-up} & \citenums{gupta.2020b, jaques.2018, yu.2022} \\
\cmidrule{3-5}
 &  & \faCheck & \faBullseye & \citenums{blumenkamp.2020, cao.2018, liang.2020} \\
\bottomrule
\end{tabularx}
\begin{tablenotes}
    \footnotesize
    \item[] Agents: \faUser{} Single, \faUserFriends{} Dual, \faUsers{} Population
    \item[] Cooperation: \faSmileBeam[regular]{} Cooperative, \faMeh[regular]{} Semi-cooperative, \faAngry[regular] Competitive
    \item[] Symmetry: \faTimes{} No, \faCheck{} Yes 
    \item[] Recipients: \faBullseye{} Target, \faIcon{volume-up}{} Broadcast
\end{tablenotes}
\end{threeparttable}
\end{table*}

An additional factor that shapes the communication setting is the type of cooperation inherent in the setup. Determining the level of cooperation or competition feasible within the setting is a fundamental decision and closely related to the choice of the language game. We derived three options - the cooperative, semi-cooperative, and competitive type. In the literature reviewed, the majority of studies adopted a fully cooperative setting approach, where agents fully share their rewards and lack individual components. The emphasis on strongly cooperative settings is justified given that AI agents utilize a common language to coordinate and will not learn to communicate if they dominate without communication~\cite{noukhovitch.2021}. Only a few publications explore semi-cooperative settings that incorporate individual rewards alongside shared rewards, introducing the challenge of balancing tasks and rewards~\cite{lowe.2019, liang.2020}. A semi-cooperative setup can be compared to a simplified social scenario with overarching societal objectives, while also encompassing additional individual interests and goals. In contrast, investigations of fully competitive settings are rare, with only one work in which agents compete for rewards without a common goal~\cite{noukhovitch.2021}. This scarcity likely arises from the fact that such settings inherently favor deceptive language as the only advantageous strategy, making its emergence improbable without any cooperative element.~\cite{noukhovitch.2021}.

The third important factor in communications settings is symmetry. Agents should treat messages similarly to regular observations; otherwise, they risk devolving into mere directives~\cite{cowenrivers.2020}. Building on this premise, the symmetry is important for promoting robust language emergence, as opposed to languages that consist primarily of directives. An illustrative example of asymmetric settings is the commonly used, and aforementioned, speaker-listener paradigm~\cite{lazaridou.2016, lazaridou.2018, korbak.2020}. Languages developed in such settings are severely limited compared to \ac{nl}, lacking the capacity for diverse discourse or even basic information exchange beyond directives~\cite{cowenrivers.2020}. Contrary to promoting informed choices by the listener, the speaker-listener approach emphasizes obedience to commands. Conversely, a symmetric setting facilitates bi-directional communication, thereby allowing for more comprehensive language development~\cite{bouchacourt.2019, dubova.2020}. For instance, symmetry may result from agents being randomly assigned roles within the interaction~\cite{bouchacourt.2019}. Additionally, symmetry can emerge from tasks that are inherently balanced, such as negotiations between equal partners where both parties have equivalent roles and objectives~\cite{cao.2018}.

At the population level, another important consideration is the choice of recipients, i.e., between targeted and broadcast communication. While broadcast communication facilitates broader information dissemination across the agent group, targeted communication promotes the development of social group dynamics and regularization~\cite{das.2018, simoes.2020}. For example, targeted communication strategies can be learned through mechanisms such as attention~\cite{das.2018}, and agents can develop minimized communication strategies that optimize group performance~\cite{wang.2019b}.

Table~\ref{tab:communication_settings_overview} provides a summary of these settings and their variations. The setting categories presented and their implementation are not inherently tied to the language itself but are crucial in determining the likelihood of meaningful language emergence and in shaping the features and experimental possibilities. These initial choices dictate the options for the language development process, the opportunities for regularization~\cite{agarwal.2019, hardinggraesser.2019}, and the requirements regarding computational resources.

\subsection{Language Games}\label{sec_taxnonomy_languagegames}
Distinct communication settings are implemented through different communication games. In this section, we provide an overview of the games used in \ac{el} literature. Specifically, we focus on a subset of these games known as language games, that emphasize explicit communication via a predefined language channel.
The literature identifies several categories of language games, such as referential games, reconstruction games, question-answer games, grid-world games, among others. Our review indicates that these categories represent the most commonly used game types. To give a comprehensive view, Table~\ref{tab:games_overview} lists the publications that focus on these game types. In the following, we offer a concise overview of each category to provide a clearer understanding of their characteristics. 

\smallskip \noindent \emph{Referential Game}: Generally, a referential game, also called signaling game, consists of two agents, a sender and a receiver~\cite{lazaridou.2016}. The objective of this game is for the receiver to correctly identify a particular sample from a set, which may include distractors, solely based on the message received from the sender. This set can consist of images~\cite{lazaridou.2016, bouchacourt.2018, denamganai.2023b}, object feature vectors~\cite{carmeli.2022}, texts~\cite{carmeli.2022}, or even graphs~\cite{sowik.2020}. To accomplish this selection task, the sender must first encode a message that contains information about the correct sample. In game design, a fundamental decision arises regarding whether the sender should only view the correct sample or also some distractors that may differ from those presented to the receiver~\cite{lazaridou.2016}. Another design decision concerns the receiver's side, specifically the number of distractors and whether to provide the original sample shown to the sender or only a similar one for selection~\cite{kang.2020}. However, only the encoded message is transmitted to the receiver, who then selects an item from their given collection. 

\smallskip \noindent \emph{Reconstruction Game}: The reconstruction game is similar to the referential game, but with a key difference: the receiver does not have a collection to choose from. Instead, the receiver must construct a sample based on the message from the sender, aiming to replicate the original sample shown to the sender as closely as possible~\cite{chaabouni.2020, rita.2022}. Consequently, this game setup resembles an autoencoder-based approach, with a latent space tailored to mimic or facilitate language~\cite{kharitonov.2019}. Therefore, the key distinction between reconstruction and referential games, often used interchangeably in early literature, lies in the collection's presence (referential) or absence (reconstruction) for the receiver to select from~\cite{guo.2019}.

\smallskip \noindent \emph{Question-Answer Game}: The question-answer game is a variant of the referential game, but without strict adherence to previously established rules. It operates as a multi-round referential game, allowing for iterative and bilateral communication~\cite{bouchacourt.2019}. Unlike referential and reconstruction games, the question-answer game explicitly incorporates provisions for multiple rounds with follow-up or clarifying queries from the receiver~\cite{das.2017, agarwal.2019}. Question-answer games have introduced intriguing inquiries and avenues for exploring the symmetry of \ac{el}, although they are not as widely adopted~\cite{das.2017}.

\smallskip \noindent \emph{Grid World Game}: Grid world games use a simplified 2D environment to model various scenarios like warehouse path planning~\cite{blumenkamp.2020}, movement of objects~\cite{kolb.2019}, traffic junctions~\cite{simoes.2020, gupta.2020}, or mazes~\cite{sukhbaatar.2016, kalinowska.2022}. They offer design flexibility, allowing agents to be part of the environment or act as external supervisors. Design choices also include environment complexity and the extent of agents' observations. Although common in the literature surveyed, implementations of grid world games vary widely in their design choices and are thus a very heterogeneous group.

\smallskip \noindent \emph{Continuous World Game}: Continuous environments add complexity to the learning process~\cite{wang.2019b, pesce.2020b}. In \ac{el} approaches, the learning landscape involves multi-task settings where one task is tackled directly within the environment while another involves language formation. Playing continuous world games, whether in two or three dimensions, presents challenges and adds a greater sense of realism and intricacy. These environments have the potential to make it more feasible to deploy \ac{el} agents in real-world scenarios compared to discrete environments~\cite{patel.2021}.

\smallskip \noindent \emph{Other}: The literature on \ac{el} also covers various other game types besides those mentioned earlier, such as matrix communication games~\cite{lowe.2019, vanneste.2022b}, social deduction games~\cite{brandizzi.2021, lipinski.2022}, or lever games~\cite{sukhbaatar.2016, li.2021b}. These game types contribute to the creation of new language emergence settings, often designed to target specific aspects or characteristics of language development. They are valuable tools to explore and understand the complexities of \ac{el} in different contexts.

\begin{table*}[!tbp]
\caption{Overview of the distribution of game types in the reviewed literature.}
\label{tab:games_overview}
\begin{tabularx}{\textwidth}{lX}
\toprule
Type & Paper  \\
\midrule
Referential & \citenums{bogin.2018, boldt.2022d, bosc.2022, botokoekila.2024, botokoekila.2024b, bouchacourt.2018, brandizzi.2023b, bullard.2020, bullard.2021, carmeli.2022, carmeli.2024, chaabouni.2022, chen.2023, choi.2018, chowdhury.2020b, chowdhury.2020c, dagan.2020, denamganai.2020, denamganai.2020b, denamganai.2023, denamganai.2023b, dessi.2019, dessi.2021, downey.2023, dubova.2020, eccles.2019, feng.2024, fernando.2020, fitzgerald.2019, fitzgerald.2020, guo.2019, guo.2019b, guo.2020, guo.2021b, gupta.2021, hagiwara.2021, hardinggraesser.2019, havrylov.2017, kang.2020, karten.2023, keresztury.2020, kharitonov.2019, korbak.2020, kubricht.2023, kucinski.2021, lazaridou.2016, lazaridou.2018, lee.2017b, li.2019, li.2020b, lipinski.2024, lipowska.2022, loreto.2016, lorkiewicz.2011, lowe.2020, luna.2020, mihai.2019, mihai.2021, mihai.2021b, mu.2021, mu.2023, nevens.2020, noukhovitch.2021, ohmer.2022, ohmer.2022b, perkins.2021, portelance.2021, qiu.2021, ren.2020, ri.2023, santamariapang.2019, sirota.2019, sowik.2020, sowik.2020b, steinertthrelkeld.2019, steinertthrelkeld.2022, tucker.2021, tucker.2022, ueda.2022, ueda.2023, vanderwal.2020b, verma.2019, verma.2021, yao.2022, yu.2022, yuan.2019} \\
Reconstruction & \citenums{ampatzis.2008, andreas.2019, auersperger.2022, baronchelli.2006, boldt.2022d, brandizzi.2023b, chaabouni.2019, chaabouni.2020, chowdhury.2020, cope.2020, denamganai.2020, eloff.2021, evtimova.2017, fernando.2020, guo.2019, guo.2020, kharitonov.2019, kharitonov.2019b, kharitonov.2020, korbak.2019, kucinski.2020, lipowska.2022, lowe.2020, mu.2021, ossenkopf.2022, resnick.2020, rita.2022, rita.2022b, taylor.2021, thomas.2021, xu.2022, yuan.2019b} \\
Question-answer & \citenums{agarwal.2019, bouchacourt.2019, brandizzi.2023b, cogswell.2019, das.2017, keresztury.2020, kottur.2017b, lei.2023b, liang.2020, vani.2021} \\
Grid World & \citenums{abdelaziz.2024, blumenkamp.2020, bogin.2018, boldt.2022, boldt.2022d, brandizzi.2023b, chaabouni.2019b, chevalierboisvert.2018, colas.2020, cowenrivers.2020, das.2018, denamganai.2023, eccles.2019, feng.2024, foerster.2018c, gupta.2020, hazra.2020, jaques.2018, jimenezromero.2023, kajic.2020, kalinowska.2022, karten.2022, karten.2023, karten.2023b, kolb.2019, li.2021b, lin.2021, lo.2022, mordatch.2017, mul.2019, nakamura.2023, resnick.2018, saha.2019, simoes.2020, simoes.2020b, sukhbaatar.2016, tucker.2022, unger.2020, vanneste.2022, vanneste.2022b, wang.2019b, woodward.2019b, yu.2023, yuan.2024, zheng.2017} \\
Continuous World & \citenums{brandizzi.2023b, hildreth.2019, li.2021b, lobostsunekawa.2022, patel.2021, pesce.2020b, simoes.2020b, wang.2019b, wu.2021, yuan.2024} \\
Other & \citenums{brandizzi.2021, cao.2018, carmeli.2022, foerster.2018c, gupta.2020b, hagiwara.2019, hagiwara.2021, li.2021b, lipinski.2022, lo.2022, lowe.2019, ropke.2021, sukhbaatar.2016, vanneste.2022, vanneste.2022b, villanger.2024, wu.2021} \\
\bottomrule
\end{tabularx}
\end{table*}

\bigskip \noindent
In summary, although many language games have been developed, comparing different games can be complex and understanding the nuances of each game can prove challenging. A promising direction would be for the research community to collectively agree on a standardized subset of these games as benchmarks. By focusing on a representative set of games from different categories, researchers could systematically explore different settings, ensuring that new approaches are rigorously tested and their results are directly comparable across studies. This would accelerate the maturation of the field of \ac{el} research, foster collaboration, and enable the community to better identify and address key challenges.

\subsection{Language Prior}\label{sec_taxnonomy_languageprior}
\ac{el} research occasionally utilizes a concept known as a \emph{language prior} to incorporate structures from human \acp{nl} into the emerging language. A language prior is used to impose specific linguistic structures on the emerging language, making it easier to align with human \ac{nl} and improve interpretability and performance. This prior can be implemented through supervised learning~\cite{das.2017, lazaridou.2016, lowe.2020}, also known as injection, or through divergence estimation~\cite{havrylov.2017}. An overview of prior usage in the literature surveyed is given in Table~\ref{tab:language_prior} in Appendix~\ref{sec:appendix_a}.

\begin{figure}[btp]%
\centering
\includegraphics[width=0.9\textwidth]{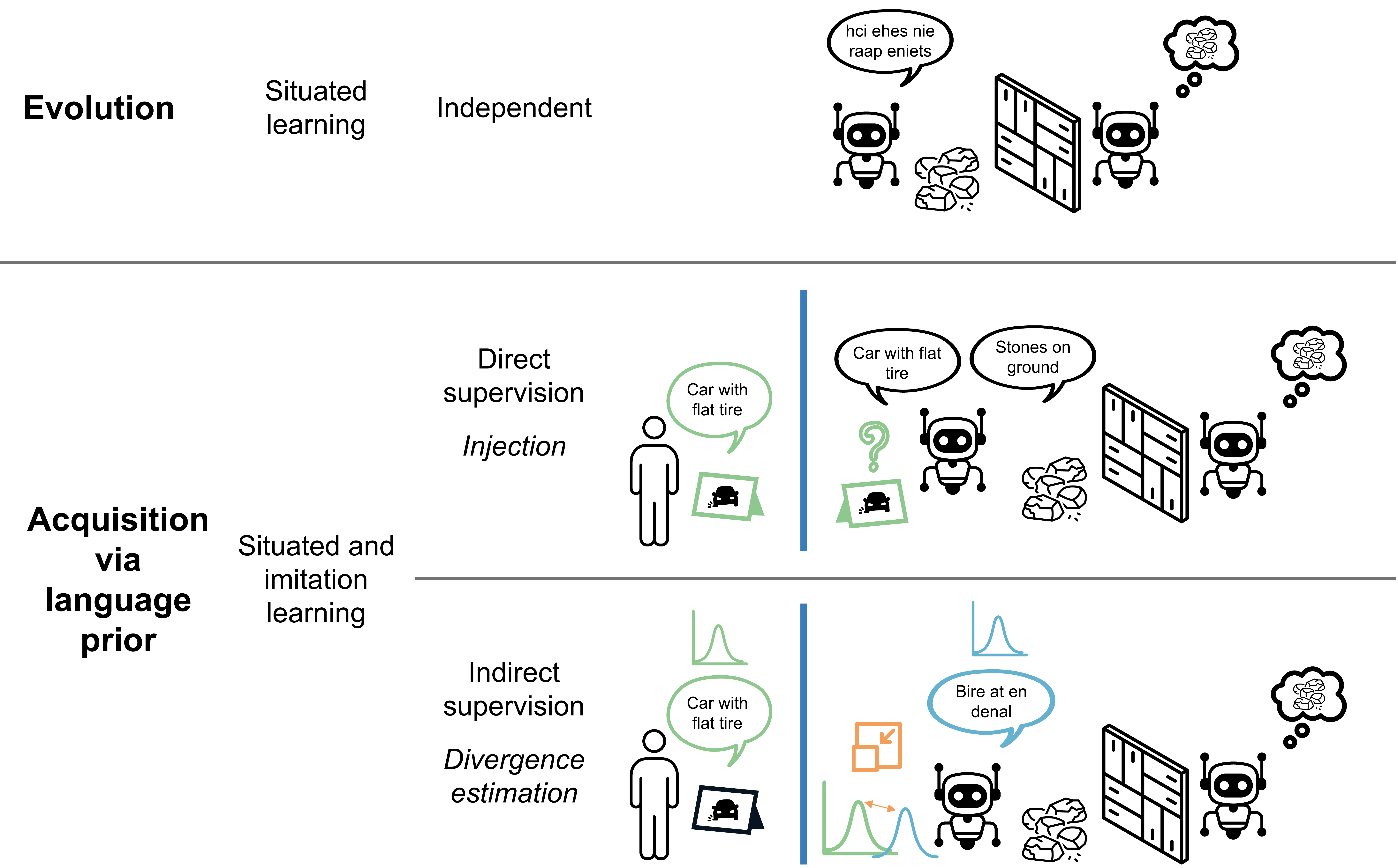}
\caption{Language evolution and the different language prior options for acquisition approaches. The \emph{evolution} of language and the guided \emph{acquisition} are contrary approaches. Language evolution is based on a situated learning environment and the \emph{independent} design of an appropriate language scheme. Acquisition, on the other hand, is based on the existence of a prior and the combination of situated and imitation learning. The prior, generally a \ac{nl}, is either introduced into the learning process via injection or divergence estimation. \emph{Injection} is a direct supervision approach that uses examples of prior usage to inform and train the language learner. \emph{Divergence estimation} is an indirect supervision method which utilizes a distribution representation of the prior and learner language. The goal is to limit the divergence of these distributions.}
\label{fig_language_prior_options}
\end{figure}

Given this context, research on \ac{el} can be divided into two main areas. The first area focuses on independent situated learning and does not use priors, so that communication and language emerge spontaneously~\cite{mordatch.2017}. The second area explores imitation learning-based approaches, which aim to replicate \ac{nl} behavior in artificial agents using priors~\cite{colas.2020}. However, it is important to note that these approaches differ from \acp{llm} because language acquisition in \ac{el} is generally task-oriented. In academic literature, the independent situated learning environment is often referred to as the evolution-based approach, while the imitation learning-related approach is commonly known as the acquisition-based approach. The term \emph{evolution} implies starting from scratch, while \emph{acquisition} involves learning an existing language~\cite{lazaridou.2018}. The terminology and different approaches are depicted in Figure~\ref{fig_language_prior_options}.

In addition, the concepts of community and generational learning are closely related~\cite{agarwal.2019, dagan.2020, hardinggraesser.2019}. In these methods, language emerges through iterative learning across and within agent sub-groups called communities. Generational learning additionally involves older generations of agents training younger ones using previously developed communication as a foundation~\cite{cogswell.2019, ren.2020}. Language transfer across groups or generations can be interpreted as an iterative prior. However, this method remains a fully evolutionary approach in the absence of a deliberately designed prior.

\subsection{Language Characteristics}\label{sec_taxonomy_languagecharacteristics}
As discussed in Section~\ref{sec_background_naturallanguage}, language is a complex, multifaceted system~\cite{hock.2019, pinker.1990}. Therefore, it is essential to establish a comprehensive taxonomy of its properties to provide a unified framework for \ac{el} research. This taxonomy will not only facilitate the unambiguous categorization of metrics used in \ac{el} studies (cf. Section~\ref{sec_metrics}) but will also enhance the comparability and comprehensibility of approaches and results within the field.
As shown previously in Figure~\ref{fig_major_levels_of_linguistic_structure}, \ac{nl} can be divided hierarchically into distinct characteristics~\cite{ieeecomputersociety.2005, chandler.2007, brinton.2010}. The following sections provide a categorization of the reviewed publications along these characteristics, occasionally breaking them down into smaller sub-characteristics if relevant.

\subsubsection{Phonetics}\label{sec_taxonomy_phonetics}
The phonetics of a language inherently represents its medium, delineating the constraints of the specific communication channel~\cite{brinton.2010, steels.1997}. These media or channels can be either discrete or continuous; for example, an audio channel is continuous, while a symbolic channel is typically discrete. Regardless of the type, they lay the foundation for the nature of communication.
However, for \ac{el} research the discrete case is of particular importance, as it closely mirrors \ac{nl} as we understand it~\cite{lazaridou.2020}. Although humans use a continuous phonetic medium for communication, some degree of discretization is essential to establish a common ground for efficient communication~\cite{chandler.2007}.

Table~\ref{tab:discrete_or_continuous_overview} provides an overview of the reviewed papers, categorized according to the continuous or discrete approach. Notably, some papers explore both approaches, providing valuable insights for researchers interested in the basic aspects of phonetics research in \ac{el}.

\begin{table*}[!tb]
\caption{Overview of the use of channel types in the reviewed literature.}
\label{tab:discrete_or_continuous_overview}
\begin{tabularx}{\textwidth}{lX}
\toprule
Type & Paper  \\
\midrule
Discrete & \citenums{abdelaziz.2024, agarwal.2019, ampatzis.2008, andreas.2019, auersperger.2022, bogin.2018, boldt.2022d, bosc.2022, botokoekila.2024, bouchacourt.2018, bouchacourt.2019, brandizzi.2021, buck.2018, bullard.2021, cao.2018, carmeli.2024, chaabouni.2019, chaabouni.2019b, chaabouni.2020, chaabouni.2022, chen.2023, chevalierboisvert.2018, choi.2018, chowdhury.2020, chowdhury.2020b, chowdhury.2020c, cogswell.2019, colas.2020, cope.2020, cowenrivers.2020, dagan.2020, das.2017, denamganai.2020b, denamganai.2023, denamganai.2023b, dessi.2019, dessi.2021, downey.2023, dubova.2020, eccles.2019, evtimova.2017, feng.2024, fitzgerald.2019, fitzgerald.2020, guo.2019, guo.2019b, guo.2020, guo.2021b, gupta.2020, gupta.2020b, gupta.2021, hagiwara.2019, hagiwara.2021, hardinggraesser.2019, havrylov.2017, hazra.2020, jaques.2018, jimenezromero.2023, kajic.2020, kalinowska.2022, kang.2020, karten.2023, keresztury.2020, kharitonov.2019, kharitonov.2019b, kharitonov.2020, kolb.2019, korbak.2019, korbak.2020, kottur.2017b, kubricht.2023, kucinski.2020, kucinski.2021, lazaridou.2016, lazaridou.2018, lee.2017b, li.2019, li.2020b, liang.2020, lin.2021, lipinski.2022, lipinski.2024, lipowska.2022, lobostsunekawa.2022, loreto.2016, lorkiewicz.2011, lowe.2019, lowe.2020, luna.2020, mihai.2019, mihai.2021, mordatch.2017, mu.2021, mu.2023, mul.2019, nakamura.2023, nevens.2020, ohmer.2022, ohmer.2022b, perkins.2021, portelance.2021, ren.2020, resnick.2020, ri.2023, rita.2022, rita.2022b, saha.2019, santamariapang.2019, santamariapang.2020, sirota.2019, sowik.2020, sowik.2020b, steinertthrelkeld.2019, steinertthrelkeld.2022, thomas.2021, tucker.2022, ueda.2022, ueda.2023, unger.2020, vanderwal.2020b, vani.2021, vanneste.2022, vanneste.2022b, verma.2019, verma.2021, wang.2019b, xu.2022, yao.2022, yu.2022, yuan.2019, yuan.2019b} \\
Continuous & \citenums{baronchelli.2006, blumenkamp.2020, boldt.2022, bullard.2020, das.2018, fernando.2020, hildreth.2019, lei.2023b, lo.2022, mihai.2021b, ossenkopf.2022, patel.2021, pesce.2020b, qiu.2021, resnick.2018, simoes.2020, simoes.2020b, sukhbaatar.2016, taylor.2021, tieleman.2019, wu.2021, yuan.2024} \\
Both & \citenums{botokoekila.2024b, brandizzi.2023b, carmeli.2022, eloff.2021, karten.2022, karten.2023b, lazaridou.2020, li.2021b, noukhovitch.2021, tucker.2021, villanger.2024, yu.2023} \\
\bottomrule
\end{tabularx}
\end{table*}

\subsubsection{Phonology}\label{sec_taxonomy_phonology}
Phonology encompasses the actively used vocabulary and determines the part of the medium that is utilized for communication. We identified five different types of vocabulary actively researched, however, some of them are rare to find in the literature. Table~\ref{tab:vocabulary_type_used} summarizes the results of our survey regarding vocabulary types in \ac{el} research.
One commonly used phonological type in \ac{el} is a binary encoding, while an even more prominent type is a token-based vocabulary. However, these two phonological classes are not always distinct, as a token-based vocabulary often builds upon a binary encoded representation~\cite{lin.2021}.

The other three types, which are distinct from the two most prominent, are rarely mentioned in the literature reviewed. 
One of these types involves using \ac{nl} vocabulary, such as all the words from an English dictionary. While this approach enforces the \ac{nl} resemblance of the \ac{el}, it also drastically limits the emergence and associated benefits~\cite{das.2017}. Essentially, this phonological preset strips the agents of the possibility to shape phonology and morphology.
The other two vocabulary types being referred to are sound and graphics. The former enables agents to produce and process sound~\cite{ampatzis.2008}, while the latter focuses on enabling agents to draw and analyze graphical representations~\cite{mihai.2021b, qiu.2021}. Both mediums present challenges in ensuring discretization, which may be the reason why they are not as extensively researched in \ac{el}.

\begin{table*}[!tb]
\caption{Overview of the use of vocabulary types in the reviewed literature.}
\label{tab:vocabulary_type_used}
\begin{tabularx}{\textwidth}{lX}
\toprule
Vocabulary & Paper  \\
\midrule
Binary & \citenums{boldt.2022, boldt.2022d, bosc.2022, bouchacourt.2018, brandizzi.2021, brandizzi.2023b, bullard.2021, carmeli.2022, chowdhury.2020b, cope.2020, cowenrivers.2020, dessi.2021, dubova.2020, eccles.2019, evtimova.2017, guo.2019, gupta.2020, hardinggraesser.2019, jimenezromero.2023, karten.2022, karten.2023b, kharitonov.2019b, lazaridou.2016, lee.2017b, li.2021b, lin.2021, lipinski.2022, lipowska.2022, nakamura.2023, resnick.2020, saha.2019, sirota.2019, steinertthrelkeld.2019, thomas.2021, tucker.2022, vanneste.2022, vanneste.2022b, verma.2021, wang.2019b, yuan.2019, yuan.2019b} \\
Token & \citenums{abdelaziz.2024, andreas.2019, auersperger.2022, baronchelli.2006, blumenkamp.2020, bogin.2018, botokoekila.2024, botokoekila.2024b, bouchacourt.2019, brandizzi.2021, brandizzi.2023b, buck.2018, bullard.2020, cao.2018, carmeli.2022, carmeli.2024, chaabouni.2019, chaabouni.2019b, chaabouni.2020, chaabouni.2022, chen.2023, chevalierboisvert.2018, choi.2018, chowdhury.2020, chowdhury.2020c, cogswell.2019, dagan.2020, das.2017, das.2018, denamganai.2020b, denamganai.2023, denamganai.2023b, dessi.2019, eloff.2021, feng.2024, fitzgerald.2019, fitzgerald.2020, guo.2019, guo.2019b, guo.2020, guo.2021b, gupta.2020b, hagiwara.2019, hagiwara.2021, havrylov.2017, hildreth.2019, jaques.2018, kajic.2020, kalinowska.2022, kang.2020, karten.2023, karten.2023b, keresztury.2020, kharitonov.2019, kharitonov.2020, kolb.2019, korbak.2019, korbak.2020, kottur.2017b, kubricht.2023, kucinski.2020, kucinski.2021, lazaridou.2018, li.2019, li.2020b, li.2021b, liang.2020, lipinski.2024, lipowska.2022, lo.2022, lobostsunekawa.2022, loreto.2016, lorkiewicz.2011, lowe.2019, lowe.2020, luna.2020, mihai.2019, mihai.2021, mordatch.2017, mu.2021, mu.2023, mul.2019, nevens.2020, noukhovitch.2021, ohmer.2022, ohmer.2022b, ossenkopf.2022, patel.2021, perkins.2021, pesce.2020b, portelance.2021, ren.2020, resnick.2018, ri.2023, rita.2022, rita.2022b, santamariapang.2019, santamariapang.2020, simoes.2020, simoes.2020b, sowik.2020, sowik.2020b, sukhbaatar.2016, tieleman.2019, tucker.2021, ueda.2022, ueda.2023, unger.2020, vanderwal.2020b, verma.2019, villanger.2024, wu.2021, xu.2022, yao.2022, yu.2022, yu.2023, yuan.2024} \\
NL & \citenums{agarwal.2019, brandizzi.2023b, colas.2020, das.2017, downey.2023, gupta.2021, havrylov.2017, hazra.2020, steinertthrelkeld.2022, vani.2021} \\
Sound & \citenums{ampatzis.2008, brandizzi.2023b, eloff.2021} \\
Picture & \citenums{brandizzi.2023b, fernando.2020, lei.2023b, mihai.2021b, qiu.2021} \\
\bottomrule
\end{tabularx}
\end{table*}

\subsubsection{Morphology}\label{sec_taxonomy_morphology}
Morphology governs the rules for constructing words and sentences, meaning the overall ability to combine individual elements, also called tokens, into words and to combine those words into sentences~\cite{brinton.2010}. This is particularly relevant in the field of \ac{el} due to the prominent division of existing work based on morphological setup and options. The most significant differentiation is between the use of a fixed or flexible message length. Table~\ref{tab:fixed_or_variable_message_length_tab} demonstrates that much of the existing work employs fixed message lengths, despite this setup not being comparable to \ac{nl}~\cite{havrylov.2017}. For instance, \ac{nl} users, such as humans, have the ability to adjust the length of their message to fit their intention, which may vary depending on the audience, medium, or communicative goal. When communicating with colleagues, they may use shorter sentences to be efficient, while more detailed explanations may be used when conversing with friends.

\begin{table*}[!tb]
\caption{Overview of the characteristics of message length in the literature reviewed.}
\label{tab:fixed_or_variable_message_length_tab}
\begin{tabularx}{\textwidth}{lX}
\toprule
Length & Paper  \\
\midrule
Fixed & \citenums{abdelaziz.2024, ampatzis.2008, andreas.2019, auersperger.2022, baronchelli.2006, blumenkamp.2020, boldt.2022, boldt.2022d, botokoekila.2024, botokoekila.2024b, bouchacourt.2018, bouchacourt.2019, brandizzi.2021, bullard.2021, cao.2018, chaabouni.2020, chaabouni.2022, chowdhury.2020, chowdhury.2020b, chowdhury.2020c, cogswell.2019, cope.2020, cowenrivers.2020, das.2018, dessi.2019, dessi.2021, dubova.2020, eccles.2019, evtimova.2017, feng.2024, fitzgerald.2019, fitzgerald.2020, guo.2019, guo.2019b, guo.2020, guo.2021b, gupta.2020, hagiwara.2019, hagiwara.2021, hardinggraesser.2019, hazra.2020, hildreth.2019, jaques.2018, jimenezromero.2023, kajic.2020, kalinowska.2022, karten.2022, karten.2023b, kharitonov.2020, kolb.2019, korbak.2019, kottur.2017b, kubricht.2023, kucinski.2020, kucinski.2021, lazaridou.2016, lei.2023b, li.2019, li.2021b, liang.2020, lin.2021, lipinski.2022, lipinski.2024, lo.2022, lobostsunekawa.2022, lorkiewicz.2011, lowe.2019, lowe.2020, mihai.2021b, mul.2019, nakamura.2023, nevens.2020, noukhovitch.2021, ossenkopf.2022, patel.2021, perkins.2021, pesce.2020b, qiu.2021, ren.2020, resnick.2018, resnick.2020, ri.2023, rita.2022, saha.2019, santamariapang.2019, santamariapang.2020, simoes.2020, simoes.2020b, sirota.2019, sowik.2020, sowik.2020b, steinertthrelkeld.2019, sukhbaatar.2016, thomas.2021, tieleman.2019, tucker.2021, tucker.2022, ueda.2022, unger.2020, vani.2021, vanneste.2022, vanneste.2022b, verma.2019, verma.2021, villanger.2024, wu.2021, yu.2022, yu.2023, yuan.2019, yuan.2019b, yuan.2024} \\
Variable & \citenums{agarwal.2019, bogin.2018, bosc.2022, buck.2018, bullard.2020, carmeli.2024, chaabouni.2019, chaabouni.2019b, chen.2023, chevalierboisvert.2018, choi.2018, colas.2020, dagan.2020, das.2017, denamganai.2020b, denamganai.2023, denamganai.2023b, downey.2023, eloff.2021, gupta.2020b, gupta.2021, havrylov.2017, kang.2020, karten.2023, keresztury.2020, kharitonov.2019, kharitonov.2019b, korbak.2020, lazaridou.2018, li.2020b, luna.2020, mihai.2019, mihai.2021, mordatch.2017, mu.2021, mu.2023, ohmer.2022, ohmer.2022b, portelance.2021, rita.2022b, steinertthrelkeld.2022, ueda.2023, vanderwal.2020b, wang.2019b, xu.2022, yao.2022} \\
Both & \citenums{carmeli.2022, lee.2017b} \\
\bottomrule
\end{tabularx}
\end{table*}

Accordingly, this characteristic can be measured using metrics that assess word formation and vocabulary. Based on the metrics found in the literature, distinct features of language morphology can be quantified. Specifically, this refers to the compression of language and the presence of redundancy or ambiguity.

\paragraph{Compression}\label{sec_taxonomy_compression}
Compression~\cite{cogswell.2019}, also known as combinatoriality~\cite{loreto.2016}, refers to the ability of a communication system to combine a small number of basic elements to create a vast range of words that can carry meaning. This feature of discrete communication is crucial in producing comprehensive and flexible communication with limited resources, and is an essential characteristic of \ac{nl}. We assume that using compressed language is generally favorable for language learners as it reduces the burden of learning~\cite{cogswell.2019}.

\paragraph{Redundancy or Ambiguity}\label{sec_taxonomy_redundancy}
In \ac{nl}, words and phrases can have redundant or ambiguous meanings. Redundancy occurs when multiple words convey the same meaning, while ambiguity arises from a limited vocabulary~\cite{lazaridou.2016, luna.2020}. The addition of this characteristic in the morphology subsection rather than the semantics subsection may be controversial. We argue that any metric measuring redundancy or ambiguity provides more useful information about the morphology, encompassing the form and size of the vocabulary, than it does about the semantic range and capabilities of the language. However, to quantify redundancy or ambiguity, we must establish semantic meaning first.

\subsubsection{Syntax}\label{sec_taxonomy_syntax}
The syntax of a language establishes the grammatical rules that govern sentence formation. Consequently, syntax plays a central role in establishing a functional correspondence between emerged language and \ac{nl}~\cite{lee.2019}. This specific characteristic of a discrete language is underrepresented in current \ac{el} literature.
However, we found two examples in the body of literature discussing syntax in \ac{el}. Ueda~et~al.~\cite{ueda.2022} introduced a method to examine the syntactic structure of an \ac{el} using categorial grammar induction (CGI), which is based on the induction of categorial grammars from sentence-meaning pairs. This method is straightforward in simple referential games. Additionally, van~der~Wal~et~al.~\cite{vanderwal.2020b} introduced unsupervised grammar induction (UGI) techniques for syntax analysis in \ac{el} research. We discuss the methods they use to measure and analyze syntax in an \ac{el} briefly in Section~\ref{sec_metrics_syntax}.

\subsubsection{Semantics}\label{sec_taxonomy_semantics}
\begin{figure}[btp]%
\centering
\includegraphics[width=0.9\textwidth]{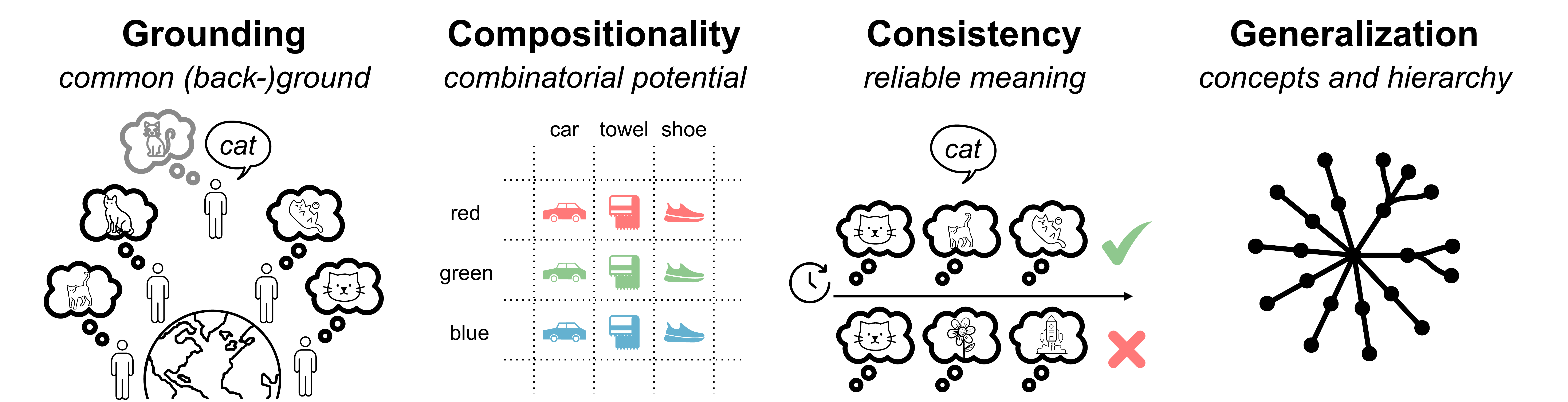}
\caption{Semantics features of language addressed in \ac{el} research: \emph{grounding}, \emph{compositionality}, \emph{consistency}, and \emph{generalization}.}
\label{fig_semantics_features}
\end{figure}

Semantics is concerned with the literal meaning of language constructs and is a dominant topic in current \ac{el} research, as shown in Table~\ref{tab:characteristics_distribution} in Appendix~\ref{sec:appendix_a}. \ac{el} studies often focus on establishing useful and meaningful communication between agents, making semantics a central feature~\cite{lazaridou.2020}. It serves as a crucial tool for distinguishing actual information exchange from mere noise utterances~\cite{bouchacourt.2019, lazaridou.2018}. Given the complexity of capturing the meaning of literal language in a single metric, several features have been introduced to measure the semantics of \ac{el}. In particular, these features include grounding, compositionality, consistency, and generalization, as shown in Figure~\ref{fig_semantics_features}. Table~\ref{tab:semantic_features} provides an overview of the literature addressing the individual semantic features in \ac{el}.

\begin{table*}[!tbp]
\caption{Semantic features discussed in the reviewed literature.}
\label{tab:semantic_features}
\begin{tabularx}{\textwidth}{lX}
\toprule
Feature & Paper  \\
\midrule
Grounding & \citenums{agarwal.2019, bouchacourt.2018, brandizzi.2023b, cao.2018, chowdhury.2020b, cowenrivers.2020, das.2017, denamganai.2023, dessi.2021, downey.2023, eloff.2021, gupta.2020, gupta.2021, havrylov.2017, keresztury.2020, kottur.2017b, kubricht.2023, lazaridou.2016, lei.2023b, li.2019, li.2021b, lipinski.2024, luna.2020, mihai.2021b, mordatch.2017, mu.2023, patel.2021, qiu.2021, santamariapang.2020, tieleman.2019, unger.2020, yao.2022, yu.2023} \\
Compositionality & \citenums{agarwal.2019, andreas.2019, auersperger.2022, bosc.2022, brandizzi.2023b, carmeli.2024, chaabouni.2020, chaabouni.2022, choi.2018, cogswell.2019, dagan.2020, denamganai.2020, denamganai.2020b, denamganai.2023b, eloff.2021, feng.2024, guo.2019, guo.2019b, guo.2020, havrylov.2017, hazra.2020, kajic.2020, kang.2020, keresztury.2020, kharitonov.2020, korbak.2019, korbak.2020, kottur.2017b, kucinski.2020, kucinski.2021, lazaridou.2018, lazaridou.2020, li.2019, loreto.2016, luna.2020, mordatch.2017, mu.2021, ohmer.2022, ossenkopf.2022, perkins.2021, ren.2020, resnick.2020, ri.2023, rita.2022, rita.2022b, santamariapang.2019, sowik.2020, sowik.2020b, ueda.2022, ueda.2023, verma.2019, xu.2022, yao.2022, yu.2023} \\
Consistency & \citenums{agarwal.2019, andreas.2019, bogin.2018, boldt.2022, bouchacourt.2019, brandizzi.2023b, choi.2018, dagan.2020, dessi.2019, dessi.2021, dubova.2020, eccles.2019, gupta.2020, hazra.2020, jaques.2018, kang.2020, korbak.2019, korbak.2020, kucinski.2021, liang.2020, lipinski.2024, lo.2022, lorkiewicz.2011, lowe.2019, luna.2020, mihai.2019, mordatch.2017, mu.2021, mul.2019, ohmer.2022, qiu.2021, rita.2022, santamariapang.2020, unger.2020, verma.2019, yu.2022} \\
Generalization & \citenums{auersperger.2022, brandizzi.2023b, bullard.2020, bullard.2021, chaabouni.2020, chaabouni.2022, choi.2018, colas.2020, denamganai.2020b, denamganai.2023b, eloff.2021, feng.2024, guo.2019, guo.2020, guo.2021b, hildreth.2019, korbak.2019, korbak.2020, kottur.2017b, kucinski.2021, lazaridou.2016, lazaridou.2018, luna.2020, mu.2021, mul.2019, ohmer.2022, qiu.2021, ren.2020, ri.2023, rita.2022, rita.2022b, santamariapang.2020, sowik.2020, taylor.2021, tucker.2021, vani.2021, verma.2019, xu.2022, yao.2022, yu.2022} \\
\bottomrule
\end{tabularx}
\end{table*}

\paragraph{Grounding}\label{sec_taxonomy_grounding}
A language is considered grounded when it is deeply intertwined with the environment, for example, when it is tightly bound to environmental concepts and objects~\cite{santamariapang.2019, denamganai.2023b, clark.2012}. Grounding is essential for the interoperability of individuals and is particularly important in \ac{nl} communication, where meaningful interaction requires shared understanding~\cite{galke.2022, bogin.2018, lin.2021}. While in theory, an \ac{el} can establish a unique form of grounding using self-emerged concepts distinct from those in \ac{nl}, deriving a useful metric for such a scenario proves challenging. This difficulty arises from the need to compare \acp{el} to existing and comprehensible grounding principles typically found in \acp{nl}~\cite{denamganai.2020, kottur.2017b, lin.2021}.

\paragraph{Compositionality}\label{sec_taxonomy_compositionality}
When a language exhibits compositionality, its components can be rearranged or replaced by conceptually equivalent words without changing the overall meaning~\cite{chaabouni.2020, kharitonov.2020}. Compositionality facilitates the construction of higher-level concepts, using conceptual foundations to enable efficient language expression~\cite{keresztury.2020, korbak.2020, auersperger.2022}. For example, \acp{nl} partition concepts such as objects and their attributes to allow compositional constructions~\cite{mordatch.2017, guo.2019}. As a result, we can describe variations of a single object using different words from the same semantic concept, such as \textquoteleft{blue towel} and \textquoteleft{red towel} for the object towel and the semantic concept of color. Similarly, we can attribute specific properties to different objects using the same phrase, as in \textquoteleft{green towel} and \textquoteleft{green car}. Ultimately, compositionality is beneficial for the learning process~\cite{chaabouni.2020, galke.2022} and promotes efficient and rich language use, even in systems with limited memory capacity~\cite{mordatch.2017, resnick.2020, liang.2020}.

\paragraph{Consistency}\label{sec_taxonomy_consistency}
Merely having grounded words in a language does not necessarily guarantee its semantic quality. In addition, consistency is essential for a language to convey meaningful and practical information effectively~\cite{hockett.1960, galke.2022}. If the words within a language lack consistency in their literal meanings, they will not facilitate effective communication. Therefore, even if a language is semantically grounded and compositional, its utility is compromised if the words exhibit inconsistent literal meanings~\cite{kottur.2017b}. While words can change their general meaning to fit the context, their literal meaning should remain consistent to keep their usefulness~\cite{bogin.2018}.

\paragraph{Generalization}\label{sec_taxonomy_generalization}
Generalization serves as a cornerstone of \ac{nl}, allowing humans to communicate about topics ranging from simple to complex, broad to specific, and known to unknown, all with a relatively limited vocabulary~\cite{auersperger.2022, chaabouni.2022}. A language that excels at generalization enables its users to navigate different levels of complexity, facilitating hierarchical descriptions of concepts and relationships~\cite{mu.2021}. Consequently, generalization and compositionality are closely related, as they both contribute to the flexibility and expressiveness of language~\cite{galke.2022, lazaridou.2018, ren.2020}. This ability to generalize not only enriches communication but also underscores the adaptability and robustness of human language.

\subsubsection{Pragmatics}\label{sec_taxonomy_pragmatics}
The final dimension of \ac{el} research is pragmatics. This field of study examines how language is employed in context, particularly in interactions, and how it conveys information~\cite{brinton.2010, bard.2020, kang.2020}. By evaluating the pragmatics of the linguistic structure, we can ascertain whether \ac{el} is itself useful and utilized effectively. While this assessment may be feasible based on rewards in a standard \ac{rl} setting, integrating communication into such environments increases the complexity. This is because most setups do not separate the agent's environment interaction from its communication capabilities, thereby expanding the network's capacity, and making it difficult to attribute an increase in reward directly to \ac{el}~\cite{lazaridou.2020}.

\begin{figure}[btp]%
\centering
\includegraphics[width=0.9\textwidth]{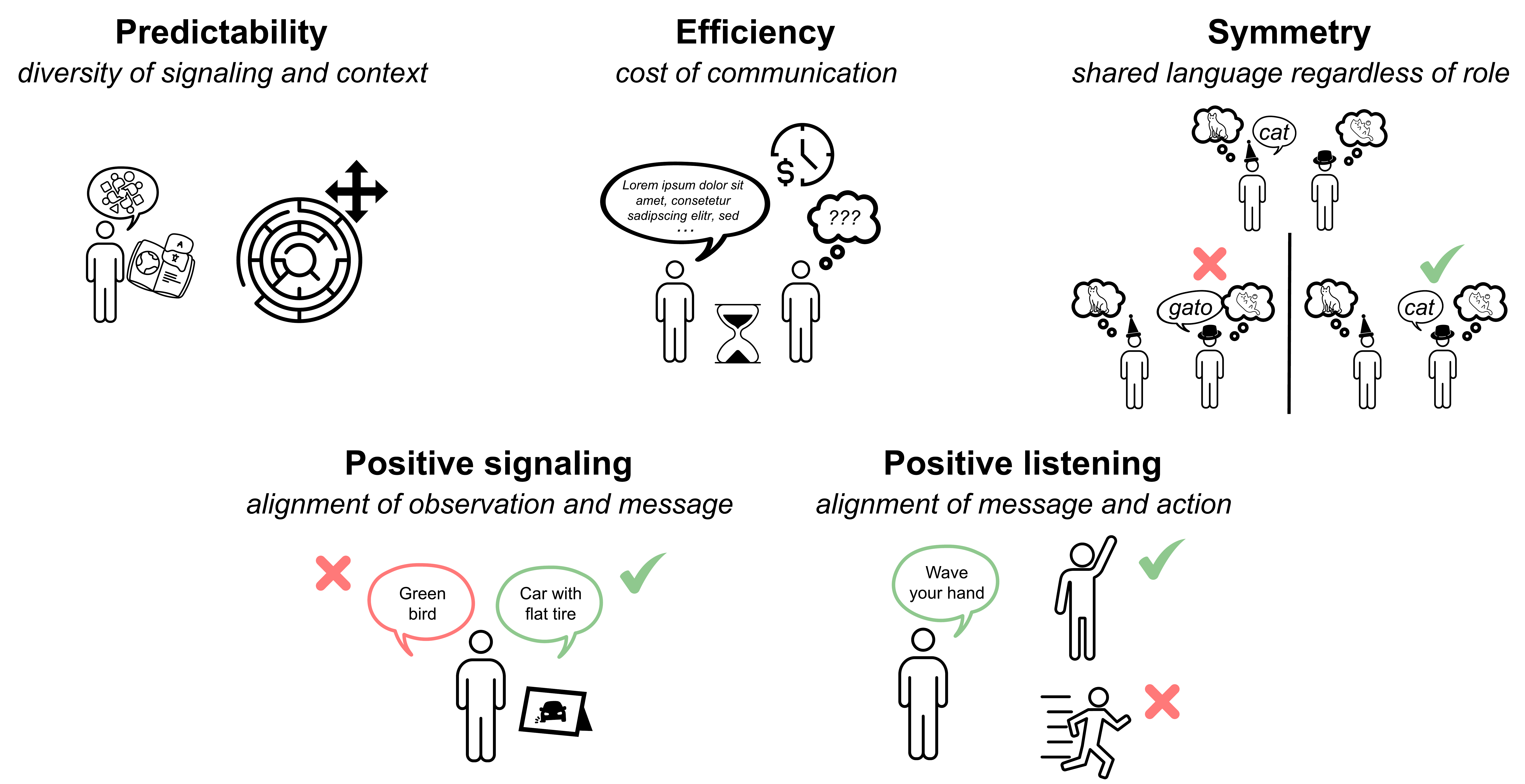}
\caption{Pragmatics features of language in \ac{el} research. These illustrations are intended to promote an intuitive understanding of the categories \emph{predictability}, \emph{efficiency}, \emph{positive signaling}, \emph{positive listening}, and \emph{symmetry}.}
\label{fig_pragmatics_features}
\end{figure}

As outlined in Table~\ref{tab:pragmatic_features} and depicted in Figure~\ref{fig_pragmatics_features}, five distinct features have been identified: predictability, efficiency, positive signaling, positive listening, and symmetry. These features are essential for assessing the constructive impact and utilization of \ac{el}. Understanding how agents employ language is crucial in evaluating its effectiveness and overall benefit.

\begin{table*}[!tbp]
\caption{Pragmatic features discussed in the reviewed literature.}
\label{tab:pragmatic_features}
\begin{tabularx}{\textwidth}{lX}
\toprule
Feature & Paper  \\
\midrule
Predictability & \citenums{dubova.2020} \\
Efficiency & \citenums{kalinowska.2022} \\
Positive signaling & \citenums{ampatzis.2008, bouchacourt.2019, brandizzi.2023b, cao.2018, cope.2020, cowenrivers.2020, dubova.2020, eccles.2019, jaques.2018, lazaridou.2020, li.2021b, liang.2020, lowe.2019, noukhovitch.2021, patel.2021, portelance.2021, sowik.2020, sowik.2020b, zhu.2024} \\
Positive listening & \citenums{ampatzis.2008, bouchacourt.2019, brandizzi.2023b, cope.2020, cowenrivers.2020, dubova.2020, eccles.2019, jaques.2018, lazaridou.2020, li.2021b, liang.2020, lin.2021, lowe.2019, mul.2019, noukhovitch.2021, patel.2021, sowik.2020, sowik.2020b, zhu.2024} \\
Symmetry & \citenums{cogswell.2019, dubova.2020} \\
\bottomrule
\end{tabularx}
\end{table*}

\paragraph{Predictability}\label{sec_taxonomy_predictability}
Predictability is concerned with the assessment of the complexity of the context, including the action space within the environment. When actions exhibit less diversity, it becomes more feasible to coordinate without communication~\cite{dubova.2020}. For instance, in a simple grid-based environment where agents have only two possible actions — moving left or right — agents can often achieve their objectives without the need for communication. In such a scenario, the limited action space reduces the necessity for \ac{el}, as agents can predict each other's movements based on past behavior or simple rules. However, in a more complex environment where agents have multiple actions, such as navigating a maze with numerous paths and obstacles, the need for effective communication increases. Here, \ac{el} can significantly enhance coordination by allowing agents to share information about their positions, plans, or discoveries, thus improving their overall performance in navigating the maze. Therefore, it is essential to compare the diversity of signaling and context attributes to evaluate the potential benefit of \ac{el}.

\paragraph{Efficiency}\label{sec_taxonomy_efficiency}
Efficiency is a critical aspect considered whenever communication entails a cost. This is particularly true in the context of modeling the emergence of \ac{nl} and the broader objective of employing \ac{el} for \ac{hci}. In \ac{el} settings, the achievement of concise communication is contingent upon the presence of an opportunity cost~\cite{kalinowska.2022}. Without such a cost, there is no incentive to communicate concisely, making \ac{el} ineffective as an intermediary for \ac{hci}. When communication is accompanied by a cost the necessity for efficiency in communication becomes paramount. In such scenarios, the objective is to minimize the cost while maximizing the effectiveness of communication within a given task.

\paragraph{Positive Signaling}\label{sec_taxonomy_positivesignalling}
The concept of positive signaling is concerned with the degree of alignment between the observations, knowledge, and experience of the message producer and their communication output~\cite{lowe.2019}. The objective is to guarantee the transmission of useful information, or at the very least, information that the speaker can discern through knowledge or observation~\cite{portelance.2021}. This feature assumes that all communication should be relevant to something observable, known or tangible to the speaker. Thus, it stipulates that the situational information content of the produced signal is crucial for a language that can be used in a contextually meaningful way.

\paragraph{Positive Listening}\label{sec_taxonomy_positivelistening}
Positive listening, in contrast to positive signaling, focuses on the role of the message receiver, to evaluate the active processing of incoming information~\cite{lowe.2019}. 
From a pragmatic point of view, it makes sense to process incoming messages in a meaningful way. However, the definition of meaningful processing is broad. Positive listening as defined in this taxonomy, contrary to earlier work~\cite{lowe.2019, bouchacourt.2019}, does not necessarily require a connection between the incoming message and the subsequent action; it is much more about active processing, which may or may not lead to inclusion in the choice of action. Thus, active engagement followed by rejection or disregard is also considered positive listening in the context of this taxonomy.

\paragraph{Symmetry}\label{sec_taxonomy_symmetry}
Symmetry in \ac{el} is defined as the consistency in language usage among participating agents~\cite{dubova.2020, cogswell.2019}. This concept applies to \ac{marl} settings where agents can assume multiple roles, such as message producer and message receiver. Symmetry plays a crucial role in achieving convergence on a shared and aligned \ac{el}. For instance, if an agent employs language differently depending on whether it is sending or receiving messages so that words have varying meanings based on the assigned role the \ac{el} setting is considered asymmetric. In such instances, rather than learning a collectively and contextually grounded language, agents develop individual protocols specific to their respective roles~\cite{dubova.2020}. This would suggest that there is no common language, but rather separate codes that can only be applied to specific combinations and conditions. For this reason, this pragmatic feature is particularly relevant, since the aim of \ac{el} is a common language.

\subsection{Summary of the Taxonomy}\label{sec_taxonomy_summary}
Our proposed taxonomy systematically categorizes the key features of \ac{el} systems, including communication settings, language games, language priors, and language characteristics. The latter is particularly detailed, with sub-characteristics and their features aligned with the major levels of linguistic structure, as previously illustrated in Figure~\ref{fig_major_levels_of_linguistic_structure}. This comprehensive taxonomy enables a standardized comparison of approaches in the \ac{el} literature, highlighting the opportunities and properties associated with individual options and topics in \ac{el} research. Specifically, by applying this taxonomy, especially in terms of language characteristics, we can uncover the capabilities and potentials of various \ac{el} approaches. This facilitates a more detailed, comparable, and insightful analysis of \ac{el}.


\section{Metrics}\label{sec_metrics}
This section provides a comprehensive categorization and review of existing metrics used in EL research. The section is organized along the same categorization used in Section~\ref{sec_taxonomy_languagecharacteristics}. Note that the categories of phonetics and phonology are excluded from this discussion, as these aspects are predetermined settings in the current \ac{el} literature and thus not yet targeted by metrics.

We begin by introducing the notational system used for all metrics to ensure consistency and facilitate ease of use. We then describe the metrics within each category, detailing the individual metric and adapting it to our notation. For each metric, we provide references to both original sources and additional literature, if available, to enable further exploration beyond the scope of this work. Figure~\ref{fig_metrics_mindmap} provides a visual summary of the existing metrics and their correspondence to the language characteristics. An extended version including all references for the individual metrics is provided in Figure~\ref{fig_metrics_mindmap_with_sources} in Appendix~\ref{sec:appendix_b}

{
\hypersetup{hidelinks}
    \tikzset{octagon/.style={regular polygon,regular polygon sides=8},}
    \begin{figure}
    \centering
    \resizebox{\columnwidth}{!}{%
        \begin{tikzpicture}[mindmap, grow cyclic, every node/.style=concept, concept color=JP-lighter_grey,
            level 1 concept/.append style={level distance=4.2cm,sibling angle=-(360/4)), minimum size=2.0cm},
            level 2 concept/.append style={level distance=2.8cm, sibling angle=-45, minimum size=1.7cm},
            level 3 concept/.append style={level distance=2.5cm, sibling angle=-40, minimum size=1.0cm, outer sep=-0.1em, concept/.append style={octagon}}]
        \node[style={minimum size=3cm, text width=3cm}]{Emergent\\ Language\\ Metrics}
            child [concept color=JP-purple, rotate = -20] { 
                node{\hyperref[sec_metrics_morphology]{Morphology}}
                child { 
                    node{\hyperref[sec_metrics_compression]{Compression}}
                    child [concept color=JP-purple_50p, rotate = 10] { node {\hyperref[sec_metrics_compression_distinct_appearances]{Distinct Appearances}}}
                    child [concept color=JP-purple_50p, rotate = 10] { node {\hyperref[sec_metrics_compression_average_message_length]{Average Message Length}}}
                    child [concept color=JP-purple_50p, rotate = 10] { node {\hyperref[sec_metrics_compression_active_words]{Active Words}}}
                }
                child [rotate = -10] { 
                    node {\hyperref[sec_metrics_redundancy_or_ambiguity]{Redundancy or \mbox{Ambiguity}}}
                    child [concept color=JP-purple_50p] { node {\hyperref[sec_metrics_redundancy_or_ambiguity_perplexity]{Perplexity}}}
                    child [concept color=JP-purple_50p] { node {\hyperref[sec_metrics_redundancy_or_ambiguity_svd]{SVD}}}
                    child [concept color=JP-purple_50p] { node {\hyperref[sec_metrics_redundancy_or_ambiguity_message_distinctness]{Message Distinctness}}}
                }
            }
            child [concept color=JP-red, rotate = 20] { 
                node {\hyperref[sec_metrics_syntax]{Syntax}}
                child [concept color=JP-red_50p, level distance=2.5cm, outer sep=-0.1em] { node[octagon, text width=2.5em]{\hyperref[sec_metrics_syntax_syntax_tree]{Syntax \\ Tree}}}
                child [concept color=JP-red_50p, level distance=2.5cm, outer sep=-0.1em] { node[octagon, text width=2.5em]{\hyperref[sec_metrics_syntax_CGI]{CGI}}}
            }
            child [concept color=JP-orange, rotate = 20, sibling angle=-60] { 
                node {\hyperref[sec_metrics_semantics]{Semantics}}
                child { 
                    node {\hyperref[sec_metrics_grounding]{Grounding}}
                    child [concept color=JP-orange_50p, rotate = 30] { node {\hyperref[sec_metrics_grounding_divergence]{Divergence}}}
                    child [concept color=JP-orange_50p, rotate = 30] { node {\hyperref[sec_metrics_grounding_purity]{Purity}}}
                    child [concept color=JP-orange_50p, rotate = 30] { node {\hyperref[sec_metrics_grounding_rsa]{Represen\-tational \mbox{Similarity} Analysis}}}
                }
                child [rotate = -12, level distance = 3.4cm] { 
                    node {\hyperref[sec_metrics_compositionality]{Composi\-tionality}}
                    child [concept color=JP-orange_50p, rotate = 10] { node {\hyperref[sec_metrics_compositionality_topsim]{Topo\-graphic Similarity}}}
                    child [concept color=JP-orange_50p, rotate = 10] { node {\hyperref[sec_metrics_compositionality_posdis]{Positional Disentanglement}}}
                    child [concept color=JP-orange_50p, rotate = 10] { node {\hyperref[sec_metrics_compositionality_bosdis]{Bag of Symbols Disentanglement}}}
                    child [concept color=JP-orange_50p, rotate = 10] { node {\hyperref[sec_metrics_compositionality_tre]{Tree Reconstruct Error}}}
                    child [concept color=JP-orange_50p, rotate = 10] { node {\hyperref[sec_metrics_compositionality_conflictcount]{Conflict Count}}}
                }
                child [rotate = -55, level distance = 3.4cm] { 
                    node {\hyperref[sec_metrics_consistency]{Consistency}}
                    child [concept color=JP-orange_50p, rotate = 30] { node {\hyperref[sec_metrics_consistency_mutualinformation]{Mutual Information}}}
                    child [concept color=JP-orange_50p, rotate = 30] { node {\hyperref[sec_metrics_consistency_correlation]{Correlation}}}
                    child [concept color=JP-orange_50p, rotate = 30] { node {\hyperref[sec_metrics_consistency_coherence]{Coherence}}}
                    child [concept color=JP-orange_50p, rotate = 30] { node {\hyperref[sec_metrics_consistency_entropy]{Entropy}}}
                    child [concept color=JP-orange_50p, rotate = 30] { node {\hyperref[sec_metrics_consistency_similarity]{Similarity}}}
                }
                child [rotate = -60] { 
                    node {\hyperref[sec_metrics_generalization]{General\-ization}}
                    child [concept color=JP-orange_50p] { node {\hyperref[sec_metrics_generalization_zero_shot]{Zero Shot Evaluation}}}
                    child [concept color=JP-orange_50p] { node {\hyperref[sec_metrics_generalization_ETL]{Ease and Transfer Learning}}}
                }
            }
            child [concept color=JP-yellow, rotate = -20] { 
                node {\hyperref[sec_metrics_pragmatics]{Pragmatics}}
                child { 
                    node {\hyperref[sec_metrics_symmetry]{Symmetry}}
                    child [concept color=JP-yellow_50p] { node {\hyperref[sec_metrics_symmetry_interagent]{Inter-Agent Divergence}}}
                    child [concept color=JP-yellow_50p] { node {\hyperref[sec_metrics_symmetry_withinagent]{Within-Agent Divergence}}}
                }
                child { 
                    node {\hyperref[sec_metrics_pragmatics_predictability]{Pre\-dictability}}
                    child [concept color=JP-yellow_50p] { node {\hyperref[sec_metrics_pragmatics_predictability_behavioral_divergence]{\mbox{Behavioral} \mbox{Divergence}}}}
                }
                child { 
                    node {\hyperref[sec_metrics_pragmatics_efficiency]{Efficiency}}
                    child [concept color=JP-yellow_50p] { node {\hyperref[sec_metrics_pragmatics_efficiency_sparsity]{Sparsity}}}
                }
                child { 
                    node {\hyperref[sec_metrics_pragmatics_positive_signaling]{Positive Signaling}}
                    child [concept color=JP-yellow_50p] { node {\hyperref[sec_metrics_pragmatics_positive_signaling_speaker_consistency]{Speaker Consistency}}}
                }
                child { 
                    node {\hyperref[sec_metrics_pragmatics_positive_listening]{Positive Listening}}
                    child [concept color=JP-yellow_50p] { node {\hyperref[sec_metrics_pragmatics_positive_listening_instantaneous_coordination]{Instan\-taneous Coordination}}}
                    child [concept color=JP-yellow_50p] { node {\hyperref[sec_metrics_pragmatics_positive_listening_message_effect]{Message Effect}}}
                    child [concept color=JP-yellow_50p] { node {\hyperref[sec_metrics_pragmatics_positive_listening_causal_influence]{Causal \mbox{Influence} of Communication}}}
                }
            };
        \end{tikzpicture}
    }
    \caption{
        Graph presenting a visual representation of the metrics identified in the surveyed literature, sorted by language characteristics. Each node contains a link to the corresponding section that describes the metric in detail, allowing for convenient navigation.
        \\
        \tikzcircle[JP-purple]{}\tikzcircle[JP-red]{}\tikzcircle[JP-orange]{}\tikzcircle[JP-yellow]{}: Language characteristics (inner nodes)
        \\
        \tikzoctagon[JP-purple_50p]{}\tikzoctagon[JP-red_50p]{}\tikzoctagon[JP-orange_50p]{}\tikzoctagon[JP-yellow_50p]{}: Individual metrics (outermost nodes)
    }
    \label{fig_metrics_mindmap}
    \end{figure}
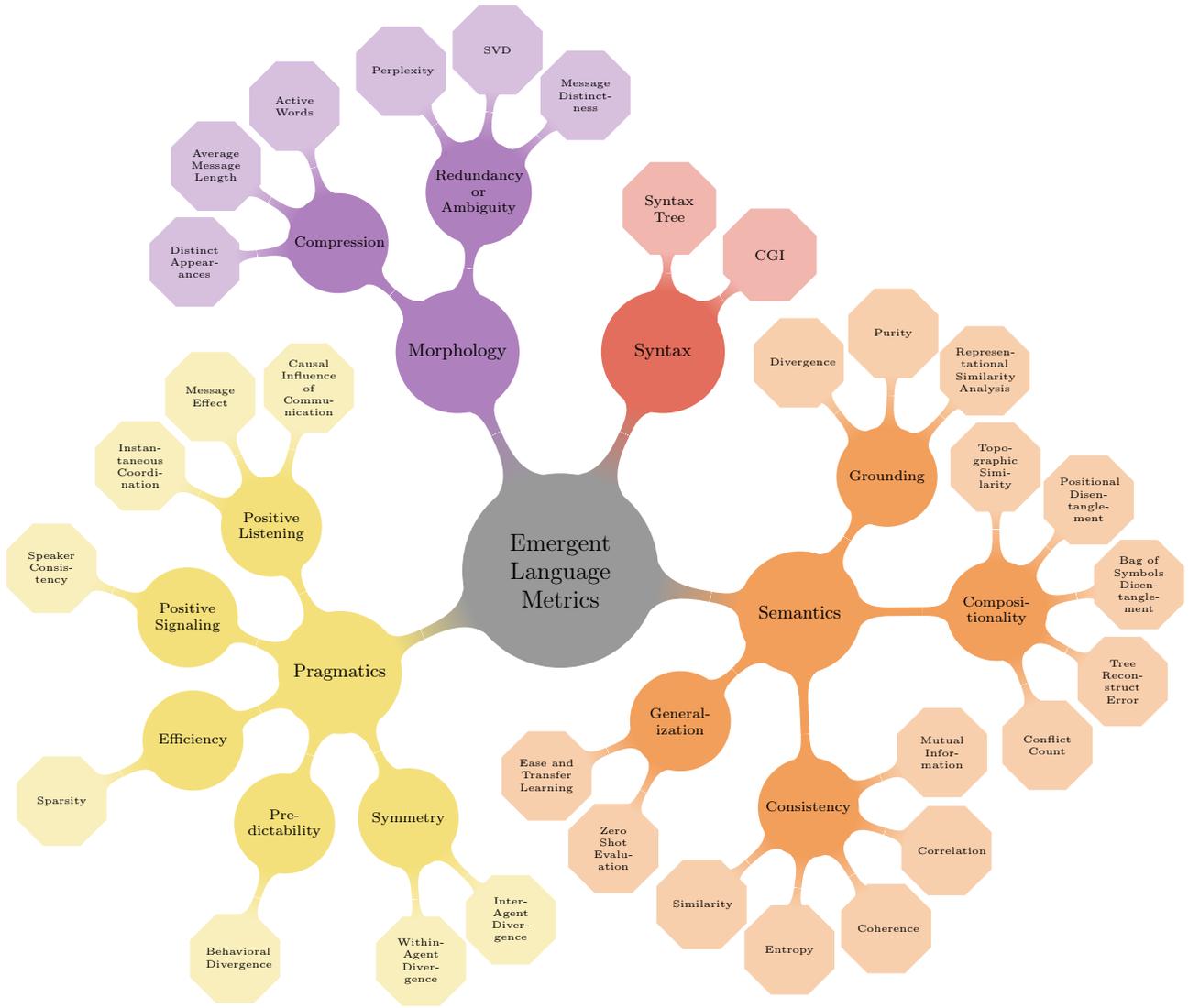
}

\subsection{Notation}\label{sec_metrics_notation}
Given the complexity and variability within the \ac{el} field, it is crucial to establish a unified and coherent notation system. In this section we present a standardized mathematical notation designed to be consistent across the various aspects of \ac{el} research, thereby facilitating clearer communication and comparison of results within the community. This approach aligns with our broader goal of advancing the field through a common taxonomy that supports the development of measurable and interpretable \acp{el}. Throughout this section we focus on finite and discrete languages, although some of the definitions and metrics discussed here are also applicable to continuous languages. These languages offer a more straightforward mapping to \acp{nl}, making them particularly relevant to the study of \ac{el} systems.

\subsubsection{Definition}\label{sec_metrics_notation_definition}
\begin{figure}[btp]%
\centering
\includegraphics[width=0.9\textwidth]{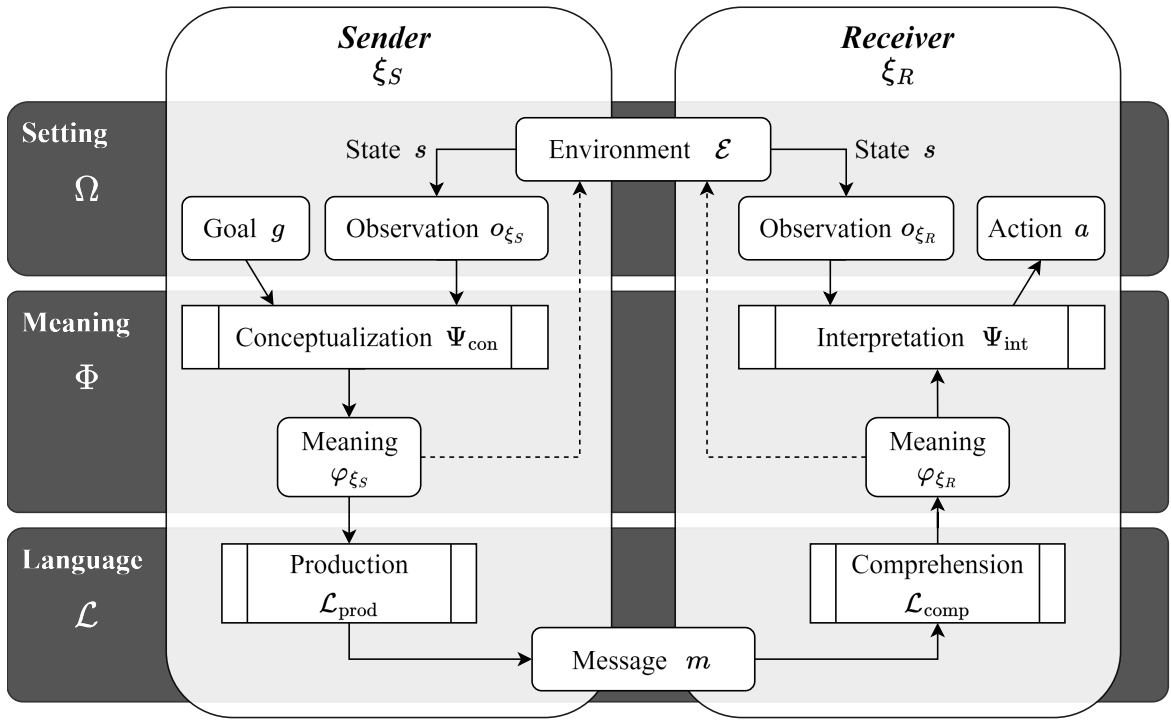}
\caption{The proposed notation is integrated within the semiotic cycle framework (cf. Figure~\ref{fig_semiotic_cycle}). The language $\mathcal{L}$, along with its associated mapping functions $\mathcal{L}{\text{prod}}$ (production) and $\mathcal{L}{\text{comp}}$ (comprehension), forms the core of \ac{el} research. These linguistic components, however, are underpinned by the representations learned in the meaning space $\Phi$, which play a crucial role in guiding and shaping effective communication. All elements in the sender or receiver domain are entity-specific, but we only index duplicate symbols to keep it concise. Adapted from~\protect\cite{bleys.2015, vaneecke.2020}.}
\label{fig_semiotic_cycle_EL_Notation}
\end{figure}

In alignment with the semiotic cycle introduced in Section~\ref{sec_background_naturallanguage}, our notation is organized into three interconnected spaces: setting, meaning, and language. The setting space encompasses the typical elements of \ac{rl}, providing the foundational environment in which agents operate. The meaning space incorporates a representation learning endeavor, whereby sensory input is integrated with decision-relevant information to generate a coherent internal representation. Finally, the language space encompasses both the production and comprehension of discrete messages, encapsulating the communication process. These components, illustrated in Figure~\ref{fig_semiotic_cycle_EL_Notation}, will be introduced and explored in detail in the following paragraphs.

\paragraph{Setting}
The overall setting, consisting of the environment, actions, goals, and other typical \ac{rl} elements, is denoted by $\Omega$. Let $\xi$ denote the set of all entities in the system, with an individual entity represented as $\xi_i \in \xi$. Each entity can assume specific roles, such as the sender ($S$) or receiver ($R$) in a communication scenario. An entity can assume several roles over the course of the entire communication scenario. However, for an individual message exchange, an entity assumes one specific role. We represent the role of an individual entity $i$ by $\xi_{i,j} \in \xi_i$, where $j$ specifies the role (e.g., $j = S$ or $j = R$). 

Entities interact with their environment $\mathcal{E}$ through actions, denoted as $a$, which belong to the set of possible actions $A$, such that $a \in A$. The action taken by a specific entity $\xi_i$ is represented as $a_{\xi_i}$. The state of the environment at any given time is denoted by $s$, which is an element of the state space $S$, so that $s \in S$. As the system progresses over time, denoted by discrete points in time $\left[ 0, \dots, t \right]$, the sequence of states and actions forms a trajectory $\tau$, generally expressed as $\tau = \left\{ s^0, a^0, \dots, s^t, a^t \right\}$. It is important to note that the entities described here do not necessarily correspond to autonomous agents in the traditional sense; they could also represent ground truth models, human participants, or abstract constructs that lack the direct interaction capabilities typically associated with agents. Despite this distinction, for the sake of clarity and consistency, we will refer to these entities as agents in the following sections.

Given the importance of partial observability in \ac{el} research~\cite{lobostsunekawa.2022, denamganai.2020, mordatch.2017, taylor.2021}, it is essential to consider that agents only have access to their own observations, denoted as $o_{\xi}$, which are derived from the underlying state $s$. An individual observation $o_{\xi}$ is an element of the collection of observations of an agent $O_{\xi}$, which is a subset of the observation space $O$, so that $o_{\xi} \in O_{\xi} \subseteq O$. In our framework, an observation $o_{\xi}$ effectively replaces the \enquote*{world model} component from the traditional semiotic cycle, highlighting the localized and subjective nature of an agent's perception in partially observable environments.

Referential games (cf. Table~\ref{tab:games_overview}) are frequently employed in \ac{el} literature. They often operate on individual, static samples that are drawn from a corresponding dataset or distribution. In doing so, they differ from traditional \ac{rl} setups that emphasize sequential decision-making and environmental interactions over time. In such cases, rather than speaking of a state $s$ or an observation $o$, we use the term sample $k$, which is an element of the collection of all samples $K$, so that $k \in K$. The specific nature of a sample depends on the environment; for example, in an image-based sender-receiver game, the sample would be an image. Each sample is represented by its feature vector $f$, which belongs to the feature space $F$, so that $f \in F$. The feature vector corresponding to a specific sample $k$ is denoted by $f_k$.

In \ac{el} settings, the communicative goal $g$ of an agent may differ from the (reinforcement) learning task goal. In addition, depending on the game, the sender and receiver may have distinct goals. These are important factors to consider when evaluating the communicative behavior.

\paragraph{Meaning}
In our notation, the meaning space, denoted by $\Phi$, serves as the critical intermediary between the setting space and the language space. The meaning space represents the semantic connections derived from the provided information. Each element within this space, represented by a specific meaning vector $\varphi \in \Phi$, captures the essence of concepts or objects as understood by the agent. These meaning vectors are critical to the processes of language comprehension and production, as well as to the processes of conceptualization and interpretation, that allow an agent to effectively use inputs and generate outputs in the setting space (cf. Figure~\ref{fig_semiotic_cycle_EL_Notation}).

The representation mappings $\Psi$ within the meaning space are agent-specific and referred to as $\Psi_{\text{con}}$ and $\Psi_{\text{int}}$, given in Equation~\ref{eq_meaning_mappings_psi}. These mappings enable the transition between an arbitrary space $\chi$, such as sensory inputs or raw data, and the meaning space, where the data acquires semantic meaning. $\Psi_{\text{con}}$ refers to the conceptualization process that transforms raw, uninterpreted data into meaningful representations within $\Phi$. Conversely, $\Psi_{\text{int}}$ denotes the interpretation process that translates these meaning vectors back into the arbitrary space that can represent any external or internal stimuli. These mappings are critical to the agent's ability to both understand its environment and communicate effectively within it through language that is both grounded in and reflective of the underlying reality with which the agents interact.

\begin{equation}\label{eq_meaning_mappings_psi}
    \Psi = 
    \begin{cases*}
        \Psi_{\text{con}} : \chi \to \Phi \\
        \Psi_{\text{int}} : \Phi \to \chi
    \end{cases*}
\end{equation}

\paragraph{Language}
In our proposed framework, a message $m$ belongs to the message space $M$, such that $m \in M$. Each message encapsulates semantic and pragmatic content, serving as a vehicle for meaningful communication between agents. A message is composed of individual words $w$, which are elements of a finite collection $W$, commonly referred to as vocabulary, lexicon, or dictionary. In this context, each word is considered a semantic unit that carries (intrinsic) meaning. At the lowest level, a word is composed of characters or symbols $\upsilon \in \Upsilon$. These atomic characters, while essential for constructing words, do not independently carry semantic meaning. Instead, they function as elements of a finite set $\Upsilon$ from which any number of meaningful words can be composed.

Building on the formalization from~\cite{resnick.2020}, we describe the message space $M_{\xi}$ of an agent $\xi$, which represents the agent's language capabilities from a compositional standpoint. The message space $M_{\xi} \subseteq M$ is composed of a set of messages or strings $m_{\xi}$, each constructed from words within $W_{\xi}$, as shown in Equation~\ref{eq_entity_messages_to_words}. Further, each $w_{\xi} \in m_{\xi}$ is composed of a set of characters $\upsilon_{\xi} \in \Upsilon_{\xi} \subseteq \Upsilon$ utilized by the agent, given by Equation~\ref{eq_entity_words_to_characters}.

\begin{equation}\label{eq_entity_messages_to_words}
\begin{aligned}
    m_{\xi} 
    & \subseteq M_{\xi} \\
    & =
        \left\{ 
            w_{\xi}
            \mid
             w_{\xi} \in W_{\xi} \subseteq W
             \land \lvert w_{\xi} \rvert \geq 0
        \right\}
\end{aligned}
\end{equation}

\begin{equation}\label{eq_entity_words_to_characters}
\begin{aligned}
    w_{\xi} 
    & \subseteq W_{\xi} \\
    & =
        \left\{ 
            \upsilon_{\xi}
            \mid
             \upsilon_{\xi} \in \Upsilon_{\xi} \subseteq \Upsilon
             \land \lvert \upsilon_{\xi} \rvert \geq 0
        \right\}
\end{aligned}
\end{equation}

A language $\mathcal{L}$ encompasses a set of mapping functions that facilitate the transformation between the message space $M$ and other arbitrary spaces $\chi$. These mappings are agent-specific and enable both the production of messages, denoted as $\mathcal{L}_{\text{prod}}$, and the comprehension of messages, denoted as $\mathcal{L}_{\text{comp}}$. This framework aligns with the linguistic level description of the semiotic cycle presented in Figure~\ref{fig_semiotic_cycle}. Within this context, we formally define a language $\mathcal{L}$ in Equation~\ref{eq_language_definition_prod_comp}.

\begin{equation}\label{eq_language_definition_prod_comp}
    \mathcal{L} = 
    \begin{cases*}
        \mathcal{L}_{\text{prod}} : \chi \to M \\
        \mathcal{L}_{\text{comp}} : M \to \chi 
    \end{cases*}
\end{equation}

These emerging mapping functions are not necessarily injective, meaning that distinct inputs from the space $\chi$ could potentially be mapped to an identical message within $M$~\cite{ren.2020, guo.2021b}. Conversely, distinct messages within $M$ could also be mapped to the same value in $\chi$. While this non-injectivity adds a layer of complexity to the expressiveness of the language, it also introduces a degree of flexibility that can be advantageous in certain communication scenarios. For example, it allows for synonymy (where different messages convey the same meaning), which can provide redundancy and flexibility in communication, and homonymy (where the same message may have multiple interpretations depending on context), which can facilitate more nuanced and context-dependent communication. These natural phenomena, though challenging, are well-documented in \acp{nl} and are of particular interest in the design and evaluation of artificial communication systems~\cite{brinton.2010, lipowska.2018}. However, managing these complexities effectively is crucial, as unchecked non-injectivity could lead to ambiguities that complicate communication rather than simplifying it.

\subsubsection{Important Notes}\label{sec_metrics_notation_important_notes}
The notation presented here is designed to be comprehensible and thorough; however, it may not be directly applicable in all cases to existing works, as these employ different wordings. a lot of existing work uses the term \enquote*{word}, which in our notation describes element carrying semantic meaning, and \enquote*{symbols}, which in our notation serve as fundamental building blocks without inherent semantic meaning, interchangeably~\cite{das.2017, dessi.2019, lazaridou.2020, sowik.2020, andreas.2019}. Furthermore, a considerable proportion of existing literature utilizes a multitude of different definitions for concepts such as \enquote*{meaning space}~\cite{portelance.2021, perkins.2021b, bosc.2022}, \enquote*{ground-truth oracle}~\cite{andreas.2019, denamganai.2023}, and other pivotal elements. In our endeavor to establish a unified framework, we have occasionally adopted terminology that differs from that used by the original authors. While this may initially lead to some confusion, we intend to mitigate this by providing transparent and detailed descriptions. Our objective is a consistent application of these concepts across the field of \ac{el} research, thereby promoting coherence between different studies. The following sections attempt to align existing research and metrics with the proposed framework. While this alignment has required some linguistic adjustments to existing terminology and procedures, it is important to note that no substantive changes have been made to the underlying methodologies.

\subsection{Morphology}\label{sec_metrics_morphology}
Morphological metrics aim to evaluate the structure and formation of words within a language, as well as the richness and diversity of its vocabulary. The identified metrics focus on aspects such as language compression, redundancy, and ambiguity. The morphology of a language significantly influences the complexity of language based tasks~\cite{park.2021}. Therefore, the evaluation of morphological features is a crucial component for understanding and evaluating the effectiveness of \acp{el}.

\subsubsection{Compression}\label{sec_metrics_compression}
The concept of compression within a language refers to its ability to efficiently combine and reuse a limited set of characters to generate a large collection of words or meanings~\cite{loreto.2016, cogswell.2019}. Several metrics can be used to quantify compression in \acp{el}. A straightforward approach for these metrics is to use statistical measures, as shown in the following paragraphs. These metrics provide insight into the efficiency of the language, indicating how well it minimizes redundancy while maximizing expressiveness. Efficient compression is a key indicator of a communication system, especially in scenarios where resources (such as memory or bandwidth) are constrained.

\paragraph{Distinct Appearances}\label{sec_metrics_compression_distinct_appearances}
The metric of distinct appearances ($\operatorname{DA}$) was proposed by Loreto et~al.~\cite{loreto.2016}. It is formalized in Equation~\ref{eq_morphology_compression_distinct_appearance} and designed to quantify the capacity of a communication system to name a diverse set of objects or categories using its available symbols~\cite{loreto.2016}. Specifically, this metric evaluates how frequently characters $\upsilon \in \Upsilon$ are reused across different words or names $w$ within the lexicon $W$. By examining the set $W_{\upsilon}$, which includes all words containing a given character $\upsilon$, we can assess the system's flexibility in recombining basic units to generate a broad spectrum of expressions.

A high $\operatorname{DA}$ value, approaching $1$, indicates that the characters are highly versatile and reused extensively across different words, thereby reflecting a flexible communication system. Conversely, a low $\operatorname{DA}$ value suggests limited reuse of characters, which may imply constraints in the system's expressiveness or a less efficient use of its symbolic resources. This metric provides insights into how efficiently a system can balance the trade-off between a compact character set and the richness of its vocabulary.

\begin{equation}\label{eq_morphology_compression_distinct_appearance}
    \operatorname{DA} 
        = \frac{\sum_{\upsilon \in \Upsilon} \left( \lvert W_{\upsilon} \rvert - 1 \right)}
                { \left( \lvert W \rvert - 1 \right) \lvert \Upsilon \rvert } 
        \quad \text{with} \quad
        W_{\upsilon} = \left\{ w \mid \upsilon \in w \land w \in W \ \right\}
\end{equation}

\paragraph{Average Message Length}\label{sec_metrics_compression_average_message_length}
Another way to assess the degree of compression achieved by agents in their communication is to analyze the average message length~\cite{chaabouni.2019, choi.2018, kharitonov.2019b, lei.2023b, luna.2020, vanderwal.2020b}. This metric, which appears for the first time in Choi et~al.~\cite{choi.2018}, captures the typical length of generated messages and provides insight into the efficiency of the \ac{el} in terms of information density~\cite{luna.2020}. By tracking the average number of words in the messages, we can quantify how effectively the agents compress their language. This metric is computed at the word level, meaning each word within a message is counted. The average message length $\overline{ \lvert m \rvert }$ for a set of messages $M$ is calculated as follows:

\begin{equation}
    \overline{\lvert m \rvert} = \frac{1}{\lvert M \rvert} \sum_{m \in M} \lvert m \rvert
    \quad \text{with} \quad
    \lvert m \rvert = \sum_{w \in m} 1
\end{equation}

\paragraph{Active Words}\label{sec_metrics_compression_active_words}
The active words metric, introduced by Lazaridou et~al.~\cite{lazaridou.2016}, complements the average message length by quantifying the diversity of word usage within the vocabulary~\cite{luna.2020}. Specifically, this metric measures the variety and utilization of distinct words in a communication system. A high number of active words indicates a diverse vocabulary, reflecting a more complex or redundant \ac{el}. Conversely, a lower number suggests that the communication system relies on a limited set of words, which may indicate a more efficient and compressed language with less synonyms~\cite{mordatch.2017}. This metric is widely used in the literature~\cite{boldt.2022d, carmeli.2022, chaabouni.2019, chowdhury.2020b, dagan.2020, dessi.2021, lazaridou.2016, luna.2020, mordatch.2017, resnick.2020, sowik.2020b, yu.2022}. Mathematically, the active word value $\operatorname{AW}$ for an agent $\xi_{i}$ can be defined as the size of the collection of words actively used by the agent $W_{\xi_{i}}$, as given in Equation~\ref{eq_metrics_morphology_compression_active_words}. In multi-agent setups, this metric can be averaged across all agents to provide a collective measure of vocabulary diversity within the joint system.

\begin{equation}\label{eq_metrics_morphology_compression_active_words}
    \operatorname{AW} \left( \xi_{i} \right) = \lvert W_{\xi_{i}} \rvert \quad \text{with} \quad W_{\xi_{i}} \subset W
\end{equation}

\subsubsection{Redundancy or Ambiguity}\label{sec_metrics_redundancy_or_ambiguity}
Redundancy in language occurs when multiple words are associated with the same meaning, providing alternative expressions for the same concept. Conversely, ambiguity occurs when a single word is associated with multiple meanings, creating the potential for different interpretations depending on the context. Both redundancy and ambiguity are characteristic features of \acp{nl}, reflecting the complexity and flexibility inherent in human communication~\cite{chaabouni.2019b, luna.2020}.

\paragraph{Perplexity}\label{sec_metrics_redundancy_or_ambiguity_perplexity}
Perplexity, introduced by Havrylov and Titov~\cite{havrylov.2017}, measures how often a word was used in a message to describe the same object~\cite{choi.2018, luna.2020}. \enquote{A lower perplexity shows that the same \textins{words} are consistently used to describe the same objects.}~\cite{luna.2020}. 
Mathematically, $P \left( w | \varphi \right)$ represents the probability or score of a word for a specific concept or meaning, e.g., derived from an affine transformation of the sender's hidden state~\cite{luna.2020} or from a ground truth label~\cite{choi.2018}. Thus, perplexity, given in Equation~\ref{eq_metrics_morphology_redundancy_perplexity}, quantifies the predictability of word usage, with lower values reflecting a less redundant communication system. It is usually calculated based on a sampled set of meanings $\Phi_{\text{test}}$ for which the word probability can be generated.

\begin{equation}\label{eq_metrics_morphology_redundancy_perplexity}
    \operatorname{Ppl} 
    = \exp \left( 
            - \sum_{w \in W} \left[ P \left( w | \varphi \right) \cdot \log \left( P \left( w | \varphi \right) \right) \right] 
        \right)
        \quad \forall \varphi \in \Phi_{\text{test}} \subseteq \Phi
\end{equation}

\paragraph{Singular Value Decomposition}\label{sec_metrics_redundancy_or_ambiguity_svd}
Another approach to quantitatively assess the redundancy of the vocabulary used in a communication system is outlined by Lazaridou et~al.~\cite{lazaridou.2016}. This method involves constructing a matrix where the rows correspond to distinct meanings, the columns represent individual words, and the matrix entries indicate the frequency with which each word is used for a given meaning. The rows are thus constructed based on a predefined ground truth classification. By applying Singular Value Decomposition (SVD) to this matrix, we can examine the dimensionality of the underlying communication strategy. If the communication system relies on a limited set of highly synonymous words, we would expect the SVD to reveal a low-dimensional structure. Conversely, a higher-dimensional decomposition would indicate a more diverse use of vocabulary, reflecting a potentially less synonymous and more redundant language.

\paragraph{Message Distinctness}\label{sec_metrics_redundancy_or_ambiguity_message_distinctness}
Message distinctness evaluates the linguistic representation of distinct features and thus aims to quantify ambiguity~\cite{lazaridou.2018, choi.2018, lorkiewicz.2011, luna.2020}. The metric, first suggested in Lazaridou et~al.~\cite{lazaridou.2018} and Choi et~al.~\cite{choi.2018}, quantifies the diversity of messages generated by the agent by assessing how well it differentiates between various inputs. Specifically, message distinctness $\operatorname{MD}$ is calculated as the ratio of the number of unique messages generated within a batch (cf.~\ref{eq_metrics_morphology_ambiguity_message_distinctness_unique_messages}) to the batch size (cf.~\ref{eq_metrics_morphology_ambiguity_message_distinctness}). A higher message distinctness indicates less ambiguity of the language.

\begin{equation}\label{eq_metrics_morphology_ambiguity_message_distinctness_unique_messages}
    M_{\text{unique}} = 
        \left\{ 
            m_{i} \mid m_{i} \in  M_{\text{test}} \land m_{i} \neq m_{j} \ \forall \ m_{j} \in M_{\text{test}}, i \neq j
        \right\}
\end{equation}

\begin{equation}\label{eq_metrics_morphology_ambiguity_message_distinctness}
    \operatorname{MD} 
        = \frac{\lvert M_{\text{unique}} \rvert}{\lvert M_{\text{test}} \rvert}
\end{equation}

\subsection{Syntax}\label{sec_metrics_syntax}
Despite the significance of structural properties in \acp{el}, particularly regarding their syntax and its relation to semantics, research in this area remains limited~\cite{lee.2019}. Recurrent syntactical patterns are central to the robustness and versatility of \acp{nl}~\cite{denamganai.2020}. Exploring these properties within the context of \ac{el} could provide valuable insights into their development and alignment with \ac{nl}.

\paragraph{Syntax Tree}\label{sec_metrics_syntax_syntax_tree}
Van der Wal et~al.~\cite{vanderwal.2020b} introduced unsupervised grammar induction (UGI) techniques for syntax analysis in \ac{el} research, describing a two-stage approach to deriving grammar and syntax. The first phase involves the induction of unlabeled constituent tree structures, explained below, and the labeling of these structures. The second phase extracts a probabilistic context-free grammar (PCFG) from the labeled data. Two methods were compared for constituency structure induction: the Common Cover Link (CCL), a pre-neural statistical parser that makes assumptions about \ac{nl} such as the Zipfian distribution, and the Deep Inside-Outside Recursive Auto-encoder (DIORA), a neural parser. For the labeling process, Van der Wal et~al.~\cite{vanderwal.2020b} used Bayesian Model Merging (BMM), to consolidate probabilistic models to label the induced syntax trees.

In syntax trees, the structure of the language is represented in a hierarchical manner, where nodes represent grammatical constructs (such as sentences, phrases, and words) and edges represent the rules or relationships that connect these constructs. Analysis of these trees helps to understand how well grammar induction methods match the true syntactic nature of \acp{el}. There are several metrics associated with syntax trees that are used to measure the complexity of the grammar~\cite{vanderwal.2020b}. First, tree depth measures the maximum distance from the root of the tree to its deepest leaf. Tree depth reflects the hierarchical complexity of the grammar. Shallow trees indicate a simpler grammar, while deeper trees suggest a more complex syntactic structure. Second, the number of unique preterminal groups is a metric that counts the different sets of preterminals (intermediate symbols) that appear to the right of production rules in a grammar. A larger number of unique preterminal groups indicates a richer and more diverse syntactic organization, suggesting that the grammar can generate a greater variety of structures.

\paragraph{Categorical Grammar Induction}\label{sec_metrics_syntax_CGI}
Ueda et~al.~\cite{ueda.2022} proposed a novel approach for analyzing the syntactic structure of \acp{el} using Categorial Grammar Induction (CGI). This technique focuses on deriving categorial grammars from message-meaning pairs, making it particularly well-suited for simple referential or signaling games.

In this method, derivation trees are constructed using lexical entries and application rules, mapping messages to atomic syntactical representations. Given that multiple derivations might exist for a single message, \enquote{the most likely derivation [is selected] using a log-linear model}~\cite{ueda.2022}. CGI is particularly valuable for assessing the syntactic structure of an \ac{el} using the generated trees.

\subsection{Semantics}\label{sec_metrics_semantics}
Capturing the semantic properties of \acp{el} is inherently complex, making it difficult to encapsulate nuances in a single metric. To address this, several key features have been introduced, including grounding, compositionality, consistency, and generalization. These are important because agents can develop representations that are well aligned with task performance but fail to capture the underlying conceptual properties~\cite{bouchacourt.2018}. Thus, an \ac{el} might enable successful task completion without truly encoding semantic meaning. Therefore, evaluating these semantic features is essential to evaluate the value and validity of the \ac{el}.

\subsubsection{Grounding}\label{sec_metrics_grounding}
Grounding is essential for the development of meaning and for systematic generalization to novel combinations of concepts~\cite{suglia.2024}. It forms the basis of human-agent communication~\cite{bogin.2018}, and without proper grounding, meaningful communication cannot be effectively learned~\cite{lin.2021}. However, in general dialog settings, grounding does not emerge naturally without specific regularization techniques~\cite{kottur.2017b}. The grounding problem, which concerns how words acquire semantic meaning, is central to this challenge~\cite{santamariapang.2019}.
Thus, grounding metrics are vital as they largely define the usability of a language. However, a significant limitation of these metrics is their reliance on some form of oracle or a \ac{nl}-grounded precursor~\cite{andreas.2019, patel.2021, gupta.2021, havrylov.2017}.

\paragraph{Divergence}\label{sec_metrics_grounding_divergence}
Havrylov and Titov~\cite{havrylov.2017} proposed a weak form of grounding. Weak grounding means that the same word can correspond to completely different concepts in the induced \ac{el} and \ac{nl}. They used the Kullback-Leibler divergence $\operatorname{D_{KL}}$ (cf. Equation~\ref{eq_kl_divergence}) of an \ac{el} and a \ac{nl} distribution to ensure that the statistical properties of \ac{el} messages resemble those of \ac{nl}. They introduced this approach as an indirect supervision measure during training but it can also serve as a metric for evaluating the alignment between \ac{el} and \ac{nl}. For a given sample $k$ and the message $m_{\xi_{S}}$ produced by the sender, the grounding divergence $\operatorname{G_{Div}}$ calculation is shown in Equation~\ref{eq_semantics_grounding_divergence_calculation}. Since the true \ac{nl} distribution $P_{\text{NL}} \left( m_{\xi_{S}} \right)$ is inaccessible, a language model is trained to approximate this distribution. The KL divergence yields a value in the range $\left[ 0, \infty \right)$, with lower values indicating a closer resemblance between the generated messages and \ac{nl}.

\begin{equation}\label{eq_kl_divergence}
    \operatorname{D_{KL}}(P||Q)= \sum_{x} P(x) \log \left( \frac{P(x)}{Q(x)} \right)
\end{equation}

\begin{equation}\label{eq_semantics_grounding_divergence_calculation}
    \operatorname{G_{Div}} 
        = \operatorname{D_{KL}} 
            \left( 
                P \left( m_{\xi_{S}} \mid k \right) \parallel P_{\text{NL}} \left( m_{\xi_{S}} \right) 
            \right)
\end{equation}

\paragraph{Purity}\label{sec_metrics_grounding_purity}
Purity, proposed by Lazaridou et~al.~\cite{lazaridou.2016}, is a metric used to assess the alignment between predefined semantic categories and those observed in an \ac{el}. It measures the effectiveness of a communication system in consistently mapping signals or words to specific concepts~\cite{yu.2023}. Thus, purity quantifies the extent to which the clustering of words reflects meaningful and coherent categories, as determined by ground-truth labels. To assess purity, we first form clusters by grouping samples based on the most frequently activated words to describe them. The quality of these clusters is then evaluated using the purity metric, which calculates the proportion of labels in each cluster that match the majority category of that cluster. A higher purity score indicates that the sender is producing words that are semantically aligned with predefined categories, as opposed to arbitrary or agnostic symbol usage, as demonstrated in~\cite{lazaridou.2016}. However, this metric requires the existence of predefined ground-truth labels, limiting its applicability in scenarios where such labels are unavailable or ambiguous.

Formally, given a set of clusters $\{ C_{k} \}$ where each cluster of samples $C_{k}$ has a corresponding majority ground-truth label $c_{k}$, the purity of a cluster $C_{k}$ is defined as:

\begin{equation} \label{eq_semantics_grounding_purity}
    \operatorname{purity} \left( C_{k} \right) = \frac{\lvert \{ w_{c} \mid w_{c} \in C_{k} \land w_{c} = c_{k} \} \rvert}{\lvert \{ w \mid w \in C_{k}\} \rvert}
\end{equation}

Here, $\{ w \mid w \in C_{k} \}$ is the collection of all words used to describe the samples in the cluster and $\{ w_{c} \mid w_{c} \in C_{k} \land w_{c} = c_{k} \}$ is the collection of words within the cluster that fit the majority label of that cluster. The purity metric ranges from 0 to 1, where a value of 1 indicates perfect alignment with the ground-truth categories.

\paragraph{Representational Similarity Analysis}\label{sec_metrics_grounding_rsa}
Representational Similarity Analysis (RSA) emerged in the field of neuroscience and was proposed by Kriegeskorte~et~al.~\cite{kriegeskorte.2008}.  It has since been adapted for the evaluation of the similarity of neural representations across different modalities, including computational models and brain activity patterns. This technique has been effectively applied in \ac{el} research~\cite{bouchacourt.2018, tieleman.2019, luna.2020}, where the focus shifts from analyzing neural activity to exploring the structural relationships between different embedding spaces. For example, RSA has been employed to compare the similarity of embedding space structures between input, sender, and receiver in a referential game~\cite{bouchacourt.2018, luna.2020}. By calculating pairwise cosine similarities within these spaces and then computing the Spearman correlation between the resulting similarity vectors, we can calculate an RSA score that measures the global agreement between these spaces, independent of their dimensionality. The agreement of an agent's embedding space with the input embedding space as such provides an intuitive measure of the grounding of the \ac{el}.

This approach offers the advantage of being applicable to heterogeneous agents and arbitrary input spaces. In our framework, this corresponds to any ground truth structured embedding $e \left( o_{\xi} \right)$  of an agent's observation $o_{\xi}$ and its internal meaning representation $\varphi_{\xi}$. Nevertheless, a significant limitation is the necessity for an embedding, which provides a structured description of the observation oriented towards a ground truth, for example, based on a \ac{nl} model. Furthermore, RSA is not directly applicable to the language itself, particularly for discrete languages. Instead, it operates at the level of earlier meaning representations. Despite this, RSA provides valuable insights into whether the \ac{el} can be grounded by evaluating the grounding of the meaning space.

The methodology of~\cite{bouchacourt.2018} utilizes a collection $K$ of samples, comprising $k$ observations, images, or feature vectors, to compute representational similarities between input and meaning space. First, we generate input or ground truth embeddings $e_{GT} = e \left( o_{\xi} \right)$ using an appropriate model and generate the corresponding internal representations $\varphi_{\xi}$ from the appropriate architecture part of agent $\xi$. Next, we compute pairwise similarities within each embedding space, denoted as $\operatorname{S_{e}}$ for the ground truth embeddings and $\operatorname{S_{\varphi}}$ for the agent representations, typically using cosine similarity $S_{cos}$ as defined in Equation~\ref{eq_semantics_grounding_RSA_cossim}. This yields a similarity vector of size $N \cdot (N - 1)$ for each embedding space. The vectors are converted into rank vectors $\operatorname{R}\left( \operatorname{S_{e}} \right)$ and $\operatorname{R}\left( \operatorname{S_{\varphi}} \right)$. Finally, we calculate the Spearman rank correlation $\rho$~\cite{spearman.1904} between the ranked similarity vectors, using the covariance $\operatorname{cov}$ and standard deviation $\sigma$, to assess the alignment between the input and agent representation spaces (cf. Equation~\ref{eq_semantics_grounding_RSA_spearman}). The correlation coefficient $\rho$ takes on values between $-1$ and $1$. A high absolute value of this coefficient indicates a strong alignment between the two variables.

\begin{equation}\label{eq_semantics_grounding_RSA_cossim}
    \operatorname{S_{e}}
        = \operatorname{S_{cos}} 
            \left(
                e_i, \varphi_i
            \right)
            \quad \forall i, j \in k, \, i \neq j \\
        \quad \text{with} \quad
            \operatorname{S_{cos}} \left( a, b \right)
                = \frac{a \cdot b}{\lvert\lvert a \rvert\rvert \cdot \lvert\lvert b \rvert\rvert}
\end{equation}

\begin{equation}\label{eq_semantics_grounding_RSA_spearman}
    \rho = \frac{\operatorname{cov} \left( \operatorname{R}\left( \operatorname{S_{e}} \right), \operatorname{R}\left( \operatorname{S_{\varphi}} \right) \right)}
                {\sigma_{\operatorname{R}\left( \operatorname{S_{e}} \right)} \sigma_{\operatorname{R}\left( \operatorname{S_{\varphi}} \right)}}
\end{equation}

\subsubsection{Compositionality}\label{sec_metrics_compositionality}
In \ac{el} research, achieving compositionality often requires deliberate guidance, as it does not naturally arise without specific interventions~\cite{resnick.2020}. For instance, training models on diverse tasks and varying environmental configurations can facilitate the development of compositional structures. This occurs as atomic concepts, learned in simpler contexts, are recombined in more complex scenarios~\cite{mordatch.2017}. When a language is truly compositional, its components can be systematically rearranged or substituted with conceptually equivalent components without altering the overall meaning~\cite{chaabouni.2020, kharitonov.2020}.

The formalization of compositionality can be framed using the comprehension $ \mathcal{L}_{\text{comp}} $ or production $ \mathcal{L}_{\text{prod}} $ function that map expressions from a language $ \mathcal{L} $ to a space of meanings $ \Phi $ or vice versa~\cite{bosc.2022}. For example, the function $ \mathcal{L}_{\text{comp}} : \mathcal{L} \rightarrow \Phi $ reflects \enquote{all the things that the language can denote}~\cite{bosc.2022}. A language is compositional if these functions act as a homomorphism, e.g., there exist binary operators $ \circ $ on $ \mathcal{L}_{\text{comp}} $ and $ \times $ on $ \Phi $ such that for any expression composed of two constituents $ m_1 $ and $ m_2 $ in $ \mathcal{L} $, the following condition holds:

\begin{equation}
    \mathcal{L}_{\text{comp}}(m_1 \circ m_2) = \mathcal{L}_{\text{comp}}(m_1) \times \mathcal{L}_{\text{comp}}(m_2)
\end{equation}

\paragraph{Topographic similarity}\label{sec_metrics_compositionality_topsim}
Topographic similarity (topsim), originally proposed by Brighton and Kirby~\cite{brighton.2006} and first applied to \ac{el} by Lazaridou et~al.~\cite{lazaridou.2018}, is a metric designed to quantify the structural alignment between the internal representations of meanings and the corresponding generated messages in a communication system. Unlike RSA (cf. Section~\ref{sec_metrics_grounding_rsa}), which compares the meaning space against a ground truth, topsim focuses on the internal alignment within an agent's meaning and message spaces. \enquote{The intuition behind this measure is that semantically similar objects should have similar messages}~\cite{lazaridou.2018}. It has become a widely used metric in the study of \ac{el}, as depicted in Figure~\ref{fig_metrics_mindmap_with_sources} in Appendix~\ref{sec:appendix_b}).

To compute topsim, we start by sampling $k$ meaning representations denoted by $\varphi$, typically embedded feature vectors, from the meaning space $\Phi$. Let $\phi = \{ \varphi_{1}, \dots , \varphi_{k} \}$ denote the collection of these samples, with $\varphi \in \Phi$. Using the sender's policy $\pi_{\xi_{S}}^{M}$, we generate corresponding messages $m_{i} = \pi_{\xi_{S}}^{M}(\varphi_i)$ for each sample $\varphi_{i} \in \phi$. We then compute distances within the meaning and language spaces using suitable distance functions for language $\Delta_{\mathcal{L}}$ and meaning $\Delta_{\Phi}$ space.
The choice of distance function $\Delta$ depends on the nature of the spaces involved. For discrete communication, typical choices include Hamming~\cite{hamming.1950} or Levenshtein~\cite{levenshtein.1966} distance, whereas for continuous spaces, cosine or Euclidean distance are often used~\cite{korbak.2020}. Finally, we compute the Spearman rank correlation $\rho$~\cite{spearman.1904} using the ranked distances to get the topsim value of the language:

\begin{equation}\label{eq_semantics_compositionality_topsim_spearman}
    \rho = 
        \frac{
                \operatorname{cov}
                \left(
                    \operatorname{R}\left( \Delta_{\mathcal{L}} \left( m_{i} \right) \right), 
                    \operatorname{R}\left( \Delta_{\Phi} \left( \varphi_{i} \right) \right) 
                \right)
            }
            {
                \sigma_{\operatorname{R}\left( \Delta_{\mathcal{L}} \left( m_{i} \right) \right)} 
                \sigma_{\operatorname{R}\left( \Delta_{\Phi} \left( \varphi_{i} \right) \right)}
            }
    \quad \forall \varphi_{i} \in \phi
\end{equation}

\paragraph{Positional Disentanglement}\label{sec_metrics_compositionality_posdis}
Positional Disentanglement (posdis) was introduced by Chaabouni et~al.~\cite{chaabouni.2020} as a metric to evaluate the extent to which words in specific positions within a message uniquely correspond to particular attributes of the input. This metric operates on an order-dependent strategy, which is normalized by the message length and calculated as the ratio of mutual information to entropy. The underlying assumption is that the language leverages positional information to disambiguate words, such that \enquote{each position of the message should only be informative about a single attribute}~\cite{chaabouni.2020}. Thus, \enquote{posdis assumes a message whose length equals the number of attributes in the input object, and where each message token, in a specific position, represents a single attribute}~\cite{perkins.2021}. This order-dependence is a characteristic feature of \ac{nl} structures and is essential for the emergence of sophisticated syntactic patterns~\cite{chaabouni.2020}.

The metric begins by identifying each word $w_p$ at position $p$ in a message $m$, where $f$ represents the feature vector of the ground truth. The mutual information $I(w_p, f_i)$ between $w_p$ and a specific feature $f_i$ is calculated to determine how informative the position $p$ is about the attribute $f_i$ (cf. Equation~\ref{eq_metrics_compositionality_posdis_mutualinformation}). The two most informative features $f_i^{1}$ and $f_i^{2}$ are then identified based on the mutual information value (cf. Equation~\ref{eq_metrics_compositionality_posdis_mostinformative}). To quantify positional disentanglement, the mutual information difference between the two most informative features is normalized by the entropy $H(w_p)$ of the word at position $p$, as defined in Equation~\ref{eq_metrics_compositionality_posdis_positionposdisentropy} and Equation~\ref{eq_metrics_compositionality_posdis_positionposdis}.
Finally, the overall posdis value for a language is calculated by averaging the posdis scores across all positions in the messages within the dataset. For messages of varying lengths, the posdis score is normalized by the average message length $ 
\overline{\lvert m \rvert} $, as given in Equation~\ref{eq_metrics_compositionality_posdis_posdis}.

\begin{equation}\label{eq_metrics_compositionality_posdis_mutualinformation}
    I \left( w_{p}, f_{i} \right) 
        = \sum_{ w_{p} \in m } \sum_{ f_{i} \in f } P \left( w_{p}, f_{i} \right) 
                \log \left( \frac{ P \left( w_{p}, f_{i} \right) }{ P \left(w_{p}) P(f_{i}) \right) } \right)
\end{equation}

\begin{equation}\label{eq_metrics_compositionality_posdis_mostinformative}
    f_{i}^{1} = \argmax_{ f_{i} \in f } 
                    I \left( w_{p}, f_{i} \right) 
    \quad \text{and} \quad 
    f_{i}^{2} = \argmax_{ f_{i} \in f \land f_{i} \neq f_{i}^{1} }
                    I \left( w_{p}, f_{i} \right) 
\end{equation}

\begin{equation}\label{eq_metrics_compositionality_posdis_positionposdisentropy}
    H \left( w_{p} \right) = - \sum_{ w_{p} \in m } P \left( w_{p} \right) \log \left( P \left( w_{p} \right) \right)
\end{equation}

\begin{equation}\label{eq_metrics_compositionality_posdis_positionposdis}
    \operatorname{posdis}_{p} 
        = \frac { I \left( w_{p}, f_{i}^{1} \right) - I \left( w_{p}, f_{i}^{2} \right) }{ H \left( w_{p} \right) }
\end{equation}

\begin{equation}\label{eq_metrics_compositionality_posdis_posdis}
    \operatorname{posdis} = \frac{ 1 }{ \overline{\lvert m \rvert} } \sum_{p} \operatorname{posdis}_{p}
    \quad \text{with} \quad 
    \overline{\lvert m \rvert} = \frac{ 1 }{ \lvert M \rvert } \sum_{m \in M} \lvert m \rvert
\end{equation}

\paragraph{Bag of Symbols Disentanglement}\label{sec_metrics_compositionality_bosdis}
Bag of Symbols Disentanglement (bosdis) is a metric introduced by Chaabouni et~al.~\cite{chaabouni.2020} to assess the degree to which words in a language unambiguously correspond to different input elements, regardless of their position within a message. While positional disentanglement (posdis) relies on the assumption that positional information is crucial for disambiguating words (cf. Section~\ref{sec_metrics_compositionality_posdis}), bosdis relaxes this assumption and captures the intuition behind a permutation-invariant language. In such a language, the order of words is irrelevant, and only the frequency of words carries meaning~\cite{chaabouni.2020}. The metric normalizes the mutual information between symbols and input features by the entropy summed over the entire vocabulary. 
This approach maintains the requirement that each symbol uniquely refers to a distinct meaning, but shifts the focus to symbol counts as the primary informative element.

\begin{equation}\label{eq_metrics_semantics_compositionality_bosdis_1}
    I \left( w, f_{i} \right) = \sum_{w \in m} \sum_{f_{i} \in f} P \left( w, f_{i} \right) 
        \log \left( \frac{ P \left( w, f_{i} \right) }{ P \left( w \right) P \left( f_{i} \right) } \right)
\end{equation}

\begin{equation}\label{eq_metrics_semantics_compositionality_bosdis_2}
    f_{i}^{1} = \argmax_{ f_{i} \in f } 
                    I \left( w, f_{i} \right) 
    \quad \text{and} \quad 
    f_{i}^{2} = \argmax_{ f_{i} \in f \land f_{i} \neq f_{i}^{1} }
                    I \left( w, f_{i} \right) 
\end{equation}

\begin{equation}\label{eq_metrics_semantics_compositionality_bosdis_3}
    H \left( w \right) = - \sum_{ w \in m } P \left( w \right) \log \left( P \left( w \right) \right)
\end{equation}

\begin{equation}\label{eq_metrics_semantics_compositionality_bosdis_4}
    \operatorname{bosdis}_{w} 
    = \frac{ I \left( w, f_{i}^{1} \right) - I \left( w, f_{i}^{2} \right) }
            { H \left( w \right)}
\end{equation}

\begin{equation}\label{eq_metrics_semantics_compositionality_bosdis_5} 
    \operatorname{bosdis} = \frac{ 1 }{ \lvert W \rvert } \sum_{ w \in W } \operatorname{bosdis}_{w}
\end{equation}

\paragraph{Tree Reconstruct Error}\label{sec_metrics_compositionality_tre}
Tree Reconstruct Error (TRE) assumes prior knowledge of the compositional structure within the input data, enabling the construction of tree-structured derivations~\cite{andreas.2019}. As defined by Andreas~\cite{andreas.2019}, a language is considered compositional if it functions as a homomorphism from inputs to their representations. The compositionality of a language should be evaluated by identifying representations that allow an explicitly compositional language to closely approximate the true underlying structure~\cite{andreas.2019}. One metric for this assessment is TRE, which quantifies the discrepancy between a compositional approximation and the actual structure, using a composition function and a distance metric. A TRE value of zero indicates perfect reproduction of compositionality.

The compositional nature of a sender's language is affirmed if there exists an assignment of representations to predefined primitives (e.g., categories, concepts, or words) such that for each input, the composition of primitive representations according to the oracle's derivation precisely reproduces the sender's prediction~\cite{andreas.2019}. TRE specifically measures the accuracy with which a given communication protocol can be reconstructed while adhering to the compositional structure of the derivation or embedding of the input $e \in E$~\cite{korbak.2020}.

One of the key advantages of the TRE framework is its flexibility across different settings, whether discrete or continuous. It allows for various choices of compositionality functions, distance metrics, and other parameters. However, this flexibility comes with challenges, including the requirement for an oracle-provided ground truth and the necessity of pre-trained continuous embeddings.

It is defined in a way that allows the choice of the distance metric $\delta$ and the compositionality function $\circ$ to be determined by the evaluator~\cite{andreas.2019}. When the exact form of the compositionality function is not known a priori, it is common to define $\circ$ with free parameters, as suggested by Andreas~\cite{andreas.2019}, treating these parameters as part of the learned model and optimizing them jointly with the other parameters $\eta$. However, care must be taken when learning the compositional function to avoid degenerate solutions~\cite{andreas.2019}.

Given a data sample $k$ from the dataset $K$ ($k \in K$) and a corresponding message $m$ from the set of all possible messages $M$ ($m \in M$), TRE requires a distance function $\delta$ and learnable parameters $\eta$. Additionally, it employs a compositionality function $\circ$ and pre-trained embeddings of ground truth, denoted by $e \in E$, which can be obtained using models like word2vec.

The functions involved in the TRE calculation are as follows:
\begin{itemize}
    \item Pre-trained ground truth oracle (e.g., word2vec): $\mathcal{E} : K \to E$
    \item Learned language speaker: $\xi_S : K \to M$
    \item Learnable approximation function for TRE: $\widetilde{f}_{\eta} : E \to M$
\end{itemize}

In the discrete message setting, which is the focus here, a discrete distance metric such as $L_1$ is typically chosen, along with a compositional function $\circ$ defined by a weighted linear combination~\cite{andreas.2019, korbak.2020}:
\begin{equation} 
    m_{1} \circ m_{2} = A m_{1} + B m_{2}
    \quad \text{with} \quad 
    \eta = \left\{ A , B \right\}
\end{equation}

To compute the TRE, an optimized approximation function $\widetilde{f}_{\eta}$ is required. This function must satisfy two key properties: embedding consistency, meaning that the learned parameters $\eta$ are specific to an embedding, and compositionality, which ensures that the function behaves according to:
\begin{equation*}
    \widetilde{f}_{\eta} \left( e_{i} \right) = \eta_{i} 
    \quad \text{and} \quad
    \widetilde{f}_{\eta} \left( \langle e_{i}, e_{j} \rangle \right) = \widetilde{f}_{\eta} \left( e_{i} \right) \circ \widetilde{f}_{\eta} \left( e_{j} \right)
\end{equation*}

The optimization process involves minimizing the distance between the output of the learned language speaker $\xi_{S}(k_{i})$ and the approximation function $\widetilde{f}_{\eta}(e_{i})$, based on the ground truth:
\begin{equation} 
    \eta^{\ast} =  \argmin_{\eta} \sum_{i} \delta 
        \left( \xi_{S} \left( k_{i} \right),
            \widetilde{f}_{\eta} \left( e_{i} \right) \right)
    \quad \text{with} \quad
    \mathcal{E} \left( k_{i} \right) = e_{i}
\end{equation}

With the optimized parameters $\eta^{\ast}$, TRE can be calculated at two levels: the datum level, which assesses individual instances:
\begin{equation} 
    \operatorname{TRE}(k_i) = \delta \left( \xi_S(k_i) , \widetilde{f}_{\eta^{\ast}}(e_i) \right)
    \quad \text{with} \quad
    \mathcal{E}(k_i) = e_i
\end{equation}

and the dataset level, which measures the overall communication performance across the dataset:
\begin{equation} 
    \operatorname{TRE}(K) = \frac{1}{|K|} \sum_{k \in K} \operatorname{TRE}(k)
\end{equation}

\paragraph{Conflict Count}\label{sec_metrics_compositionality_conflictcount}
Conflict count, introduced by Kuci\'nski~et~al.~\cite{kucinski.2020}, is designed to quantify the extent to which the assignment of features to words in a language deviates from the word's principal meaning. This metric is particularly useful in scenarios where the language employs synonyms, as it accounts for the possibility of multiple words referring to the same concept.

The conflict count metric operates under the assumption that the number of concepts or features $f_{i}$ given in a feature vector $f$ of a sample $k$ in the collection of samples $K$ is equal to the message length $\lvert m \rvert$, and that there exists a one-to-one mapping between a concept $f_i \in f$ and a word $w \in W$. The metric counts how frequently this one-to-one mapping is violated, with a value of $0$ indicating no conflicts and, therefore, high compositionality. An advantage of this metric is its ability to accommodate redundancy in the language. However, it also has limitations, such as the assumption that the number of features or attributes equals the message length, i.e., $\lvert f \rvert = \lvert m \rvert$. Additionally, because conflict count assumes the number of concepts in a derivation to be equal to the message length, it becomes undefined for languages or protocols that violate this assumption, such as those involving negation or context-sensitive constructions presented in~\cite{korbak.2020}.

The primary objective of conflict count is to quantify the number of times the mapping from a word $w$ to its principal meaning $\varphi_w$ is violated. This requires the assumption that a mapping $\alpha$ exists from the position $p$ of word $w$ in message $m$ to an individual feature in feature vector $f$, such that:
\begin{equation} 
    \alpha = 
        \left\{
            1, \dots, \lvert m \rvert
        \right\}
        \to
        \left\{
            1, \dots, \lvert f \rvert
        \right\}
\end{equation}

In this framework, the meaning of a word, denoted by $\varphi_{w}$, is determined by both the word $w$ itself and its position $p$ within the message. This meaning corresponds to a specific instance $j$ of a particular feature $i$ within the feature vector $f$, such that $f_{i,j} = \varphi(w,p)$.

The process of calculating the conflict count begins by identifying the principal meaning of each word-position pair:
\begin{equation} 
    \varphi(w,p:\alpha) = \argmax_{f_{i,j}} \operatorname{count}(w,p,f_{i,j}:\alpha)
    \quad
    \forall f_{i,j} \in f
\end{equation}

using the count function:
\begin{equation} 
    \operatorname{count}(w,p,f_{i,j}:\alpha) 
        = \sum_{k \in K} 
            \lvert
            \left\{ 
                w \mid w \in m \left( k \right) \land \operatorname{pos}_{m} \left( w \right) = p \land f_{i,j} \in k
            \right\}
            \rvert
\end{equation}
where $m \left( k \right)$ is the message produced for sample $k$ and $\operatorname{pos}_{m} \left( w \right)$ computes the position of word $w$ in message $m$.

Finally, the conflict count value $\operatorname{conf}$ is determined by finding the mapping $\alpha$ that minimizes the score:
\begin{equation} 
    \operatorname{conf} = \argmin_{\alpha} \sum_{w,p} \operatorname{score} \left( w,p:\alpha \right)
\end{equation}
where the score function is defined as:
\begin{equation} 
    \operatorname{score} \left( w,p:\alpha \right) = \sum_{f_{i,j} \neq \varphi \left( w , p \right)} \operatorname{count}\left( w,p,f_{i,j}:\alpha \right)
\end{equation}

\subsubsection{Consistency}\label{sec_metrics_consistency}
For a language to be effective, the meaning of each word must be consistent across different contexts. Inconsistent word meanings can render a language practically useless, even if the language is semantically grounded and exhibits compositional properties~\cite{kottur.2017b}. In dialogue settings, particularly in the absence of explicit regularization mechanisms, words often fail to maintain consistent groundings across different instances, leading to ambiguity and reduced communicative effectiveness~\cite{kottur.2017b}. Thus, it is crucial to carefully monitor this language characteristic in \ac{el} settings.

\paragraph{Mutual Information}\label{sec_metrics_consistency_mutualinformation}
Consistency in language can be quantitatively assessed by examining the mutual information between messages and their corresponding input features. Ideally, a consistent language will exhibit a high degree of overlap between messages and features, leading to a high mutual information value, indicating strong correspondence~\cite{dessi.2019}. 

Formally, mutual information between two random variables, say $X$ and $Y$, with joint distribution $P_{(X,Y)}$ and marginal distributions $P_X$ and $P_Y$, is defined as the Kullback–Leibler divergence $\operatorname{D_{KL}}$ (see Equation~\ref{eq_kl_divergence}) between the joint distribution and the product of the marginals:

\begin{equation}
    I(X;Y) = \operatorname{D_{KL}} \left( P_{(X,Y)} \parallel P_{X} \otimes P_{Y} \right)
\end{equation}

In the context of discrete communication, where both messages and sample features are represented as discrete variables, the mutual information between the set of messages $M$ and the set of features $F$ is computed using a double summation over all possible message-feature pairs:

\begin{equation}
  \operatorname{I}\left( M ; F \right) 
    = \sum_{m \in M} \sum_{f \in F}
            P_{\left( M , F \right) } \left( m , f \right) 
                \log \left( 
                        \frac{ P_{ \left( M , F \right) } \left( m , f \right) }
                                {P_{M} \left( m \right) P_{F} \left( f \right)} 
                    \right) 
\end{equation}

where $P_{ \left( M , F \right) } \left( m , f \right)$ is the joint probability of message $m$ and feature $f$, and $P_{M} \left( m \right)$ and $P_{F} \left( f \right)$ are the marginal probabilities of $m$ and $f$, respectively.

\paragraph{Correlation}\label{sec_metrics_consistency_correlation}
Various studies employ different statistical techniques to measure consistency using correlations~\cite{dagan.2020, luna.2020, mihai.2019, mul.2019, santamariapang.2019, verma.2019}.
For example, consistency within a language system can be quantified by analyzing the variability of words produced for a given sample $k$. Specifically, given the set of all words representing $k$, a heatmap is generated using the mean of this set. The sharpness of the heatmap is then quantified by computing the Variance of the Laplacian (VoL). The average consistency score is obtained by dividing the VoL of the heatmap by the count of all samples considered, as introduced by Verma and Dhar~\cite{verma.2019}.

Additionally, Mul et~al.~\cite{mul.2019} explored the correlation between messages and actions as well as between messages and salient properties of the environment. The analysis reveals correlations by examining the conditional probability distribution of actions given the messages produced by a pretrained or fine-tuned receiver. This distribution, denoted as $P \left(a \mid m \right)$, was visualized using bin bar plots to highlight the prominent correlations~\cite{mul.2019}. Similarly, the relationship between input and messages is analyzed by examining the conditional distribution of a pretrained sender's messages given the observational input, represented as $P \left( m \mid o \right)$~\cite{mul.2019}.

\paragraph{Coherence}\label{sec_metrics_consistency_coherence}
Coherence is often assessed through context independence, a metric initially proposed by Bogin et~al.~\cite{bogin.2018}. Context independence examines whether words within a language maintain consistent semantics across varying contexts. However, context independence may be considered restrictive, particularly in languages where synonyms are prevalent~\cite{lowe.2019, korbak.2019}. The context independence metric aims to measure the alignment between words $ w \in W $ and features $ f \in F $ of the input samples by analyzing their probabilistic associations. Specifically, $ P \left(w \mid f \right) $ denotes the probability that a word $ w $ is used when a feature $ f $ is present, while $ P \left( f \mid w \right) $ represents the probability that a feature $ f $ appears when a word $ w $ is used. For each feature $ f $, we identify the word $ w_{f} $ most frequently associated with it by maximizing $ P \left( f \mid w \right) $:
\begin{equation}
    w_{f} \coloneqq \argmax_{w} P \left( f \mid w \right)
\end{equation}
The context independence or coherence metric $ \operatorname{CI} $ is then computed as the average product of these probabilities across all features:
\begin{equation}
    \operatorname{CI} \left( w_{f} , f \right) 
        = \frac{1}{\lvert F \rvert} 
            \sum_{f \in F} P \left( w_{f} \mid f \right) P \left( f \mid w_{f} \right)
\end{equation}
This metric ranges from $0$ to $1$, with $1$ indicating perfect alignment, meaning that each word retains its meaning consistently across different contexts and is thus used coherently.

\paragraph{Entropy}\label{sec_metrics_consistency_entropy}
Entropy metrics are instrumental in analyzing the variability and predictability within linguistic systems. The most fundamental use of entropy involves marginal probabilities, which capture the variability in the number of words in a language~\cite{luna.2020, liang.2020}. More advanced applications of entropy focus on sender language entropy, which examines the conditional entropy of messages given features and vice versa~\cite{rita.2022, yu.2022}. Specifically, low conditional entropy $ H \left( M \mid F \right) $ indicates that a unique message is used for a specific feature, whereas high $ H \left( M \mid F \right) $ reflects the generation of synonyms for the same feature~\cite{rita.2022}.

Recent approaches further extend this analysis by combining conditional entropies~\cite{ohmer.2022, mu.2021}. For example, $ H \left( M \mid F \right) $ quantifies the uncertainty remaining about messages after knowing the concepts, while $ H \left( F \mid M \right) $ measures the uncertainty about concepts given the messages. A negative correlation between these measures and agent performance is expected~\cite{ohmer.2022}. However, a notable limitation of these entropy-based methods is that they focus on complete messages rather than individual words, which can limit the evaluation of more complex languages.

For example, Ohmer et~al.~\cite{ohmer.2022} provide the following comprehensive evaluation approach.
First, the conditional entropy of messages given features $H \left(M \mid F \right)$, see Equation~\ref{eq_metrics_semantics_consistency_entropy_conditionalentropy}, and $H \left(F \mid M \right)$ are calculated. Additionally, the marginal entropies are calculated using Equation~\ref{eq_metrics_semantics_consistency_entropy_marginalentropy}, where $ X $ represents either messages $ M $ or features $ F $.

\begin{equation}\label{eq_metrics_semantics_consistency_entropy_conditionalentropy}
    H \left(M \mid F \right) = - \sum_{m \in M} \sum_{f \in F} P \left( f , m \right) 
                                    \log \left( 
                                        \frac{ P \left( f , m \right) }{ P \left( f \right) } 
                                    \right)
\end{equation}
\begin{equation}\label{eq_metrics_semantics_consistency_entropy_marginalentropy}
    H \left( X \right) = - \sum_{x \in X} P \left( x \right) \log \left( P \left( x \right) \right)
\end{equation}

Using these entropies, $\operatorname{consistency}$, see Equation~\ref{eq_metrics_semantics_consistency_entropy_consistency}, measures how much uncertainty about the message is reduced when the feature is known, with lower values indicating more consistent message usage. $\operatorname{effectiveness}$, on the other hand, see Equation~\ref{eq_metrics_semantics_consistency_entropy_effectiveness}, evaluates the reduction in uncertainty about the feature when the message is known, with lower values reflecting more unique messages for individual features.

\begin{equation}\label{eq_metrics_semantics_consistency_entropy_consistency}
    \operatorname{consistency}(F, M) = 1 - \frac{H(M \mid F)}{H(M)}
\end{equation}
\begin{equation}\label{eq_metrics_semantics_consistency_entropy_effectiveness}
    \operatorname{effectiveness}(F, M) = 1 - \frac{H(F \mid M)}{H(F)}
\end{equation}

Finally, the normalized mutual information $ \operatorname{NI} $ provides a combined score:
\begin{equation}
    \operatorname{NI}(F, M) = \frac{H(M) - H(M \mid F)}{0.5 \cdot (H(F) + H(M))}
\end{equation}
A high $ \operatorname{NI} $ score indicates a strong predictive relationship between messages and features, reflecting high consistency.

\paragraph{Similarity}\label{sec_metrics_consistency_similarity}
The Jaccard similarity coefficient is a another metric for evaluating the consistency of language usage among agents~\cite{choi.2018, dagan.2020}. It quantifies the similarity between two sets by comparing the size of their intersection to the size of their union~\cite{dagan.2020}. To measure language consistency, the Jaccard similarity is computed by sampling messages for each input and averaging the similarity scores across the population~\cite{dagan.2020}. This approach reflects how consistently words are used across different messages. Specifically, Jaccard similarity $ J(M_{\xi_{i}}, M_{\xi_{j}}) $ is defined in Equation~\ref{eq_metrics_semantics_consistency_jaccard_similarity}, where $ M_{\xi_{i}} $ and $ M_{\xi_{j}} $ represent sets of messages generated by different agents based on the same input. The similarity ranges from $0$ to $1$, with $1$ indicating complete overlap and thus perfect similarity.

In practice, Jaccard similarity helps to assess the coherence of languages emerging from agent-based systems. For instance, in referential game experiments, high perplexity (cf. Section~\ref{sec_metrics_redundancy_or_ambiguity_perplexity}) and low Jaccard similarity have been observed, suggesting that agents assign unique but incoherent strings to object types to gain an advantage in the game without producing a consistent language~\cite{choi.2018}. However, Jaccard similarity is only applicable to scenarios where multiple agents generate messages about the same set of objects. Thus, its application is limited to cases where the goal is to compare the overlap of message sets between agents attempting to convey similar meanings.

\begin{equation}\label{eq_metrics_semantics_consistency_jaccard_similarity}
    J(M_{\xi_{i}}, M_{\xi_{j}})
        = \frac{ \lvert M_{\xi_{i}} \cap M_{\xi_{j}} \rvert }
                { \lvert M_{\xi_{i}} \cup M_{\xi_{j}} \rvert } 
        = \frac{ \lvert M_{\xi_{i}} \cap M_{\xi_{j}} \rvert }
                { \lvert M_{\xi_{i}} \rvert + \lvert M_{\xi_{j}} \rvert - \lvert M_{\xi_{i}} \cap M_{\xi_{j}} \rvert }
\end{equation}

\subsubsection{Generalization}\label{sec_metrics_generalization}
A language's ability to generalize is crucial for describing objects and concepts at different levels of complexity, allowing for effective clustering and hierarchical representation. Generalization in \acp{el} reflects their ability to extend beyond specific training instances to novel situations. \enquote{If the emergent languages can be generalised, we then could say that these languages do capture the structure of meaning spaces}~\cite{guo.2019}. Research shows that languages capable of generalization tend to emerge only when the input is sufficiently varied~\cite{chaabouni.2020}. In contrast, a large dictionary size often indicates a lack of generalization~\cite{liang.2020}.
Human languages have evolved under the pressure of a highly complex environment, fostering their generalization capabilities~\cite{chaabouni.2020}. However, deep learning models often exploit dataset-specific regularities rather than developing systematic solutions~\cite{keresztury.2020}. To address this, much research is being done on the systematic generalization abilities of \acp{el}.

\paragraph{Zero Shot Evaluation}\label{sec_metrics_generalization_zero_shot}
Zero-shot evaluation, which assesses the ability of an agent to generalize to novel stimuli~\cite{choi.2018, lazaridou.2018}, has become a standard metric in the study of \ac{el} as illustrated in Figure~\ref{fig_metrics_mindmap_with_sources} in Appendix~\ref{sec:appendix_b}. This evaluation is critical to understand the generalization capabilities of an agent. Zero-shot evaluation can be done in two different scenarios, one with unseen input and the other with an unseen partner.

In the unseen input scenario, models are tested on a zero-shot test set consisting of samples with feature combinations not encountered during training. Performance, such as accuracy, is reported for these unseen samples~\cite{kottur.2017b, lazaridou.2018, korbak.2019, chaabouni.2020, luna.2020}. Different methods for constructing novel inputs include exposing models to objects that resemble training data but have unseen properties or entirely novel combinations of features~\cite{lazaridou.2018}. Moreover, a more drastic approach may involve moving to entirely new input scenarios, such as testing the ability of agents to generalize across different game types~\cite{mu.2021}.

The unseen partner scenario, also known as cross-play or zero-shot coordination, evaluates models by pairing agents that did not communicate during training. Again, performance is measured, typically in terms of accuracy~\cite{hu.2020, bullard.2020}.

However, these approaches also have drawbacks. The unseen input scenario requires a ground truth oracle to withhold feature combinations, which is necessary to accurately define novel combinations. Meanwhile, the unseen partner setup can introduce inefficiencies by requiring additional resources to train novel communication partners for testing.

\paragraph{Ease and Transfer Learning}\label{sec_metrics_generalization_ETL}
Ease and Transfer Learning (ETL), as proposed by Chaabouni et~al.~\cite{chaabouni.2022}, evaluates how easily new listeners can adapt to an \ac{el} on distinct tasks. ETL extends the concept of ease-of-teaching~\cite{li.2019} by assessing how effectively a deterministic language, developed by a fixed set of speakers, can be transferred to new listeners who are trained on tasks different from the original one for which the language was optimized~\cite{chaabouni.2022}. This metric not only gauges the language's generality but also its transferability across tasks~\cite{chaabouni.2022}.

To measure ETL, after convergence, a fixed number of speakers produce a deterministic language by selecting symbols using an $\argmax$ operation over their distributions. This language is then used to train newly initialized listeners on a new task. The training curve is tracked to observe how quickly and accurately the listeners learn the task, which may involve more challenging objectives than the former training tasks~\cite{chaabouni.2022, feng.2024}.

\subsection{Pragmatics}\label{sec_metrics_pragmatics}
Pragmatics is a critical aspect of language that examines how context influences meaning~\cite{yuan.2019}. It goes beyond the literal interpretation of words and requires the listener to infer the speaker's intentions, beliefs, and mental states, an ability known as Theory of Mind (ToM)~\cite{yuan.2019}. In human interactions, this contextual reasoning is essential for predicting and understanding behavior. In the context of \ac{el}, pragmatics focuses on how effectively agents use the communication ability in their environment. Empirical studies have shown that agents may initially fail to use communication meaningfully, but, once they do communicate, they can reach a locally optimal solution to the communication problem~\cite{eccles.2019}. Thus, evaluating the pragmatics of \ac{el} is essential to determining its utility and effectiveness in real-world applications.

\subsubsection{Predictability}\label{sec_metrics_pragmatics_predictability}
Predictability evaluates the complexity of an environment and its effect on the need for communication. Thus, it is a central metric for the probability of emergence and the use of \ac{el}. In simple environments with limited actions, agents can often coordinate without communication~\cite{dubova.2020}.

\paragraph{Behavioral Divergence}\label{sec_metrics_pragmatics_predictability_behavioral_divergence}
Behavioral divergence, introduced by Dubova et~al.~\cite{dubova.2020}, posits that less diversity in actions or messages correlates with more predictable behavior, potentially reducing the need for communication. To quantify this, we calculate Behavioral Action Predictability $\operatorname{BAP}$ and Behavioral Message Predictability $\operatorname{BMP}$. Both use the Jensen-Shannon Divergence (JSD) (see Equation~\ref{eq_jensen_shannon_divergence}) which itself uses the Kullback-Leibler Divergence $\operatorname{D_{KL}}$ (cf. Equation~\ref{eq_kl_divergence}).

\begin{equation}\label{eq_jensen_shannon_divergence}
    \operatorname{D_{JS}} \left( P \parallel Q \right) = \frac{1}{2} \operatorname{D_{KL}} \left( P \parallel M \right) + \frac{1}{2} \operatorname{D_{KL}} \left( Q \parallel M \right)
    \quad \text{with} \quad M = \frac{ P + Q }{ 2 }
\end{equation}

BAP (see Equation~\ref{eq_metrics_pragmatics_predictability_bap}) and BMP (see Equation~\ref{eq_metrics_pragmatics_predictability_bmp}) both use a uniform distribution $Q$ for comparison. BAP further uses the distribution of actions by the agent $P \left( a_{\xi} \right)$ while BMP uses the distribution of messages by the agent $P \left( m_{\xi} \right)$. Based on that, these metrics provide a robust measure of how predictable agent behaviors and messages are, with higher values indicating less predictability and greater need for beneficial communication~\cite{dubova.2020}.

\begin{equation}\label{eq_metrics_pragmatics_predictability_bap}
    \operatorname{BAP} = \operatorname{D_{JS}} \left( P \left( a_{\xi} \right) \parallel Q \right)
\end{equation}

\begin{equation}\label{eq_metrics_pragmatics_predictability_bmp}
    \operatorname{BMP} = \operatorname{D_{JS}} \left( P \left( m_{\xi} \right) \parallel Q \right)
\end{equation}

\subsubsection{Efficiency}\label{sec_metrics_pragmatics_efficiency}
In \ac{el} settings, efficient communication arises only when there is an opportunity cost~\cite{kalinowska.2022}. Without such a cost, there is no drive towards brevity, which limits the effectiveness and efficiency of \ac{el} in \ac{hci}.

\paragraph{Sparsity}\label{sec_metrics_pragmatics_efficiency_sparsity}
Sparsity, as proposed by Kalinowska et~al.\cite{kalinowska.2022}, measures the extent to which agents minimize their communication during task execution. This metric requires only the collection of messages exchanged per episode for computation. However, its applicability is limited to scenarios where communication is not strictly necessary for task completion, i.e., agents have the option to send no messages at all or to send messages that contain no meaningful information.
A sparsity value of $0$ indicates that an agent can solve the task using only a single message throughout an episode, reflecting a highly efficient communication strategy. Conversely, higher sparsity values indicate more frequent or verbose communication, which may indicate inefficiencies in the \ac{el}.

Communication sparsity $ \operatorname{ComSpar} $ is mathematically defined as:

\begin{equation}
    \operatorname{ComSpar} = \frac{1}{n_{\text{ep}}} \cdot 
        \sum_{ M_{\text{ep}, i} \in M }
        - \log 
        \left(
            \lvert \left\{ m \mid m \in M_{\text{ep}, i} \land m \neq 0 \right\} \rvert
        \right)
\end{equation}

In this equation, $ M_{\text{ep}, i} $ represents the set of all messages exchanged during episode $ i $, and $ n_{\text{ep}} $ is the total number of episodes observed. The collection $\left\{ m \mid m \in M_{\text{ep}, i} \land m \neq 0 \right\}$ consists of all messages $ m $ of episode $ i $ that are non-zero and thus contributing.

\subsubsection{Positive Signaling}\label{sec_metrics_pragmatics_positive_signaling}
Positive signaling evaluates the alignment between an agent's observations and its communication output~\cite{lowe.2019}. The goal is to ensure that the outgoing transmitted information is both relevant and observable by the agent~\cite{portelance.2021}.

\paragraph{Speaker Consistency}\label{sec_metrics_pragmatics_positive_signaling_speaker_consistency}
Speaker Consistency (SC), introduced by Jaques et~al.~\cite{jaques.2018}, measures how effectively an agent's messages reflect its state or trajectory, thereby ensuring the communication is meaningful. This is quantified using mutual information. For an agent $\xi_i$, the trajectory $\tau_{\xi_i}^t$ represents the sequence of states and actions up to time step $t$. The message produced at time $t$ is denoted by $m_{\xi_i}^t$. The mutual information $I(m_{\xi_i}^t, \tau_{\xi_i}^t)$ between the message and trajectory is calculated as:

\begin{equation}
\begin{aligned}
    I(m_{\xi_i}^t, \tau_{\xi_i}^t) 
        &= H(m_{\xi_i}^t) - H(m_{\xi_i}^t | \tau_{\xi_i}^t)
        \\
        &= - \sum_{m \in M_{\xi_i}} \overline{P_{\xi_i}}(m) \log \overline{P_{\xi_i}}(m)
        \\
            & \quad \quad \quad \quad 
                + \mathbb{E}_{\tau_{\xi_i}^t} 
                    \left[
                        \sum_{m \in M_{\xi_i}} P_{\xi_i}(m|\tau_{\xi_i}^t) \log P_{\xi_i}(m|\tau_{\xi_i}^t)
                    \right]
\end{aligned}
\end{equation}

Here, $H(m_{\xi_i}^t)$ is the entropy of the message distribution, $H(m_{\xi_i}^t | \tau_{\xi_i}^t)$ is the conditional entropy given the trajectory, $\overline{P_{\xi_i}}(m)$ as marginal distribution of message $m$ over all trajectories, and $P_{\xi_i}(m|\tau_{\xi_i}^t)$ as conditional distribution of message $m$ given the trajectory $\tau_{\xi_i}^t$. This way, the mutual information value reflects how much information the message carries about the agent's trajectory.

Lowe~et~al.~\cite{lowe.2019} built on this concept and provided the following formula for Speaker Consistency ($\operatorname{SC}$):

\begin{equation}
    \operatorname{SC}
    = \sum_{a \in A} \sum_{m \in M} P \left( a, m \right)
        \log \frac{ P \left( a, m \right) }{ P \left( a \right) P \left( m \right)}
\end{equation}

In this equation, $P \left( a, m \right)$ is the joint probability of action $a$ and message $m$, calculated empirically by averaging their co-occurrences across episodes. In general, SC is a valuable metric for evaluating whether the \ac{el} is both informative and aligned with the behavioral patterns of the sender.

\subsubsection{Positive Listening}\label{sec_metrics_pragmatics_positive_listening}
Positive listening evaluates the effectiveness of how a message receiver utilizes and applies incoming information~\cite{lowe.2019}. However, agents should not simply process messages similarly to other observations to avoid treating them as mere directives~\cite{cowenrivers.2020}. Nevertheless, the metrics presented in this section focus on evaluating the receiver's ability to effectively integrate and use the information received, rather than evaluating the receiver's ability to do more than just follow instructions. 

\paragraph{Instantaneous Coordination}\label{sec_metrics_pragmatics_positive_listening_instantaneous_coordination}
Instantaneous Coordination (IC), also referred to as listener consistency~\cite{lowe.2019}, was introduced by Jaques et~al.~\cite{jaques.2018} as a metric to evaluate how effectively an agent's message influences another agent's subsequent action. IC is computed similarly to Speaker Consistency (cf. Section~\ref{sec_metrics_pragmatics_positive_signaling_speaker_consistency}), but differs in that it measures the mutual information between one agent's message and the other agent's next action, averaged over episodes. This metric directly captures the receiver's immediate reaction to an incoming message, making it a measure of positive listening. However, it primarily captures situations where the receiver's action is directly changed by the sender's message, without considering the broader context or long-term dependencies~\cite{lowe.2019}. Accordingly, \enquote{IC can miss many positive listening relationships}~\cite{lowe.2019}.

Jaques et~al.~\cite{jaques.2018} proposed two specific measures for IC: One that quantifies the mutual information between the sender's message and the receiver's next action (see Equation~\ref{eq_metrics_pragmatics_positive_listening_IC_m_to_a}), and another one that measures the mutual information between the sender's current action and the receiver's next action (see Equation~\ref{eq_metrics_pragmatics_positive_listening_IC_a_to_a}). These measures are calculated by averaging over all trajectory steps and taking the maximum value between any two agents, focusing on short-term dependencies between consecutive timesteps.
\begin{equation}\label{eq_metrics_pragmatics_positive_listening_IC_m_to_a}
    IC_{m_{\xi_S} \to a_{\xi_R}}=I(m_k^t ;a_j^{t+1})
\end{equation}
\begin{equation}\label{eq_metrics_pragmatics_positive_listening_IC_a_to_a}
    IC_{a_{\xi_S} \to a_{\xi_R}}=I(a_k^t ;a_j^{t+1})
\end{equation}
A unified equation for $\operatorname{IC}$ is provided by Lowe~et~al.~\cite{lowe.2019}:
\begin{equation}
    \operatorname{SC}
    = \sum_{m_{\xi_{S}}^{t} \in M_{\xi_{S}}} \sum_{a_{\xi_{R}}^{t+1} \in A_{\xi_{R}}} 
        P \left( a_{\xi_{R}}^{t+1}, m_{\xi_{S}}^{t} \right)
        \log \frac{ P \left( a_{\xi_{R}}^{t+1}, m_{\xi_{S}}^{t} \right) }{ P \left( a_{\xi_{R}}^{t+1} \right) P \left( m_{\xi_{S}}^{t} \right)}
\end{equation}
Here, $P \left( a_{\xi_{R}}^{t+1}, m_{\xi_{S}}^{t} \right)$ is the empirical joint probability of the sender's message and the receiver's subsequent action, averaged over episodes within each epoch.

\paragraph{Message Effect}\label{sec_metrics_pragmatics_positive_listening_message_effect}
The Message Effect ($\operatorname{ME}$) metric, introduced by Bouchacourt and Baroni~\cite{bouchacourt.2019}, quantifies the influence of a message sent by one agent on the subsequent actions and messages of another agent. This metric explicitly considers bidirectional communication, so in the following we use generic agents $\xi_{A}$ and $\xi_{B}$ instead of sender and receiver. A notable challenge of this metric is the requirement for counterfactual analysis.

Given an agent $\xi_{A}$ at timestep $t$ sending a message $m_{\xi_{A}}^t$, we define $z_{\xi_{B}}^{t+1}$ as the combination of the action and message produced by agent $\xi_{B}$ at the following timestep. Accordingly, the conditional distribution $P \left( z_{\xi_{B}}^{t+1} \mid m_{\xi_A}^t \right)$ represents the response of $\xi_{B}$ to the message from $\xi_{A}$. To account for counterfactuals, which encode what might have happened had $\xi_{A}$ sent a different message $\widetilde{m}_{\xi_{A}}^t$, we define the counterfactual distribution $\widetilde{P} \left( z_{\xi_B}^{t+1} \right)$ (see Equation~\ref{metrics_pragmatics_positive_listening_ME_1}). 

The $\operatorname{ME}$ is then measured by the Kullback-Leibler divergence between the actual response and the counterfactual response (see Equation~\ref{metrics_pragmatics_positive_listening_ME_2}). The computation involves sampling $z_{\xi_{B}}^{t+1, k}$ from the conditional distribution for the actual message and sampling counterfactuals $\widetilde{m}_{\xi_{A}}^t$ to estimate $\widetilde{P} \left( z_{\xi_{B}}^{t+1, k} \right)$ (see Equation~\ref{metrics_pragmatics_positive_listening_ME_3}). The final ME is calculated as the average KL divergence over the collection of samples $K$ (see Equation~\ref{metrics_pragmatics_positive_listening_ME_4}).

\begin{equation}\label{metrics_pragmatics_positive_listening_ME_1}
    \widetilde{P} \left( z_{\xi_{B}}^{t+1} \right) 
        = \sum_{\widetilde{m}_{\xi_{A}}^t} P \left( z_{\xi_{B}}^{t+1} \mid \widetilde{m}_{\xi_{A}}^t \right)
            \widetilde{P} \left( \widetilde{m}_{\xi_{A}}^t \right)
\end{equation}

\begin{equation}\label{metrics_pragmatics_positive_listening_ME_2}
    \operatorname{ME_{\xi_{A} \to \xi_{B}}^t}
    = \operatorname{D_{KL}} 
        \left(
            P \left( z_{\xi_{B}}^{t+1} \mid m_{\xi_{A}}^t \right) 
                \parallel \widetilde{P} \left( z_{\xi_{B}}^{t+1} \right)
        \right)
\end{equation}

\begin{equation}\label{metrics_pragmatics_positive_listening_ME_3}
    \widetilde{P} \left( z_{\xi_{B}}^{t+1, k} \right) 
        = \sum_{j=1}^{J} P \left( z_{\xi_{B}}^{t+1, k} \mid \widetilde{m}_{\xi_{A}}^t \right) 
            \widetilde{P} \left( \widetilde{m}_{\xi_{A}}^t \right)
\end{equation}

\begin{equation}\label{metrics_pragmatics_positive_listening_ME_4}
    \operatorname{ME_{\xi_A \to \xi_B}^t} = \frac{1}{\lvert K \rvert} \sum_{k \in K}
        \log \frac{ P \left( z_{\xi_B}^{t+1, k} \mid m_{\xi_A}^t \right) }{ \widetilde{P} \left( z_{\xi_B}^{t+1, k} \right) }
\end{equation}

\paragraph{Causal Influence of Communication}\label{sec_metrics_pragmatics_positive_listening_causal_influence}
The Causal Influence of Communication (CIC) metric, introduced independently by Jaques et~al.~\cite{jaques.2018} and Lowe et~al.~\cite{lowe.2019}, provides a direct measure of positive listening by quantifying the causal effect that one agent's message has on another agent's behavior. Traditional methods of evaluating communication often fall short, as simply testing for a decrease in reward after removing the communication channel does not adequately capture the utility of communication~\cite{lowe.2019}.

CIC is computed using the mutual information between an agent's message and the subsequent action of the receiving agent. Unlike Instantaneous Coordination (cf. Section~\ref{sec_metrics_pragmatics_positive_listening_causal_influence}), CIC considers the probabilities $P \left( a , m \right) = \pi_{\xi_R} \left( a \mid m \right) \pi_{\xi_S} \left( m \right)$ that represent changes in the action distribution of the receiver $\xi_R$ when the message $m$ from the sender $\xi_S$ is altered. These probabilities are normalized within each game to accurately reflect the influence of messages on actions within the same context~\cite{lowe.2019}.

For multi-time-step causal influence, the CIC metric is defined as the difference between the entropy of the receiver's actions with and without communication:

\begin{equation}
    \operatorname{CIC}(\tau_{\xi_R}) = H(a_{\xi_R}^t | \tau_{\xi_R}) - H(a_{\xi_R}^t | \tau_{\xi_R}^{+M})
\end{equation}

Here, $\tau_{\xi_R}$ denotes the standard trajectory of the receiver, comprising state-action pairs, while $\tau_{\xi_R}^{+M}$ includes the communicated messages. The CIC is estimated by learning an approximate policy function $\pi(\cdot | \tau_{\xi_R})$. For more details on the multistep version, refer to Eccles~et~al.~\cite{eccles.2019}, and for the single-step version, see Jaques~et~al.~\cite{jaques.2018}.

\subsubsection{Symmetry}\label{sec_metrics_symmetry}
Symmetry in \ac{el} refers to consistent language use across agents in settings, where agents alternate between roles such as message sender and receiver~\cite{dubova.2020, cogswell.2019}. Thus, symmetry ensures convergence to a common language rather than distinct dialects~\cite{dubova.2020}.

\paragraph{Inter-Agent Divergence}\label{sec_metrics_symmetry_interagent}
Inter-Agent Divergence (IAD), introduced by Dubova et~al.~\cite{dubova.2020, dubova.2020b}, quantifies the similarity in how different agents map messages to actions. Let $a_{\xi_i}$ denote the action of agent $\xi_i$. The first step involves computing the marginal action distributions for each agent given a message $m$, represented as $P(a_{\xi_i}|m)$.

\begin{equation}
    P \left( a_{\xi_i} \mid m \right) \quad \forall \xi_i \in \xi \land m \in M
\end{equation}

The divergence between two agents, $\xi_i$ and $\xi_j$, based on their responses to the same message, is then calculated using the Jensen-Shannon Divergence (JSD) as follows:
\begin{equation}
\begin{aligned}
    \operatorname{D_{JS}} \left( \xi_i, \xi_j, m \right) 
        & = \operatorname{D_{JS}} 
            \left( 
                P\left(a_{\xi_i}|m\right) \parallel P\left(a_{\xi_j}|m\right)
            \right)
        \\
        & = \frac{1}{2} 
            \left[ 
                \operatorname{D_{KL}} \left( P \left( a_{\xi_i} \mid m \right) \parallel M\right) 
                +
                \operatorname{D_{KL}} \left( P \left( a_{\xi_j} \mid m \right) \parallel M \right) 
            \right]
        \\
        & \quad \text{where} \quad M = \frac{P \left( a_{\xi_i} \mid m \right) + P \left( a_{\xi_j} \mid m \right)}{2}
\end{aligned}   
\end{equation}

Finally, the overall IAD is computed by averaging these divergences across all possible agent pairs $\left( \xi_i, \xi_j \right) \in \xi_{\text{comb}}$ and messages $m \in M$:

\begin{equation}
    \operatorname{IAD} = 
        \frac{1}{|\xi_{\text{comb}}|} 
        \frac{1}{|M|} 
        \sum_{\left( \xi_i, \xi_j \right) \in \xi_{\text{comb}}} 
        \sum_{m \in M}
        \operatorname{D_{JS}}\left(\xi_i,\xi_j,m\right)
\end{equation}

While IAD effectively captures the consistency of inter-agent communication, it may have limitations when applied to more complex languages where message-level comparisons become difficult.

\paragraph{Within-Agent Divergence}\label{sec_metrics_symmetry_withinagent}
Within-Agent Divergence (WAD), proposed by Dubova et~al.~\cite{dubova.2020, dubova.2020b}, measures the consistency of an agent's communication behavior when it changes roles, such as from sender to receiver. This metric captures the internal symmetry in an agent's behavior and is crucial in complex systems where agents can assume different roles within the same environment. To compute WAD, we again first consider the action distribution $P \left( a_{\xi_i} \mid m \right)$ for each agent $\xi_i$ over a set of messages $m \in M$. This distribution reflects how an agent's actions are conditioned on receiving or sending a specific message.

\begin{equation}
    P \left( a_{\xi_i} \mid m \right) \quad \forall \xi_i \in \xi \land m \in M
\end{equation}

Given this, the Jensen-Shannon Divergence (JSD) is used to assess the divergence between an agent's behavior when acting as a sender $\xi_{i,S}$ versus as a receiver $\xi_{i,R}$:

\begin{equation}
\begin{aligned}
    \operatorname{D_{JS}} \left( \xi_{i,S} , \xi_{i,R} , m \right) 
        & = \operatorname{D_{JS}}
            \left( 
                P \left( a_{\xi_{i,S}} \mid m \right) \parallel P \left( a_{\xi_{i,R}} \mid m \right)
            \right)
        \\
        & = \frac{1}{2} 
            \left[ 
                \operatorname{D_{KL}} \left( P \left( a_{\xi_{i,S}} \mid m \right) \parallel Q \right) 
                +
                \operatorname{D_{KL}} \left( P \left( a_{\xi_{i,R}} \mid m \right) \parallel Q \right) 
            \right]
        \\
        & \quad \text{with} \quad Q 
                            = \frac{ P \left( a_{\xi_{i,S}} \mid m \right) + P \left( a_{\xi_{i,R}} \mid m \right) }{2}
\end{aligned}   
\end{equation}

Finally, the overall WAD is computed by averaging this divergence across all agents $\xi_i \in \xi$ based on the WAD for individual agents and their messages $m \in M_{\xi_i}$:
\begin{equation}
    \operatorname{WAD} = 
        \frac{1}{|\xi|} 
        \sum_{\xi_i \in \xi} 
        \frac{1}{|M_{\xi_i}|}
        \sum_{m \in M_{\xi_i}}
        \operatorname{D_{JS}}\left(\xi_{i,S},\xi_{i,R},m\right)
\end{equation}

\subsection{Summary of the Metrics}\label{sec_metrics_summary}
While some \ac{el} features are quantifiable by multiple metrics and have been investigated in multiple studies, others remain underexplored, as illustrated in Figure~\ref{fig_metrics_mindmap_with_sources} in Appendix~\ref{sec:appendix_b}. Metrics such as \emph{topographic similarity} and \emph{zero shot evaluation}, both of which assess semantic properties, are well established and widely utilized across multiple studies. In contrast, metrics related to pragmatics, such as \emph{speaker consistency} and \emph{instantaneous coordination}, are fairly well established but are less frequently used. Morphology metrics, particularly \emph{active words} and \emph{average message length}, are more commonly used, whereas syntax remains a peripheral concern, with only two isolated metrics proposed and not adopted in subsequent research. This imbalance indicates that while semantic metrics dominate \ac{el} research, morphology and pragmatics receive moderate attention, and syntax is mostly neglected.

Furthermore, the optimality of these metrics is not straightforward. Rather than being simply minimized or maximized, their ideal values are likely to lie at a nuanced balance point that varies depending on the specific \ac{el} system and application. This uncertainty leaves the critical question of what constitutes a \enquote*{good} \ac{el} system largely unanswered. Addressing this gap will require a deeper exploration of underrepresented metrics and a more refined understanding of how to evaluate \ac{el} systems holistically.


\section{Future Work}\label{sec_map}
In this section, we outline potential future directions for the research field of \ac{el}, based on our vision outlined in Section~\ref{sec_map_vision}. We present major research opportunities, organized along key research dimensions, in Section~\ref{sec_map_dimensions}. 

Along with future research directions, we have summarized a list of open source code repositories in Table~\ref{tab:code_repositories} in Appendix~\ref{sec:appendix_a} that can serve as convenient starting points for experimenting with these directions, for example, comprehensive frameworks such as the EGG toolkit~\cite{kharitonov.2019} and BabyAI~\cite{chevalierboisvert.2018} are included.

\subsection{Vision}\label{sec_map_vision}
Our vision for \ac{el} research is grounded in a functional perspective, aiming to achieve significant breakthroughs in human-agent interaction~\cite{bogin.2018, brandizzi.2022, gupta.2021, hu.2020, lazaridou.2020, lowe.2019, mul.2019}. This means developing communication systems that enable \ac{hci} at the human level, addressing the purpose, cost, and value of communication with intuitive and effective interfaces~\cite{brandizzi.2023b, karten.2022, tucker.2021, denamganai.2023, galke.2022, karten.2023b}. A key goal is to ensure that \acp{el} are grounded in real-world contexts, allowing agents to understand and interact with human-like comprehension and vice versa~\cite{agarwal.2019, das.2017, hermann.2017, lemon.2022}. This includes creating hierarchical, compositional conceptualization capabilities that allow agents to discuss and understand novel concepts in a structured, human-relevant manner~\cite{chaabouni.2022, ri.2023, sowik.2020b, woodward.2019b}. In addition, exploring the potential for AI explainability through communication is an exciting area~\cite{tucker.2021, agarwal.2019, kottur.2017b}. Finally, in the long term, creation and creativity through \ac{el} comparable to human capabilities would be a milestone. This would allow agents to truly communicate on a human level and enhance their ability to perceive and adapt to their environment through the use of language~\cite{colas.2020}.

\subsection{Dimensions and Opportunities}\label{sec_map_dimensions}
The development, evaluation, and application of \ac{el} in communication systems can be systematically analyzed along several critical dimensions. 
Given the relative youth of the field, with the majority of research emerging within the last eight years, specific areas of focus have gained prominence, particularly in the context of semantic metrics, such as topographic similarity and zero-shot evaluation, as highlighted in Figure~\ref{fig_metrics_mindmap_with_sources} in Appendix~\ref{sec:appendix_b}.
However, Figure~\ref{fig_publication_characteristics_year_panel_bar_chart} in Appendix~\ref{sec:appendix_b} illustrates that there is no discernible chronological trend or evolution in the way the different language characteristics are addressed.
This absence of a historical trajectory is likely attributable to the relatively brief history of the field and the considerable diversity of proposed approaches and methodologies.
Despite this, we identified nine key dimensions that, to the best of our knowledge, represent the primary areas of focus in \ac{el} research.

\begin{dimensions}
\enumdimension{Evaluation Metrics}

Evaluation metrics are essential for rigorously assessing the characteristics and effectiveness of \acp{el}. As detailed in our taxonomy (cf. Section~\ref{sec_taxonomy_languagecharacteristics}), we have identified key characteristics and their associated metrics. While some \ac{el} features are quantifiable through multiple metrics and have been examined in multiple studies, others remain underexplored, as illustrated in Figure~\ref{fig_metrics_mindmap_with_sources} in Appendix~\ref{sec:appendix_b}. We emphasize the need to develop comprehensive and quantitative metrics that accurately capture these features, which are critical to determining the practical utility of \acp{el}. Previous studies have similarly highlighted this need~\cite{lowe.2019, cope.2020, keresztury.2020, chaabouni.2022, ossenkopf.2022, abdelaziz.2024, choi.2018, chen.2023, li.2021b}.

In addition, further research is needed to systematically investigate existing metrics, especially with respect to their sensitivity to variations in settings, algorithms, and agent architectures~\cite{denamganai.2020, lowe.2019}. It is imperative that these metrics be subjected to more rigorous investigation to ensure that they enable meaningful quantitative comparisons and support well-founded conclusions about the capabilities and utility of \acp{el}. Thus, we endorse more comprehensive studies, more edge case testing and, in particular, more analysis of actual human-agent interaction. We see this as a critical priority for advancing the field.

\enumdimension{Emergent Language and Natural Language Alignment}

This dimension addresses the convergence and divergence between \ac{el} and \ac{nl}. A key approach to this challenge, discussed in Section~\ref{sec_taxnonomy_languageprior}, involves leveraging language priors to guide this alignment. Achieving robust \ac{el}-\ac{nl} alignment is essential for advancing human-agent interaction. Thus, future research should explore the integration of \ac{nl}-centered metrics and regularization techniques to enhance this alignment~\cite{yao.2022, lazaridou.2016}.

However, this alignment presents a fundamental dilemma. On the one hand, agents need the autonomy to develop languages organically, tailored to their specific interactions and requirements. On the other hand, to facilitate seamless human-agent communication, these \acp{el} must closely resemble \acp{nl}, which imposes significant constraints on their development. This tension creates what we call the Evolution-Acquisition Dilemma, where the evolutionary process fosters intrinsically motivated language emergence, while the acquisition process necessitates alignment with \ac{nl}. Balancing these competing needs is a critical challenge for future research in this area.

\enumdimension{Emergent Language and Large Language Models}

The remarkable performance of \acp{llm} on various benchmarks has established them as a cornerstone of modern \ac{nlp}~\cite{lappin.2023} and potential foundation for more complex agents~\cite{wang.2023, wang.2023b, wenqizhang.2024, zhao.2024}. Despite their success, however, \acp{llm} face fundamental limitations, particularly in grounding language use in shared environments and experiences~\cite{nottingham.2023} as well as agency~\cite{sharma.2024} and truthfulness~\cite{wang.2024}. Addressing these shortcomings may require insights from \ac{el} research. A key challenge in \ac{el} is the evolution-acquisition dilemma - the need to ground language in shared, incremental experiences, which current learning systems cannot achieve due to resource and technology constraints. While most \acp{llm} applications rely on fine-tuning~\cite{malladi.2023} and scaling~\cite{zhou.2024c}, these methods do not inherently address this challenge or the broader issues of grounding and adaptability.

One promising avenue lies in the concept of agentic \acp{llm}~\cite{chen.2023d, guo.2024d, nottingham.2023} or cognitive language agents~\cite{theodoresumers.2024}, which combine the representational strength of LLMs with the adaptive, experiential learning processes of \ac{rl}. Here, looking at opportunities for \ac{el} related research, \acp{llm} might act as language priors, providing a foundation that can be iteratively refined through agentic interaction and experience~\cite{nottingham.2023, shinn.2023}. This approach mirrors human language acquisition, where teachers provide guidance based on shared experience. In the absence of such a teacher, agentic \acp{llm} provide a synthetic framework for combining supervised and \ac{el} paradigms, allowing agents to relax static supervised training regimes and develop more adaptive communication protocols.

We argue for further exploration of cognitive language agents, focusing not only on established benchmarks but also on challenges central to \ac{el} research. Bridging these fields could open up new opportunities for developing systems that combine the scalability of \acp{llm} with the adaptability and grounding capabilities of \ac{el}.

\enumdimension{Representation Learning}

\ac{el} can be viewed as a complex representation learning task, focusing on how agents encode, interpret, and construct internal representations of observations and linguistic data. While representation learning is a well-established area in artificial intelligence research, its application in the context of \ac{el} remains underexplored. This dimension is central to the analysis of meaning and language space as outlined in our framework, which is based on the semiotic cycle (cf. Figure~\ref{fig_semiotic_cycle_EL_Notation}). Advancing this dimension requires advanced latent space analyses to elucidate the relationships between \acp{el}, underlying world models, and \ac{nl} structures. In addition, evaluating the impact of discrete versus continuous representations is critical to refining our understanding of \ac{el} dynamics.

Future research directions include developing methodologies to ensure that agent representations more accurately reflect the input they receive~\cite{bouchacourt.2018}, exploring efficient representation of (multimodal) information~\cite{zhu.2024}, conducting in-depth analyses to uncover and mitigate influencing factors and biases in learned representations~\cite{keresztury.2020}, and assessing the efficacy of these representations for downstream tasks~\cite{fitzgerald.2020}.

\enumdimension{Agent Design}

Agent design is a critical aspect in \ac{el} research, directly influencing the linguistic capabilities and adaptability of artificial agents. Prominent research directions include the investigation of advanced neural network architectures tailored for \ac{el}~\cite{chaabouni.2022, li.2020b, sowik.2020}, the creation of architectures optimized for heterogeneous and dynamic agent populations, and the refinement of structures that enhance language emergence and linguistic properties~\cite{fitzgerald.2020, kucinski.2021}. In addition, modular designs rather than monolithic ones potentially offer advantages by separating language processing from other task-specific computations. Addressing these design challenges is critical to advancing both \ac{el} research and broader artificial intelligence goals.

\enumdimension{Setting Design}

The environment in which agents operate is central to shaping the \ac{el}, encompassing interaction rules, agent goals, and communication dynamics (cf. Table~\ref{tab:communication_settings_overview}). This dimension is integral to the setting space outlined in our framework (cf. Figure~\ref{fig_semiotic_cycle_EL_Notation}).
Important future research directions include scaling up experimental settings to include larger and more complex tasks~\cite{chaabouni.2022, lazaridou.2016, cowenrivers.2020, bouchacourt.2019, eccles.2019, hardinggraesser.2019} with a focus on realistic perceptually grounded game environments~\cite{dubova.2020, bouchacourt.2019}. In addition, the study of the impact of populations as such~\cite{ossenkopf.2022} and the use of heterogeneous agent populations~\cite{rita.2022} are crucial areas of research. While some benchmarks have been established and utilized~\cite{kharitonov.2019, chevalierboisvert.2018}, there remains a significant need for the development and widespread dissemination of comprehensive benchmarks in area of research.

\enumdimension{Communication Design}

The design of the communication channel in \ac{el} systems is critical, focusing on how agents exchange and structure information through the channels available to them. This aspect is directly related to the phonetics and phonology components outlined in our taxonomy (cf. Section~\ref{sec_taxonomy_phonetics} and Section~\ref{sec_taxonomy_phonology}). For discrete \acp{el}, it is essential to establish channels that support word-based communication, with considerations such as vocabulary size and variable message length being fundamental to enabling effective and scalable human-agent interaction. Future research directions in this area include the exploration of topology-aware variable communication channels, the integration of heterogeneous channels within multi-agent systems, and the evolution of communication channels over time. Moreover, the incorporation of multimodal communication channels could provide more realistic and contextually rich stimuli, which may significantly enhance the sophistication and applicability of \acp{el} in \ac{nl}-oriented human-agent coordination~\cite{chaabouni.2022}.

\enumdimension{Learning Strategies}

Learning strategies focus on how agents acquire, adapt, and refine their linguistic capabilities over time, including the development of language rules and adaptation through interactions with other agents. While \ac{marl} serves as the foundational framework, there is significant potential to enhance the learning process through strategic design choices. Future research directions include the exploration of advanced regularization techniques~\cite{jaques.2018, yao.2022}, the adoption of tailored optimization strategies~\cite{chaabouni.2022}, and the integration of supervised or self-supervised learning objectives using appropriate loss designs~\cite{havrylov.2017, dessi.2021}. Additionally, the application of meta-learning~\cite{guo.2019}, decentralized learning approaches~\cite{moulinfrier.2020}, and curriculum learning methodologies~\cite{moulinfrier.2020} offer promising avenues for optimizing the \ac{el} learning process.

\enumdimension{Human-Agent Interaction}

The final dimension focuses on the interpretability of \acp{el} by humans and the degree to which humans can shape their development. This aspect is critical for creating human-agent interaction systems where communication is intuitive and effective~\cite{zhu.2024}. To advance this dimension, future research should prioritize the integration of human-in-the-loop feedback mechanisms to ensure that \acp{el} are not only practical, but also comprehensible to human users~\cite{brandizzi.2022, mordatch.2017}. This will improve the usability and adoption of these systems in real-world applications.

Key research directions include designing experiments that create incentives for agents to develop communication strategies more closely aligned with human language~\cite{bouchacourt.2018}. Additionally, exploring the resilience of communication protocols to deception through training with competing agents can lead to more robust and realistic interactions~\cite{noukhovitch.2021}. Exploring adaptive communication strategies to optimize the sparsity and clarity of messages based on individual or group needs within human-agent teams is another promising direction~\cite{karten.2023b}.
\end{dimensions}


\section{Limitations and Discussion}\label{sec_discussion}
In this section, we critically evaluate the limitations of our survey and identify areas for future improvement. Through our review, we aimed to develop a detailed taxonomy for the field of \ac{el}, focusing on its key properties (cf. Section~\ref{sec_taxonomyofemergentlanguage}), and to analyze as well as categorize quantification approaches and metrics (cf. Section~\ref{sec_metrics}). In addition, we curated a summary of open questions and suggestions for future research (cf. Section~\ref{sec_map}). Despite considerable efforts to establish a viable taxonomy and framework in the most systematic and unbiased manner, there are several potential limitations to our research approach and methodology.

First, while we have provided an extensive overview of \litRrelevant{} scientific publications in \ac{el} research, it is important to acknowledge that our search process, despite being thorough, may have overlooked significant contributions. Consequently, we do not claim completeness. However, we are very confident that our review represents a fair and well-balanced reflection of the existing body of work and the current state of the art.

Second, our review includes sources that are not peer-reviewed, such as preprints from \href{https://arxiv.org/}{arXiv}, to ensure that our work captures the most recent developments and diverse perspectives, including those that might be controversial. While we have carefully examined each paper included in this review, we cannot guarantee that every detail in non-peer-reviewed papers is entirely accurate. Consequently, we focused on concepts, findings, and metrics that are supported by multiple studies.

Third, we have introduced a taxonomy and a comprehensive metrics categorization for \ac{el} research, a field that is still in its early stages. This effort comes with inherent challenges, and while we have addressed many of these, it is important to note that our proposed framework does not represent a consensus within the wider research community. We are transparent about this limitation and encourage further discussion and validation.

Fourth, in order to maintain focus and conciseness, we have deliberately excluded ideas that lack associated metrics. As a result, some conceptual ideas from the reviewed research literature that are difficult to quantify in this early stage may not be fully explored in this survey.

Finally, we have incorporated several existing metrics into our proposed framework. While many of these metrics are well established in the field, we acknowledge that a more rigorous and critical experimental evaluation of these metrics would be beneficial. We strongly recommend that future research conduct such evaluations to further refine and validate the tools and methods used in \ac{el} research.


\section{Conclusion}\label{sec:Conclusion}
In this paper, we present a comprehensive taxonomy of \acf{el}, an overview of applicable metrics, and a summary of open challenges and potential research directions. Additionally, we provide a list of open source code repositories of the field in Table~\ref{tab:code_repositories} in Appendix~\ref{sec:appendix_a}. Our overall goal is to create a standardised yet dynamic framework that not only facilitates progress in this area of research, but also stimulates further interest and exploration.

Section~\ref{sec_background} introduces the foundational linguistic concepts that underpin our taxonomy. Section~\ref{sec_taxonomyofemergentlanguage} offers a comprehensive taxonomy of \ac{el} based on the review of \litRrelevant{} scientific publications. Section~\ref{sec_metrics} presents a unified categorization and notation for various metrics, depicted in Figure~\ref{fig_metrics_mindmap}, ensuring consistency and clarity. Section~\ref{sec_map} provides a summary of current achievements and outlines research opportunities.

By providing a structured overview and systematic categorization of linguistic concepts relevant to \ac{el} we have created a common ground for research and discussion. The detailed presentation of metrics and their unified notation ensures readability and usability, making it easier for researchers to navigate related topics and identify potential research opportunities and blind spots of future publications and the research field as a whole. This survey provides a valuable perspective on the development and analysis of \ac{el}, serving as both a guide and a resource for advancing this area of study.

\ac{el} is a fascinating and promising way to achieve grounded and goal-oriented communication among agents and between humans and agents. Despite its significant progress in recent years, the field faces many open questions and requires further evaluation methods and metrics. Critical questions remain about the measurability of linguistic features, the validity of proposed metrics, their utility, and their necessity. Aligning \ac{el} with \acf{nlp} for \acf{hci} presents additional opportunities and challenges. We encourage continued contributions and interdisciplinary research to address these issues and advance the field.


\backmatter








\section*{Declarations}


\begin{itemize}
\item Funding: We acknowledge the funding of the internships of Arya Gopikrishnan and Gustavo Adolpho Lucas De Carvalho by the German Academic Exchange Service (DAAD) project \href{https://www.daad.de/rise/en/rise-germany/}{\enquote*{RISE Germany}}.
\item Conflict of interest/Competing interests (check journal-specific guidelines for which heading to use): Not applicable.
\item Ethics approval and consent to participate: Not applicable.
\item Consent for publication: Not applicable.
\item Data availability: Not applicable.
\item Materials availability: Not applicable.
\item Code availability: Not applicable.
\item Author contribution: J.P., H.T. and T.M. had the idea for the article. J.P. performed the literature search and data analysis. The first draft of the manuscript was written by J.P. with the continuous support of C.W.d.P. and H.T. The first draft of the metrics section was written by A.G. All authors commented on earlier versions of the manuscript and critically revised the final manuscript.
\end{itemize}







\begin{appendices}

\renewcommand\theHtable{appendix-link-\arabic{table}} 
\renewcommand\theHfigure{appendix-link-\arabic{figure}} 

\clearpage
\section{Additional Tables}\label{sec:appendix_a}


\begin{longtblr}[
  caption = {List of code repositories for the literature reviewed.},
  label = {tab:code_repositories},
]{
  colspec = {rX}, width = \linewidth,
  rowhead = 1,
}
\hline
Paper & Link  \\
\hline
\cite{agarwal.2019} & \url{https://github.com/agakshat/visualdialog-pytorch} \\
\cite{andreas.2019} & \url{https://github.com/jacobandreas/tre} \\
\cite{auersperger.2022} & \url{https://github.com/facebookresearch/EGG/tree/main/egg/zoo/compo_vs_generalization_ood } \\
\cite{bachwerk.2011} & \url{https://github.com/arski/LEW} \\
\cite{blumenkamp.2020} & \url{https://github.com/proroklab/adversarial_comms} \\
\cite{bogin.2018} & \url{https://github.com/benbogin/emergence-communication-cco/ } \\
\cite{boldt.2022} & \url{https://github.com/brendon-boldt/filex-emergent-language} \\
\cite{boldt.2022d} & \url{https://github.com/brendon-boldt/filex-emergent-language} \\
\cite{bouchacourt.2018} & \url{https://github.com/DianeBouchacourt/SignalingGame } \\
\cite{bouchacourt.2019} & \url{https://github.com/facebookresearch/fruit-tools-game} \\
\cite{brandizzi.2021} & \url{https://github.com/nicofirst1/rl_werewolf} \\
\cite{chaabouni.2019} & \url{https://github.com/facebookresearch/EGG/blob/master/egg/zoo/channel/README.md} \\
\cite{chaabouni.2019b} & \url{https://github.com/facebookresearch/brica} \\
\cite{chaabouni.2020} & \url{https://github.com/facebookresearch/EGG/blob/master/egg/zoo/compo_vs_generalization/README.md} \\
\cite{chaabouni.2022} & \url{https://github.com/deepmind/emergent_communication_at_scale} \\
\cite{chevalierboisvert.2018} & \url{https://github.com/mila-iqia/babyai/tree/master} \\
\cite{chowdhury.2020b} & \url{https://github.com/AriChow/EL} \\
\cite{cogswell.2019} & \url{https://github.com/mcogswell/evolang} \\
\cite{colas.2020} & \url{https://github.com/flowersteam/Imagine} \\
\cite{cope.2020} & \url{https://github.com/DylanCope/zero-shot-comm } \\
\cite{dagan.2020} & \url{https://github.com/gautierdag/cultural-evolution-engine} \\
\cite{das.2017} & \url{https://github.com/batra-mlp-lab/visdial-rl} \\
\cite{denamganai.2020} & \url{https://github.com/Near32/ReferentialGym} \\
\cite{denamganai.2020b} & \url{https://github.com/Near32/ReferentialGym/tree/master/zoo/referential-games\%2Bst-gs} \\
\cite{denamganai.2023} & \url{https://github.com/Near32/Regym/tree/develop-ETHER/benchmark/ETHER} \\
\cite{denamganai.2023b} & \url{https://github.com/Near32/ReferentialGym/tree/develop/zoo/referential-games\%2Bcompositionality\%2Bdisentanglement} \\
\cite{dessi.2021} & \url{https://github.com/facebookresearch/EGG/tree/main/egg/zoo/emcom_as_ssl } \\
\cite{downey.2023} & \url{https://github.com/CLMBRs/communication-translation} \\
\cite{dubova.2020} & \url{https://github.com/blinodelka/Multiagent-Communication-Learning-in-Networks} \\
\cite{evtimova.2017} & \url{https://github.com/nyu-dl/MultimodalGame} \\
\cite{fitzgerald.2020} & \url{https://github.com/jacopotagliabue/On-the-plurality-of-graphs} \\
\cite{foerster.2018c} & \url{https://github.com/alshedivat/lola } \\
\cite{guo.2019} & \url{https://github.com/Shawn-Guo-CN/EmergentNumerals} \\
\cite{guo.2020} & \url{https://github.com/Shawn-Guo-CN/GameBias-EmeCom2020} \\
\cite{guo.2021b} & \url{https://github.com/uoe-agents/Expressivity-of-Emergent-Languages} \\
\cite{hazra.2020} & \url{https://github.com/SonuDixit/gComm } \\
\cite{jimenezromero.2023} & \url{https://github.com/Meta-optimization/emergent_communication_in_agents} \\
\cite{kang.2020} & \url{https://fringsoo.github.io/pragmatic_in2_emergent_papersite/ } \\
\cite{kharitonov.2019} & \url{https://github.com/facebookresearch/EGG } \\
\cite{kharitonov.2019b} & \url{https://github.com/facebookresearch/EGG/tree/master/egg/zoo/language_bottleneck} \\
\cite{kharitonov.2020} & \url{https://github.com/facebookresearch/EGG/tree/master/egg/zoo/compositional_efficiency} \\
\cite{korbak.2019} & \url{https://github.com/tomekkorbak/compositional-communication-via-template-transfer} \\
\cite{korbak.2020} & \url{https://github.com/tomekkorbak/measuring-non-trivial-compositionality} \\
\cite{kottur.2017b} & \url{https://github.com/batra-mlp-lab/lang-emerge} \\
\cite{lee.2017b} & \url{https://github.com/facebookresearch/translagent} \\
\cite{lei.2023b} & \url{https://github.com/MediaBrain-SJTU/ECISQA} \\
\cite{li.2020b} & \url{https://github.com/cambridgeltl/ECNMT} \\
\cite{liang.2020} & \url{https://github.com/pliang279/Competitive-Emergent-Communication} \\
\cite{lin.2021} & \url{https://github.com/ToruOwO/marl-ae-comm} \\
\cite{lipinski.2022} & \url{https://github.com/olipinski/rl_werewolf} \\
\cite{lipinski.2024} & \url{https://anonymous.4open.science/r/TPG-916B} \\
\cite{lowe.2019} & \url{https://github.com/facebookresearch/measuring-emergent-comm} \\
\cite{lowe.2020} & \url{https://github.com/backpropper/s2p} \\
\cite{mihai.2021b} & \url{https://github.com/Ddaniela13/LearningToDraw} \\
\cite{mu.2021} & \url{https://github.com/jayelm/emergent-generalization} \\
\cite{noukhovitch.2021} & \url{https://github.com/mnoukhov/emergent-compete } \\
\cite{ohmer.2022} & \url{https://github.com/XeniaOhmer/hierarchical_reference_game} \\
\cite{ohmer.2022b} & \url{https://github.com/XeniaOhmer/language_perception_communication_games} \\
\cite{patel.2021} & \url{https://github.com/saimwani/CoMON} \\
\cite{perkins.2021} & \url{https://github.com/asappresearch/compositional-inductive-bias} \\
\cite{portelance.2021} & \url{https://github.com/evaportelance/emergent-shape-bias} \\
\cite{ren.2020} & \url{https://github.com/Joshua-Ren/Neural_Iterated_Learning } \\
\cite{resnick.2020} & \url{https://github.com/backpropper/cbc-emecom} \\
\cite{rita.2022b} & \url{https://github.com/MathieuRita/Population} \\
\cite{ropke.2021} & \url{https://github.com/wilrop/communication_monfg} \\
\cite{saha.2019} & \url{https://github.com/Homagn/MultiAgentRL} \\
\cite{simoes.2020} & \url{https://github.com/david-simoes-93/A3C3 } \\
\cite{simoes.2020b} & \url{https://github.com/david-simoes-93/A3C3 } \\
\cite{steinertthrelkeld.2019} & \url{https://github.com/shanest/function-words-context} \\
\cite{steinertthrelkeld.2022} & \url{https://github.com/CLMBRs/communication-translation} \\
\cite{sukhbaatar.2016} & \url{https://github.com/facebookarchive/CommNet} \\
\cite{ueda.2023} & \url{https://github.com/mynlp/emecom_SignalingGame_as_betaVAE} \\
\cite{unger.2020} & \url{https://github.com/thomasaunger/babyai_sr } \\
\cite{vanderwal.2020b} & \url{https://github.com/i-machine-think/emergent_grammar_induction} \\
\cite{wang.2019b} & \url{https://github.com/TonghanWang/NDQ} \\
\cite{wu.2021} & \url{https://github.com/jimmyyhwu/spatial-intention-maps} \\
\cite{xu.2022} & \url{https://github.com/wildphoton/Compositional-Generalization} \\
\cite{yao.2022} & \url{https://github.com/ysymyth/ec-nl } \\
\cite{zheng.2017} & \url{https://github.com/geek-ai/Magent} \\
\end{longtblr}

\begin{table*}[!ht]
\caption{Overview of language prior usage in the reviewed literature.}
\label{tab:language_prior}
\begin{tabularx}{\textwidth}{lX}
\toprule
Language prior & Paper  \\
\midrule
No - Evolution & \citenums{abdelaziz.2024, ampatzis.2008, andreas.2019, auersperger.2022, baronchelli.2006, blumenkamp.2020, bogin.2018, boldt.2022, boldt.2022d, bosc.2022, botokoekila.2024, botokoekila.2024b, bouchacourt.2018, bouchacourt.2019, brandizzi.2021, bullard.2020, bullard.2021, carmeli.2022, carmeli.2024, chaabouni.2019, chaabouni.2020, chaabouni.2022, chen.2023, choi.2018, chowdhury.2020, chowdhury.2020b, chowdhury.2020c, cogswell.2019, cope.2020, cowenrivers.2020, dagan.2020, das.2018, denamganai.2020b, denamganai.2023b, dessi.2019, dessi.2021, dubova.2020, eccles.2019, evtimova.2017, feng.2024, fitzgerald.2019, fitzgerald.2020, guo.2019, guo.2019b, guo.2020, guo.2021b, gupta.2020, gupta.2020b, hagiwara.2019, hagiwara.2021, hardinggraesser.2019, hildreth.2019, jaques.2018, jimenezromero.2023, kajic.2020, kalinowska.2022, kang.2020, karten.2022, karten.2023, karten.2023b, keresztury.2020, kharitonov.2019, kharitonov.2019b, korbak.2019, korbak.2020, kottur.2017b, kubricht.2023, kucinski.2020, kucinski.2021, lazaridou.2018, li.2019, li.2020b, li.2021b, liang.2020, lin.2021, lipinski.2022, lipinski.2024, lo.2022, lobostsunekawa.2022, lorkiewicz.2011, lowe.2019, luna.2020, mihai.2019, mihai.2021, mihai.2021b, mordatch.2017, mu.2021, mu.2023, mul.2019, nakamura.2023, noukhovitch.2021, ohmer.2022, ohmer.2022b, ossenkopf.2022, patel.2021, pesce.2020b, portelance.2021, ren.2020, resnick.2020, ri.2023, rita.2022, rita.2022b, saha.2019, santamariapang.2019, santamariapang.2020, simoes.2020, simoes.2020b, sirota.2019, sowik.2020, sowik.2020b, steinertthrelkeld.2019, sukhbaatar.2016, taylor.2021, thomas.2021, tieleman.2019, tucker.2022, ueda.2022, unger.2020, vanderwal.2020b, vanneste.2022, vanneste.2022b, verma.2021, villanger.2024, wang.2019b, xu.2022, yao.2022, yu.2022, yu.2023, yuan.2019b, yuan.2024} \\
Yes - Acquistion & \citenums{agarwal.2019, buck.2018, chaabouni.2019b, chevalierboisvert.2018, colas.2020, denamganai.2023, downey.2023, gupta.2021, hazra.2020, kharitonov.2020, kolb.2019, lee.2017b, lei.2023b, lowe.2020, nevens.2020, perkins.2021, qiu.2021, resnick.2018, steinertthrelkeld.2022, ueda.2023, vani.2021, verma.2019, woodward.2019b, wu.2021} \\
Both & \citenums{cao.2018, das.2017, eloff.2021, havrylov.2017, lazaridou.2016, tucker.2021, yuan.2019} \\
\bottomrule
\end{tabularx}
\end{table*}

\begin{table*}[!htbp]
\caption{Overview of language characteristics discussed in the reviewed literature.}
\label{tab:characteristics_distribution}
\begin{tabularx}{\textwidth}{lX}
\toprule
Characteristic & Paper  \\
\midrule
Morphology & \citenums{boldt.2022d, brandizzi.2023b, carmeli.2022, carmeli.2024, chaabouni.2019, choi.2018, chowdhury.2020b, dagan.2020, denamganai.2020, denamganai.2023b, dessi.2021, havrylov.2017, kajic.2020, kharitonov.2019b, lazaridou.2016, lazaridou.2018, lei.2023b, li.2019, loreto.2016, lorkiewicz.2011, luna.2020, mordatch.2017, ohmer.2022, resnick.2020, rita.2022, sowik.2020b, ueda.2022, vanderwal.2020b, yu.2022} \\
Syntax & \citenums{ueda.2022, vanderwal.2020b} \\
Semantics & \citenums{agarwal.2019, andreas.2019, auersperger.2022, bogin.2018, boldt.2022, bosc.2022, bouchacourt.2018, bouchacourt.2019, brandizzi.2023b, bullard.2020, bullard.2021, cao.2018, carmeli.2024, chaabouni.2020, chaabouni.2022, choi.2018, chowdhury.2020b, cogswell.2019, colas.2020, cowenrivers.2020, dagan.2020, das.2017, denamganai.2020, denamganai.2020b, denamganai.2023, denamganai.2023b, dessi.2019, dessi.2021, downey.2023, dubova.2020, eccles.2019, eloff.2021, feng.2024, guo.2019, guo.2019b, guo.2020, guo.2021b, gupta.2020, gupta.2021, havrylov.2017, hazra.2020, hildreth.2019, jaques.2018, kajic.2020, kang.2020, keresztury.2020, kharitonov.2020, korbak.2019, korbak.2020, kottur.2017b, kubricht.2023, kucinski.2020, kucinski.2021, lazaridou.2016, lazaridou.2018, lazaridou.2020, lei.2023b, li.2019, li.2021b, liang.2020, lipinski.2024, lo.2022, loreto.2016, lorkiewicz.2011, lowe.2019, luna.2020, mihai.2019, mihai.2021b, mordatch.2017, mu.2021, mu.2023, mul.2019, ohmer.2022, ossenkopf.2022, patel.2021, perkins.2021, qiu.2021, ren.2020, resnick.2020, ri.2023, rita.2022, rita.2022b, santamariapang.2019, santamariapang.2020, sowik.2020, sowik.2020b, taylor.2021, tieleman.2019, tucker.2021, ueda.2022, ueda.2023, unger.2020, vani.2021, verma.2019, xu.2022, yao.2022, yu.2022, yu.2023} \\
Pragmatics & \citenums{ampatzis.2008, bouchacourt.2019, brandizzi.2023b, cao.2018, cogswell.2019, cope.2020, cowenrivers.2020, dubova.2020, eccles.2019, jaques.2018, kalinowska.2022, lazaridou.2020, li.2021b, liang.2020, lin.2021, lowe.2019, mul.2019, noukhovitch.2021, patel.2021, portelance.2021, sowik.2020, sowik.2020b, zhu.2024} \\
\bottomrule
\end{tabularx}
\end{table*}

\clearpage
\section{Additional Figures}\label{sec:appendix_b}

%
\begin{figure}[hbp]%
\centering
\includegraphics[width=0.95\textwidth]{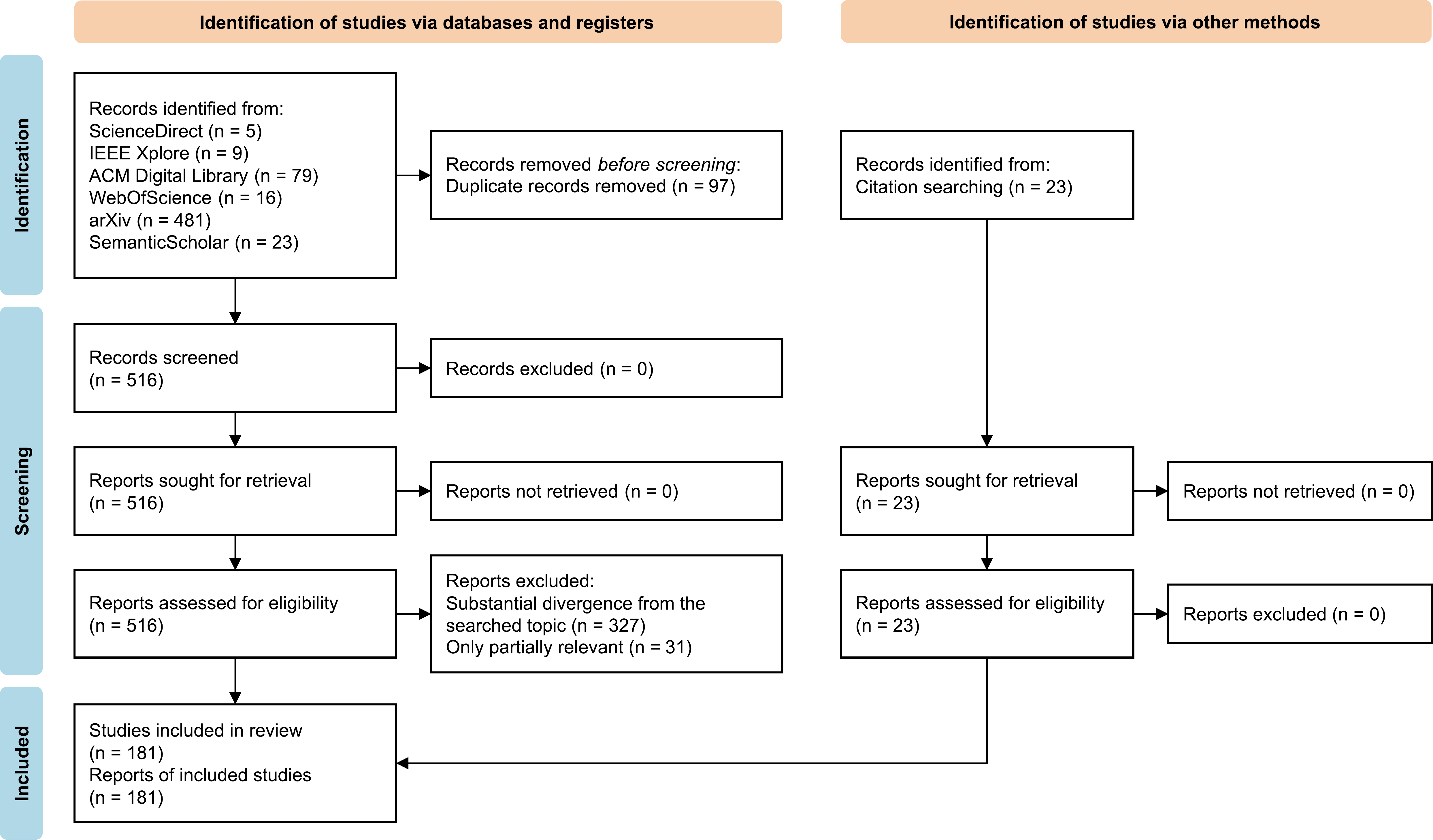}
\caption{PRISMA 2020 flow diagram for new systematic reviews for the present survey. Adapted from~\protect\cite{page.2021}.}
\label{fig_PRISMA_flowchart}
\end{figure}
%
{
    \tikzset{octagon/.style={regular polygon,regular polygon sides=8},}
    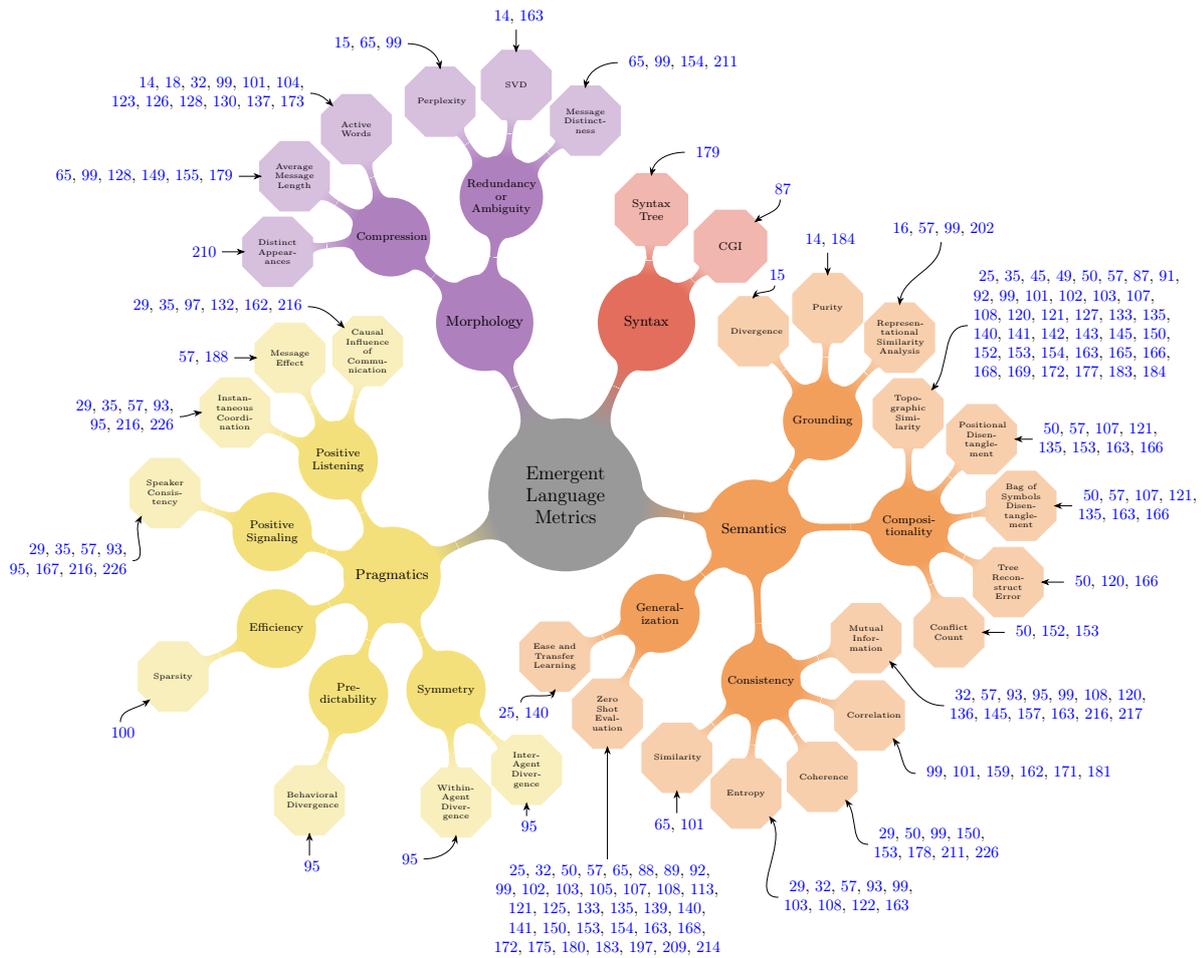
\begin{figure}[hbp]
    \centering
    \resizebox{\columnwidth}{!}{%
        \begin{tikzpicture}
        \begin{scope}[mindmap, grow cyclic, every node/.style=concept, concept color=JP-lighter_grey,
            level 1 concept/.append style={level distance=4.2cm,sibling angle=-(360/4)), minimum size=2.0cm},
            level 2 concept/.append style={level distance=2.8cm, sibling angle=-45, minimum size=1.7cm},
            level 3 concept/.append style={level distance=2.5cm, sibling angle=-40, minimum size=1.0cm, outer sep=-0.1em, concept/.append style={octagon}}]
        \hypersetup{hidelinks}
        \node[style={minimum size=3cm, text width=3cm}]{Emergent\\ Language\\ Metrics}
            child [concept color=JP-purple, rotate = -20] { 
                node{\hyperref[sec_metrics_morphology]{Morphology}}
                child { 
                    node{\hyperref[sec_metrics_compression]{Compression}}
                    child [concept color=JP-purple_50p, rotate = 10] { node [alias=mo_co_da]{\hyperref[sec_metrics_compression_distinct_appearances]{Distinct Appearances}}}
                    child [concept color=JP-purple_50p, rotate = 10] { node [alias=mo_co_aml]{\hyperref[sec_metrics_compression_average_message_length]{Average Message Length}}}
                    child [concept color=JP-purple_50p, rotate = 10] { node [alias=mo_co_aw]{\hyperref[sec_metrics_compression_active_words]{Active Words}}}
                }
                child [rotate = -10] { 
                    node {\hyperref[sec_metrics_redundancy_or_ambiguity]{Redundancy or \mbox{Ambiguity}}}
                    child [concept color=JP-purple_50p] { node [alias=mo_re_pe] {\hyperref[sec_metrics_redundancy_or_ambiguity_perplexity]{Perplexity}}}
                    child [concept color=JP-purple_50p] { node [alias=mo_re_svd]{\hyperref[sec_metrics_redundancy_or_ambiguity_svd]{SVD}}}
                    child [concept color=JP-purple_50p] { node [alias=mo_re_md]{\hyperref[sec_metrics_redundancy_or_ambiguity_message_distinctness]{Message Distinctness}}}
                }
            }
            child [concept color=JP-red, rotate = 20] { 
                node {\hyperref[sec_metrics_syntax]{Syntax}}
                child [concept color=JP-red_50p, level distance=2.5cm, outer sep=-0.1em] { node[octagon, text width=2.5em, alias=sy_tree]{\hyperref[sec_metrics_syntax_syntax_tree]{Syntax \\ Tree}}}
                child [concept color=JP-red_50p, level distance=2.5cm, outer sep=-0.1em] { node[octagon, text width=2.5em, alias=sy_cgi]{\hyperref[sec_metrics_syntax_CGI]{CGI}}}
            }
            child [concept color=JP-orange, rotate = 20, sibling angle=-60] { 
                node {\hyperref[sec_metrics_semantics]{Semantics}}
                child { 
                    node {\hyperref[sec_metrics_grounding]{Grounding}}
                    child [concept color=JP-orange_50p, rotate = 30] { node [alias=se_g_div] {\hyperref[sec_metrics_grounding_divergence]{Divergence}}}
                    child [concept color=JP-orange_50p, rotate = 30] { node [alias=se_g_pur]{\hyperref[sec_metrics_grounding_purity]{Purity}}}
                    child [concept color=JP-orange_50p, rotate = 30] { node [alias=se_g_rsa]{\hyperref[sec_metrics_grounding_rsa]{Represen\-tational \mbox{Similarity} Analysis}}}
                }
                child [rotate = -12, level distance = 3.4cm] { 
                    node {\hyperref[sec_metrics_compositionality]{Composi\-tionality}}
                    child [concept color=JP-orange_50p, rotate = 10] { node [alias=se_com_topsim]{\hyperref[sec_metrics_compositionality_topsim]{Topo\-graphic Similarity}}}
                    child [concept color=JP-orange_50p, rotate = 10] { node [alias=se_com_posdis]{\hyperref[sec_metrics_compositionality_posdis]{Positional Disentanglement}}}
                    child [concept color=JP-orange_50p, rotate = 10] { node [alias=se_com_bosdis]{\hyperref[sec_metrics_compositionality_bosdis]{Bag of Symbols Disentanglement}}}
                    child [concept color=JP-orange_50p, rotate = 10] { node [alias=se_com_tre]{\hyperref[sec_metrics_compositionality_tre]{Tree Reconstruct Error}}}
                    child [concept color=JP-orange_50p, rotate = 10] { node [alias=se_com_cc]{\hyperref[sec_metrics_compositionality_conflictcount]{Conflict Count}}}
                }
                child [rotate = -55, level distance = 3.4cm] { 
                    node {\hyperref[sec_metrics_consistency]{Consistency}}
                    child [concept color=JP-orange_50p, rotate = 30] { node  [alias=se_con_mi]{\hyperref[sec_metrics_consistency_mutualinformation]{Mutual Information}}}
                    child [concept color=JP-orange_50p, rotate = 30] { node [alias=se_con_cor] {\hyperref[sec_metrics_consistency_correlation]{Correlation}}}
                    child [concept color=JP-orange_50p, rotate = 30] { node [alias=se_con_coh]{\hyperref[sec_metrics_consistency_coherence]{Coherence}}}
                    child [concept color=JP-orange_50p, rotate = 30] { node [alias=se_con_ent]{\hyperref[sec_metrics_consistency_entropy]{Entropy}}}
                    child [concept color=JP-orange_50p, rotate = 30] { node [alias=se_con_sim]{\hyperref[sec_metrics_consistency_similarity]{Similarity}}}
                }
                child [rotate = -60] { 
                    node {\hyperref[sec_metrics_generalization]{General\-ization}}
                    child [concept color=JP-orange_50p] { node [alias=se_gen_zse]{\hyperref[sec_metrics_generalization_zero_shot]{Zero Shot Evaluation}}}
                    child [concept color=JP-orange_50p] { node [alias=se_gen_etl] {\hyperref[sec_metrics_generalization_ETL]{Ease and Transfer Learning}}}
                }
            }
            child [concept color=JP-yellow, rotate = -20] { 
                node {\hyperref[sec_metrics_pragmatics]{Pragmatics}}
                child { 
                    node {\hyperref[sec_metrics_symmetry]{Symmetry}}
                    child [concept color=JP-yellow_50p] { node [alias=pr_sym_inter] {\hyperref[sec_metrics_symmetry_interagent]{Inter-Agent Divergence}}}
                    child [concept color=JP-yellow_50p] { node [alias=pr_sym_within]{\hyperref[sec_metrics_symmetry_withinagent]{Within-Agent Divergence}}}
                }
                child { 
                    node {\hyperref[sec_metrics_pragmatics_predictability]{Pre\-dictability}}
                    child [concept color=JP-yellow_50p] { node [alias=pr_pre_bd]{\hyperref[sec_metrics_pragmatics_predictability_behavioral_divergence]{\mbox{Behavioral} \mbox{Divergence}}}}
                }
                child { 
                    node {\hyperref[sec_metrics_pragmatics_efficiency]{Efficiency}}
                    child [concept color=JP-yellow_50p] { node [alias=pr_eff_sp] {\hyperref[sec_metrics_pragmatics_efficiency_sparsity]{Sparsity}}}
                }
                child { 
                    node {\hyperref[sec_metrics_pragmatics_positive_signaling]{Positive Signaling}}
                    child [concept color=JP-yellow_50p] { node [alias=pr_ps_sc]{\hyperref[sec_metrics_pragmatics_positive_signaling_speaker_consistency]{Speaker Consistency}}}
                }
                child { 
                    node {\hyperref[sec_metrics_pragmatics_positive_listening]{Positive Listening}}
                    child [concept color=JP-yellow_50p] { node [alias=pr_pl_ic]{\hyperref[sec_metrics_pragmatics_positive_listening_instantaneous_coordination]{Instan\-taneous Coordination}}}
                    child [concept color=JP-yellow_50p] { node [alias=pr_pl_me]{\hyperref[sec_metrics_pragmatics_positive_listening_message_effect]{Message Effect}}}
                    child [concept color=JP-yellow_50p] { node [alias=pr_pl_cic]{\hyperref[sec_metrics_pragmatics_positive_listening_causal_influence]{Causal \mbox{Influence} of Communication}}}
                }
            }; 
        \end{scope}
            \node[align=right, left=5mm of mo_co_da](l_mo_co_da) {\citenums{loreto.2016}};
            \draw[-{Stealth[bend]}] (l_mo_co_da) to[out=0,in=180] (mo_co_da);
            \node[align=right, left=5mm of mo_co_aml](l_mo_co_aml) {\citenums{chaabouni.2019, choi.2018, kharitonov.2019b, lei.2023b, luna.2020, vanderwal.2020b}};
            \draw[-{Stealth[bend]}] (l_mo_co_aml) to[out=0,in=180] (mo_co_aml);
            \node[align=right, text width=50mm, above left=-2mm and 5mm of mo_co_aw](l_mo_co_aw) {
                \citenums{boldt.2022d, carmeli.2022, chaabouni.2019, chowdhury.2020b, dagan.2020, dessi.2021, lazaridou.2016, luna.2020, mordatch.2017, resnick.2020, sowik.2020b, yu.2022}
            };
            \draw[-{Stealth[bend]}] (l_mo_co_aw) to[out=0,in=135] (mo_co_aw);
            \node[align=right, above left=5mm and 2mm of mo_re_pe](l_mo_re_pe) {\citenums{choi.2018, havrylov.2017, luna.2020}};
            \draw[-{Stealth[bend]}] (l_mo_re_pe) to[out=0,in=90] (mo_re_pe);
            \node[align=right, above=5mm of mo_re_svd](l_mo_re_svd) {\citenums{lazaridou.2016, ohmer.2022}};
            \draw[-{Stealth[bend]}] (l_mo_re_svd) to[out=270,in=90] (mo_re_svd);
            \node[align=right, above right=5mm and 2mm of mo_re_md](l_mo_re_md) {\citenums{choi.2018, lazaridou.2018, lorkiewicz.2011, luna.2020}};
            \draw[-{Stealth[bend]}] (l_mo_re_md) to[out=180,in=90] (mo_re_md);
            \node[align=right, above right=5mm and 2mm of sy_tree](l_sy_tree) {\citenums{vanderwal.2020b}};
            \draw[-{Stealth[bend]}] (l_sy_tree) to[out=180,in=90] (sy_tree);
            \node[align=right, above right=5mm and 2mm of sy_cgi](l_sy_cgi) {\citenums{ueda.2022}};
            \draw[-{Stealth[bend]}] (l_sy_cgi) to[out=270,in=45] (sy_cgi);
            \node[align=left, above right=5mm and -4mm of se_g_div](l_se_g_div) {\citenums{havrylov.2017}};
            \draw[-{Stealth[bend]}] (l_se_g_div) to[out=270,in=90] (se_g_div);
            \node[align=center, above=5mm of se_g_pur](l_se_g_pur) {\citenums{lazaridou.2016, yu.2023}};
            \draw[-{Stealth[bend]}] (l_se_g_pur) to[out=270,in=90] (se_g_pur);
            \node[align=left, above right=16mm and -9mm of se_g_rsa](l_se_g_rsa) {\citenums{bouchacourt.2018, brandizzi.2023b, luna.2020, tieleman.2019}};
            \draw[-{Stealth[bend]}] (l_se_g_rsa) to[out=270,in=90] (se_g_rsa);
            \node[align=left, text width=50mm, above right=1mm and 8mm of se_com_topsim](l_se_com_topsim) {\citenums{andreas.2019, auersperger.2022, brandizzi.2023b, carmeli.2024, chaabouni.2020, chaabouni.2022, dagan.2020, denamganai.2020, denamganai.2020b, denamganai.2023b, feng.2024, guo.2019, guo.2019b, guo.2020, hazra.2020, keresztury.2020, korbak.2019, korbak.2020, kucinski.2020, kucinski.2021, lazaridou.2018, lazaridou.2020, li.2019, luna.2020, mu.2021, ohmer.2022, ossenkopf.2022, perkins.2021, ren.2020, ri.2023, rita.2022, rita.2022b, sowik.2020, ueda.2022, ueda.2023, xu.2022, yao.2022, yu.2023}};
            \draw[-{Stealth[bend]}] (l_se_com_topsim) to[out=180,in=45] (se_com_topsim);
            \node[align=left, text width=50mm, right=4mm of se_com_posdis](l_se_com_posdis) {\citenums{auersperger.2022, brandizzi.2023b, chaabouni.2020, denamganai.2023b, korbak.2020, kucinski.2021, ohmer.2022, perkins.2021}};
            \draw[-{Stealth[bend]}] (l_se_com_posdis) to[out=180,in=0] (se_com_posdis);
            \node[align=left, text width=30mm, right=4mm of se_com_bosdis](l_se_com_bosdis) {\citenums{auersperger.2022, brandizzi.2023b, chaabouni.2020, denamganai.2023b, korbak.2020, ohmer.2022, perkins.2021}};
            \draw[-{Stealth[bend]}] (l_se_com_bosdis) to[out=180,in=0] (se_com_bosdis);
            \node[align=left, right=5mm of se_com_tre](l_se_com_tre) {\citenums{andreas.2019, korbak.2020, perkins.2021}};
            \draw[-{Stealth[bend]}] (l_se_com_tre) to[out=180,in=0] (se_com_tre);
            \node[align=left, right=5mm of se_com_cc](l_se_com_cc) {\citenums{korbak.2020, kucinski.2020, kucinski.2021}};
            \draw[-{Stealth[bend]}] (l_se_com_cc) to[out=180,in=0] (se_com_cc);
            \node[align=left, text width=50mm, below right=5mm and 12mm of se_con_mi](l_se_con_mi) {\citenums{andreas.2019, brandizzi.2023b, dessi.2019, dubova.2020, eccles.2019, gupta.2020, hazra.2020, liang.2020, lipinski.2024, luna.2020, mu.2021, ohmer.2022, yu.2022}};
            \draw[-{Stealth[bend]}] (l_se_con_mi) to[out=180,in=315] (se_con_mi);
            \node[align=left, text width=50mm, below right=5mm and 5mm of se_con_cor](l_se_con_cor) {\citenums{dagan.2020, luna.2020, mihai.2019, mul.2019, santamariapang.2020, verma.2019}};
            \draw[-{Stealth[bend]}] (l_se_con_cor) to[out=180,in=315] (se_con_cor);
            \node[align=left, text width=50mm, below right=5mm and 5mm of se_con_coh](l_se_con_coh) {\citenums{jaques.2018, korbak.2019, korbak.2020, kucinski.2021, lorkiewicz.2011, lowe.2019, luna.2020, unger.2020}};
            \draw[-{Stealth[bend]}] (l_se_con_coh) to[out=180,in=315] (se_con_coh);
            \node[align=left, text width=50mm, below right=13mm and 2mm of se_con_ent](l_se_con_ent) {\citenums{boldt.2022, brandizzi.2023b, liang.2020, lowe.2019, luna.2020, mu.2021, ohmer.2022, rita.2022, yu.2022}};
            \draw[-{Stealth[bend]}] (l_se_con_ent) to[out=180,in=315] (se_con_ent);
            \node[align=center, below=5mm of se_con_sim](l_se_con_sim) {\citenums{choi.2018, dagan.2020}};
            \draw[-{Stealth[bend]}] (l_se_con_sim) to[out=90,in=270] (se_con_sim);
            \node[align=center, text width=50mm, below=25mm of se_gen_zse](l_se_gen_zse) {\citenums{auersperger.2022, brandizzi.2023b, bullard.2020, bullard.2021, chaabouni.2020, chaabouni.2022, choi.2018, colas.2020, denamganai.2020b, denamganai.2023b, eloff.2021, feng.2024, guo.2019, guo.2021b, hildreth.2019, korbak.2019, korbak.2020, kottur.2017b, kucinski.2021, lazaridou.2018, luna.2020, mu.2021, ohmer.2022, qiu.2021, ren.2020, ri.2023, rita.2022, sowik.2020, taylor.2021, tucker.2021, vani.2021, xu.2022, yao.2022, yu.2022}};
            \draw[-{Stealth[bend]}] (l_se_gen_zse) to[out=90,in=270] (se_gen_zse);
            \node[align=center, below left=5mm and -5mm of se_gen_etl](l_se_gen_etl) {\citenums{chaabouni.2022, feng.2024}};
            \draw[-{Stealth[bend]}] (l_se_gen_etl) to[out=90,in=270] (se_gen_etl);
            \node[align=center, below=3mm of pr_sym_inter](l_pr_sym_inter) {\citenums{dubova.2020}};
            \draw[-{Stealth[bend]}] (l_pr_sym_inter) to[out=90,in=270] (pr_sym_inter);
            \node[align=center, below left=5mm and 2mm of pr_sym_within](l_pr_sym_within) {\citenums{dubova.2020}};
            \draw[-{Stealth[bend]}] (l_pr_sym_within) to[out=0,in=270] (pr_sym_within);
            \node[align=center, below=5mm of pr_pre_bd](l_pr_pre_bd) {\citenums{dubova.2020}};
            \draw[-{Stealth[bend]}] (l_pr_pre_bd) to[out=90,in=270] (pr_pre_bd);
            \node[align=right, below left=5mm and 2mm of pr_eff_sp](l_pr_eff_sp) {\citenums{kalinowska.2022}};
            \draw[-{Stealth[bend]}] (l_pr_eff_sp) to[out=90,in=225] (pr_eff_sp);
            \node[align=right, text width=30mm, below left=5mm and 2mm of pr_ps_sc](l_pr_ps_sc) {\citenums{brandizzi.2023b, dubova.2020, eccles.2019, lazaridou.2020, liang.2020, lowe.2019, portelance.2021, jaques.2018}};
            \draw[-{Stealth[bend]}] (l_pr_ps_sc) to[out=0,in=225] (pr_ps_sc);
            \node[align=right, text width=30mm, below left=-9mm and 7mm of pr_pl_ic](l_pr_pl_ic) {\citenums{brandizzi.2023b, dubova.2020, eccles.2019, jaques.2018, lazaridou.2020, liang.2020, lowe.2019}};
            \draw[-{Stealth[bend]}] (l_pr_pl_ic) to[out=0,in=180] (pr_pl_ic);
            \node[align=right, left=5mm of pr_pl_me](l_pr_pl_me) {\citenums{bouchacourt.2019, brandizzi.2023b}};
            \draw[-{Stealth[bend]}] (l_pr_pl_me) to[out=0,in=180] (pr_pl_me);
            \node[align=right, above left=2mm and 8mm of pr_pl_cic](l_pr_pl_cic) {\citenums{cowenrivers.2020, lazaridou.2020, lin.2021, lowe.2019, mul.2019, eccles.2019}};
            \draw[-{Stealth[bend]}] (l_pr_pl_cic) to[out=0,in=135] (pr_pl_cic);
        \end{tikzpicture}
    }
    \caption{
        Graph presenting a visual representation of the metrics identified in the surveyed literature, sorted by language characteristics. Each node contains a link to the corresponding section that describes the metric in detail and a list of references for each metric is given at each node.
        \\
        \tikzcircle[JP-purple]{}\tikzcircle[JP-red]{}\tikzcircle[JP-orange]{}\tikzcircle[JP-yellow]{}: Language characteristics (inner nodes)
        \\
        \tikzoctagon[JP-purple_50p]{}\tikzoctagon[JP-red_50p]{}\tikzoctagon[JP-orange_50p]{}\tikzoctagon[JP-yellow_50p]{}: Individual metrics (outermost nodes)
    }
    \label{fig_metrics_mindmap_with_sources}
    \end{figure}
}

\begin{figure}[btp]%
\centering
\includegraphics[width=\textwidth]{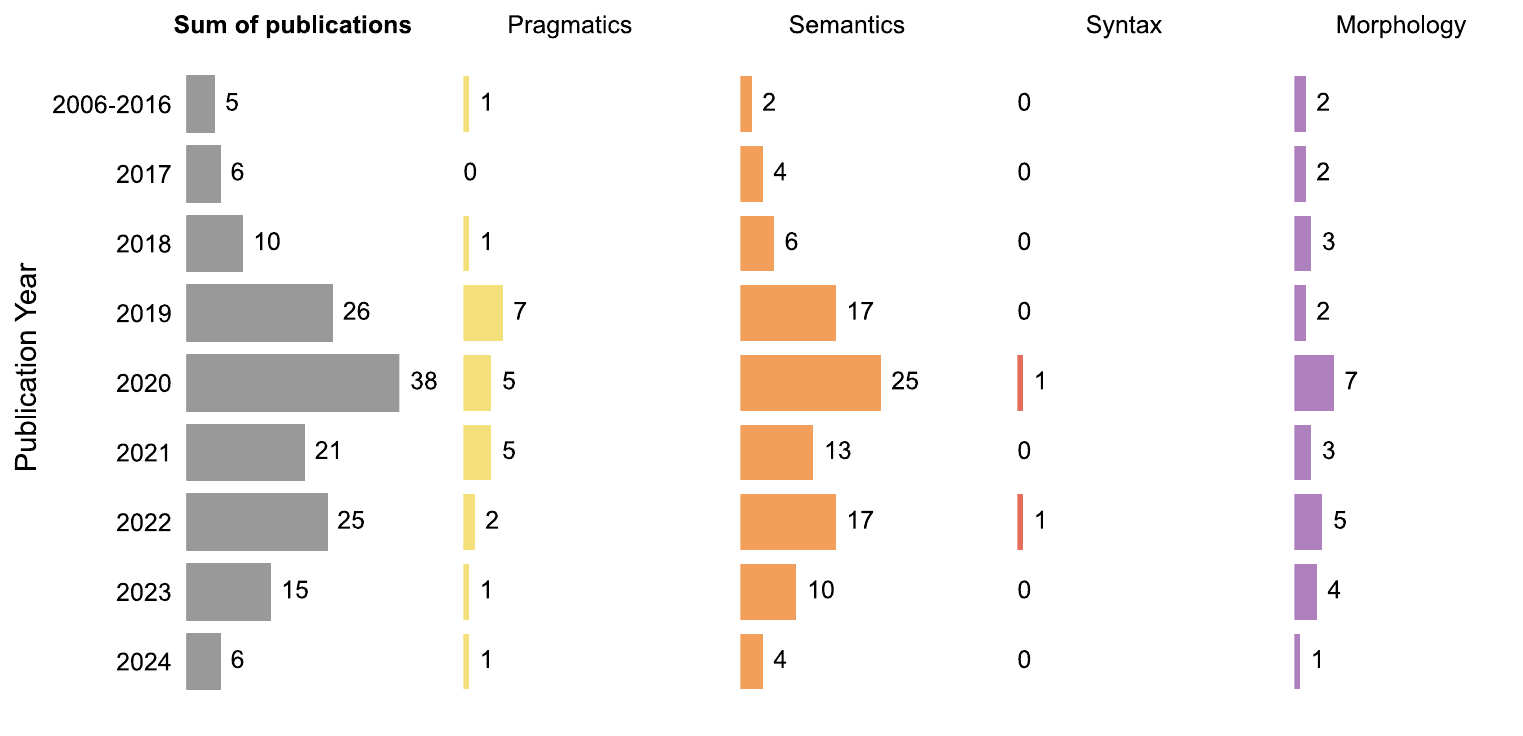}
\caption{Number of publications sorted by year and by language characteristics analyzed. The number of publications analyzing individual language characteristics per year is provided in the leftmost column, and the distribution across analyzed language characteristics is shown in the remaining columns.}
\label{fig_publication_characteristics_year_panel_bar_chart}
\end{figure}

\end{appendices}


\clearpage
\bibliography{20250122_el_survey}

\end{document}